\documentclass[fleqn,usenatbib]{mnras}

\usepackage[T1]{fontenc}

\DeclareRobustCommand{\VAN}[3]{#2}
\let\VANthebibliography\thebibliography
\def\thebibliography{\DeclareRobustCommand{\VAN}[3]{##3}\VANthebibliography}

\usepackage{subcaption}
\usepackage{fontawesome}

\usepackage{graphicx}	        
\usepackage{amsmath}	        
\usepackage{amssymb}	        
\usepackage[dvipsnames]{xcolor} 

\usepackage{newtxtext,newtxmath}

\newcommand*\dif{\mathop{}\!\mathrm{d}}

\title[Machine-learning-augmented random fields]{Fast and realistic large-scale structure from machine-learning-augmented random field simulations}

\author[D. Piras et al.]{
Davide Piras,$^{1, 2}$\thanks{\href{mailto:d.piras@ucl.ac.uk}{d.piras@ucl.ac.uk}}
Benjamin Joachimi,$^{1}$ and Francisco Villaescusa-Navarro$^{3, 4}$
\\
$^{1}$Department of Physics and Astronomy, University College London, Gower Street, London, WC1E 6BT, UK\\
$^{2}$Département de Physique Théorique, Université de Genève, 24 quai Ernest Ansermet, 1211 Genève 4, Switzerland\\
$^{3}$Center for Computational Astrophysics, Flatiron Institute, 162 5th Avenue, New York, NY, USA 10010\\
$^{4}$Department of Astrophysical Sciences, Princeton University, 4 Ivy Lane, Princeton, NJ 08544 USA}

\date{Accepted XXX. Received YYY; in original form ZZZ}

\pubyear{2022}

\begin{document}
\label{firstpage}
\pagerange{\pageref{firstpage}--\pageref{lastpage}}
\maketitle

\begin{abstract}
Producing thousands of simulations of the dark matter distribution in the Universe with increasing precision is a challenging but critical task to facilitate the exploitation of current and forthcoming cosmological surveys. Many inexpensive substitutes to full $N$-body simulations have been proposed, even though they often fail to reproduce the statistics of the smaller, non-linear scales. Among these alternatives, a common approximation is represented by the lognormal distribution, which comes with its own limitations as well, while being extremely fast to compute even for high-resolution density fields. In this work, we train a generative deep learning model, mainly made of convolutional layers, to transform projected lognormal dark matter density fields to more realistic dark matter maps, as obtained from full $N$-body simulations. We detail the procedure that we follow to generate highly correlated pairs of lognormal and simulated maps, which we use as our training data, exploiting the information of the Fourier phases. We demonstrate the performance of our model comparing various statistical tests with different field resolutions, redshifts and cosmological parameters, proving its robustness and explaining its current limitations. When evaluated on 100 test maps, the augmented lognormal random fields reproduce the power spectrum up to wavenumbers of $1 \ h \ \rm{Mpc}^{-1}$, and the bispectrum within 10\%, and always within the error bars, of the fiducial target simulations. Finally, we describe how we plan to integrate our proposed model with existing tools to yield more accurate spherical random fields for weak lensing analysis.
\end{abstract}




\section{Introduction}
\label{sec:intro}
The best current model to describe our Universe is the $\Lambda$CDM model, which prescribes the existence of a cosmological constant $\Lambda$ associated with dark energy, together with cold dark matter (CDM) and ordinary matter (baryons; see e.g.\ \citealp{Dodelson03}). In particular, the $\Lambda$CDM model predicts that dark matter is about five times more abundant than ordinary matter, with galaxies forming along the cosmic web structure woven by dark matter, made of filaments connecting different clusters, all surrounded by voids. While its gravitational effects are observed by many probes, dark matter remains a mystery, with multiple experiments still ongoing to shed light on its nature (see e.g.\ \citealp{Trimble87, Bertone05, Buchmueller17, deSwart17}, and references therein).

The most common tool to analyse and track the origin and evolution of dark matter structures are cosmological $N$-body simulations (\citealp{Holmberg41, Navarro96,Tormen97, Jenkins98, Springel05a, Springel05b, Boylan09, Angulo12, VillaescusaNavarro20b, VillaescusaNavarro20, Chacon20}, and references therein). In its basic formulation, an $N$-body simulation is run by putting a certain number of massive particles in a cubic box, imposing periodic boundary conditions and letting gravity be the only force acting on the particles through its gravitational potential, governed by the Poisson equation \citep{Springel05a}. The initial conditions of the Universe are usually approximated with a Gaussian density field,\footnote{Hereafter, ``initial conditions'' will only refer to the initial conditions of $N$-body simulations.} which can be entirely summarised by a given power spectrum, i.e.\ by the Fourier counterpart of the correlation function between different particles in the simulation. Starting from high redshift, the position and velocity of the particles are updated iteratively until today ($z=0$), while various snapshots are taken at different redshifts. 

Several methods to run an $N$-body simulation are available, with different levels of complexity, approximation, and speed \citep{Hockney88, Chacon20}. These include the direct resolution of the equation of motion for each particle \citep{Mikkola93}, approximated methods like the tree code method \citep{Barnes86, Callahan92}, or mean-field approaches like standard \citep{Klypin97} or adaptive \citep{Shea04} particle mesh. In general, though, $N$-body simulations are computationally expensive to run, and usually require access to high performance computing hardware. This limits the possibility of fully exploring the impact of different cosmological parameters on the dark matter evolution in our Universe, and hinders statistical analyses of the large-scale structure (see e.g.\ \citealp{Taylor13, Taylor14}): $N$-body simulations are essential to associate a covariance matrix to real measurements, and thousands of simulations are required to obtain accurate estimates of such matrices. 

In recent years, many cheaper approximations have been proposed, which try to capture both the large-scale structure of the cosmic web and its smaller-scale details. These approximations often rely on Lagrangian perturbation theory \citep{Buchert92, Buchert93, Buchert94}, and can produce accurate dark matter halo mock catalogues and dark matter density fields \citep{Monaco02, Monaco13, White14, Kitaura13, Chuang14, Tassev13, Tassev15, Howlett15, Rizzo17, Tosone20, Tosone21}. While being capable of capturing the large-scale-structure statistics with fewer computational resources, these methods usually fail to accurately produce the correct small-scale statistics. Although such approximate mocks are used to estimate covariance matrices in current large-volume datasets when not enough high-resolution simulations are available, to date no inexpensive exact alternative to $N$-body realisations exists.

\begin{figure*}
	\includegraphics[width=2\columnwidth]{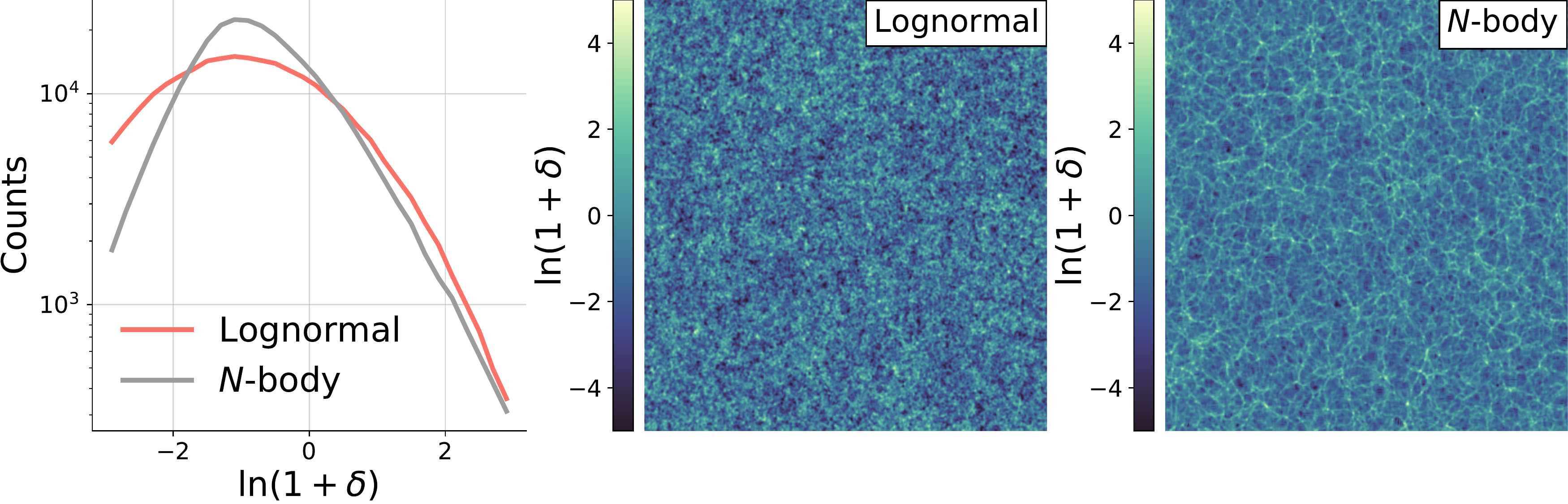}
    \caption{\textit{Left panel}: histograms of the matter overdensity $\delta$, defined in Eq.~(\ref{eq:overdensity}), for a lognormal random field (red) and an $N$-body simulation dark matter density field (grey). \textit{Middle and right panels}: square maps of a lognormal (middle) and $N$-body (right) density fields, with a side of 512 pixels, corresponding to a comoving length of $1$ $h^{-1} \ \rm{Gpc}$. The depth of these fields is $\simeq 1.9$ $h^{-1}$ Mpc. In these maps, we clipped the maximum and minimum values before applying a logarithm to reduce their dynamic range; the symbol `$\ln$' indicates the natural logarithm throughout this paper. The right-hand-side plot is a slice of a simulation from the Quijote suite \citep{VillaescusaNavarro20}, while the middle plot, obtained following the procedure described in Sect.~\ref{sec:training_data}, represents its lognormal counterpart. The goal of this paper is to train a machine learning model (described in Sect.~\ref{sec:iti_translation}) to transform the lognormal map to the more realistic $N$-body map, thus improving the statistical power of the fast lognormal approximation.}
    \label{fig:histo_with_density}
\end{figure*}

Another typical approximation to describe (dark) matter fields is found by resorting to a lognormal random field, which represents the simplest alternative to running an entire $N$-body simulation \citep{Coles91, Peebles93, Taruya02, Percival04, Hilbert11, Xavier16}. A lognormal random field can be easily obtained from a Gaussian random field (see Sect.~\ref{sec:training_data} for further details), and can be entirely described by a small number of parameters; moreover, a lognormal variable has a skewed distribution which is suited for e.g.\ the matter overdensity field, whose values range from -1 in voids to values much higher than 1 in clustered dense regions. However, as reported in \citet{Xavier16} and shown in Fig.~\ref{fig:histo_with_density}, the lognormal approximation comes with its own limitations, and fails to reproduce the correct matter density distribution.

Machine learning (ML) techniques have also been proposed to replace expensive $N$-body simulations. In \citet{Rodriguez18}, generative adversarial networks \citep[GANs,][]{Goodfellow14} were successfully trained to generate slices of $N$-body simulations, and \citet{Mustafa19} applied the same technique to weak lensing convergence maps. \citet{Perraudin19} and \citet{Feder20} then extended the application of GANs to 3-D boxes, proving that, while challenging to train, GANs can capture both large- and small-scale features, and are capable of accurately recovering the statistical information contained in the training data. \citet{He19} and \citet{Oliveira20}, on the other hand, showed that it is possible to train a U-shaped neural network architecture \citep[U-net,][]{Ronneberger15} to map simple linear initial conditions to the corresponding final evolved fields, correctly learning the non-linear growth of structures under the gravitational influence. \citet{Kaushal21} additionally used Lagrangian perturbation theory to evolve such initial conditions and only learn the difference in the density fields at $z=0$. In these latter works, it was also shown that such architectures can perform well even on input data obtained from different cosmological parameters than the training data, thus demonstrating the appealing feature of being able to extrapolate outside the training distribution. Other works have explored the use of super-resolution techniques to $N$-body simulations \citep{Ramanah20, Li21}, the application of normalising flows \citep[e.g.][]{Papamakarios19} as generative models of the large-scale structure \citep{Rouhiainen21, Dai22}, wavelet phase harmonics statistics to produce realistic 2-D density fields \citep{Allys20}, or combinations of ML-inspired techniques with more traditional methods to improve the accuracy of fast $N$-body solvers \citep{Dai18, Dai20, Dai21, Bohm20}.

While being useful, all the previous approaches still require a relatively high amount of computational resources, might not scale well to high-resolution fields, or introduce many approximations that prevent them from being used reliably in place of full $N$-body simulations. In this paper, we show that it is possible to improve the lognormal approximation by means of ML techniques, with the long-term goal of integrating our approach with the \textit{Full-sky Lognormal Astro-fields Simulation Kit} \citep[\textsc{FLASK},][]{Xavier16}, in order to be able to cheaply generate more realistic high-resolution full-sky density fields. 

For this purpose, we start from the Quijote $N$-body simulation suite \citep{VillaescusaNavarro20}, which offers thousands of realisations of a single cosmological parameterisation, as well as hundreds of simulations at different values of the cosmological parameters. We devise a pipeline to create lognormal density fields which are the approximated counterpart of the simulated density fields. By construction, these lognormal fields have the same power spectrum as the one from the fiducial $N$-body simulations, and the phases of the underlying Gaussian fields are taken from the initial conditions of the simulated fields (all details are reported in Sect.~\ref{sec:training_data}). Having the pairs of lognormal and corresponding simulated density fields, we draw from image-to-image translation techniques based on convolutional neural networks and adversarial training, in order to obtain a model that can map simple lognormal fields to more realistic density fields (see Fig.~\ref{fig:histo_with_density}). We extensively validate our model by measuring first-, second-, and higher-order statistics, obtaining good agreement, almost always within 10\%, on all scales. We additionally show that we can train our model using simulations run over a latin hypercube of cosmological parameters, obtaining a good generalisation performance over different cosmologies. While a more extensive study for different redshifts and higher resolutions will be explored in future work, these encouraging results indicate that providing the model with a lognormal field as the starting point significantly improves the model's generalisation performance. Additionally, we show that starting from lognormal maps with the correct power spectrum naturally leads to good performance at the power spectrum level, and does not allow the model to "collapse" and learn single modes in the data, a known problem of generative adversarial networks \citep{Metz17}.

The paper is structured as follows. In Sect.~\ref{sec:data} we describe the Quijote simulation data, on which this work is based. In Sect.~\ref{sec:method}, we detail the procedure that we apply to obtain the training data, and describe the image-to-image translation technique that we employ in this work. In Sect.~\ref{sec:results}, we present the results for different resolutions of the density fields, as well as for different values of redshift and cosmological parameters, and demonstrate the performance of our model through a wide range of statistical tests. We conclude in Sect.~\ref{sec:conclusions} with a summary of our work, planned improvements and an outline of possible future applications of our model.


\section{Data}
\label{sec:data}
In this work, we use the Quijote simulation suite \citep{VillaescusaNavarro20}. This set of $N$-body simulations includes $15\,000$ realisations following 512$^3$ dark matter particles in a box with comoving length of $1$ $h^{-1} \ \rm{Gpc}$, with the matter density parameter $\Omega_{\rm{m}} = 0.3175$, the baryon density parameter $\Omega_{\rm{b}} = 0.049$, the Hubble parameter $h = 0.6711$, the scalar spectral index $n_{\rm{s}} = 0.9624$, the root mean square of the matter fluctuations in spheres of radius 8 $h^{-1}$ Mpc $\sigma_8 = 0.834$, and the dark energy equation of state parameter $w = -1$; neutrinos are considered massless. These simulations were run using the TreePM code Gadget-III, which is an improved version of Gadget-II \citep{Springel05a}. We consider snapshots of both the initial conditions ($z=127$) and today ($z=0$), as well the $z=1$ snapshot and the latin-hypercube simulations at $z=0$ for further validation of our model (see Sect.~\ref{sec:zc_dep}).

In each $N$-body simulation, we convert the information on the particles' position to a continuous random field through a mass assignment scheme. We analyse the matter overdensity field $\delta(\mathbf{x})$, defined as:
\begin{equation}
    \delta(\mathbf{x})=\frac{\rho(\mathbf{x})}{\bar{\rho}} -1 \ ,
	\label{eq:overdensity}
\end{equation}
with $\rho(\mathbf{x})$ being the matter density field at each position $\mathbf{x}$, and $\bar{\rho}$ being the mean density in the volume of the simulation. 

Following \citet{Chaniotis04, Jing05, Sefusatti16}, we consider a regular grid of points in all three directions. The continuous overdensity field is obtained by interpolating the discrete overdensity field on this grid, i.e.\ by evaluating the continuous function
\begin{equation}
    \tilde{\delta}(\mathbf{x})= \int \frac{\dif \mathbf{x'}}{(2\pi)^3} W(\mathbf{x} - \mathbf{x'}) \delta(\mathbf{x'}) \ ,
	\label{eq:integral_overdensity}
\end{equation}
with $W(\mathbf{x})$ being the weight function describing the number of grid points to which every particle is assigned. We choose the piecewise cubic spline interpolation scheme, i.e.\ we explicitly write the weight function as $W(\mathbf{x}) = W_{\rm{1D}}(x_1/H) W_{\rm{1D}}(x_2/H) W_{\rm{1D}}(x_3/H)$, with $H$ being the grid spacing, $x_1$ ($x_2$, $x_3$) being the $x$ ($y$, $z$) direction, and $W_{\rm{1D}}$ being the unidimensional weight function
\begin{equation}
W_{\rm{1D}}(s) = 
    \begin{cases}
      \frac{4-6s^2+3|s|^3}{6} & \rm{if} \ 0 \leq |s| < 1 \ ;\\
      \frac{(2-|s|)^3}{6} & \rm{if} \ 1 \leq |s| < 2 \ ;\\
      0 & \rm{otherwise} \ ;
    \end{cases}
	\label{eq:pcs}
\end{equation}
we refer the reader to \citet{Sefusatti16} for more details. We consider both a grid with $N^3_{\rm{high}} = 512^3$ pixels and $N^3_{\rm{low}} = 128^3$ pixels, and present the results in Sect.~\ref{sec:highres} and Sect.~\ref{sec:lowres}, respectively.


\section{Method}
\label{sec:method}
Our goal is to obtain 2-D projected density lognormal fields corresponding to slices of the Quijote simulations, in order to train a model that can take as input a lognormal map and predict a more realistic density field with the same statistics as the simulated one. In the following sections, we describe the procedure that we follow to obtain such a dataset (Sect.~\ref{sec:training_data}), and the machine learning algorithm that we employ to learn the 
transformation (Sect.~\ref{sec:iti_translation}).

\subsection{Obtaining the training data}
\label{sec:training_data}
Since the long-term goal of the project is to increase the accuracy in large-scale structure description of random field maps on the sphere like the ones produced by \textsc{FLASK} \citep{Xavier16}, we choose to work with slices of the density field rather than the full 3-D boxes. 
We slice a given box along the third axis, and obtain multiple square density fields from a single simulation (128 in the low-resolution case, and 512 in the high-resolution case); the width of each slice is $1000 \ h^{-1} \ \rm{Mpc} /128 \simeq 7.8$ $h^{-1}$ Mpc in the former case, and $1000 \ h^{-1}\ \rm{Mpc}  /512 \simeq 1.9$ $h^{-1}$ Mpc in the latter case. Our choice of different thicknesses aims to demonstrate that our approach can work at different resolutions and different projection depths. Since we consider 800 simulations in the low-resolution case, and 200 in the high-resolution case, we are left with $102\,400$ maps in both instances. We also consider the initial conditions of the 3-D boxes, namely the $N$-body simulations at $z=127$, which we slice in the same way. 

\begin{figure*}
	\includegraphics[width=2\columnwidth]{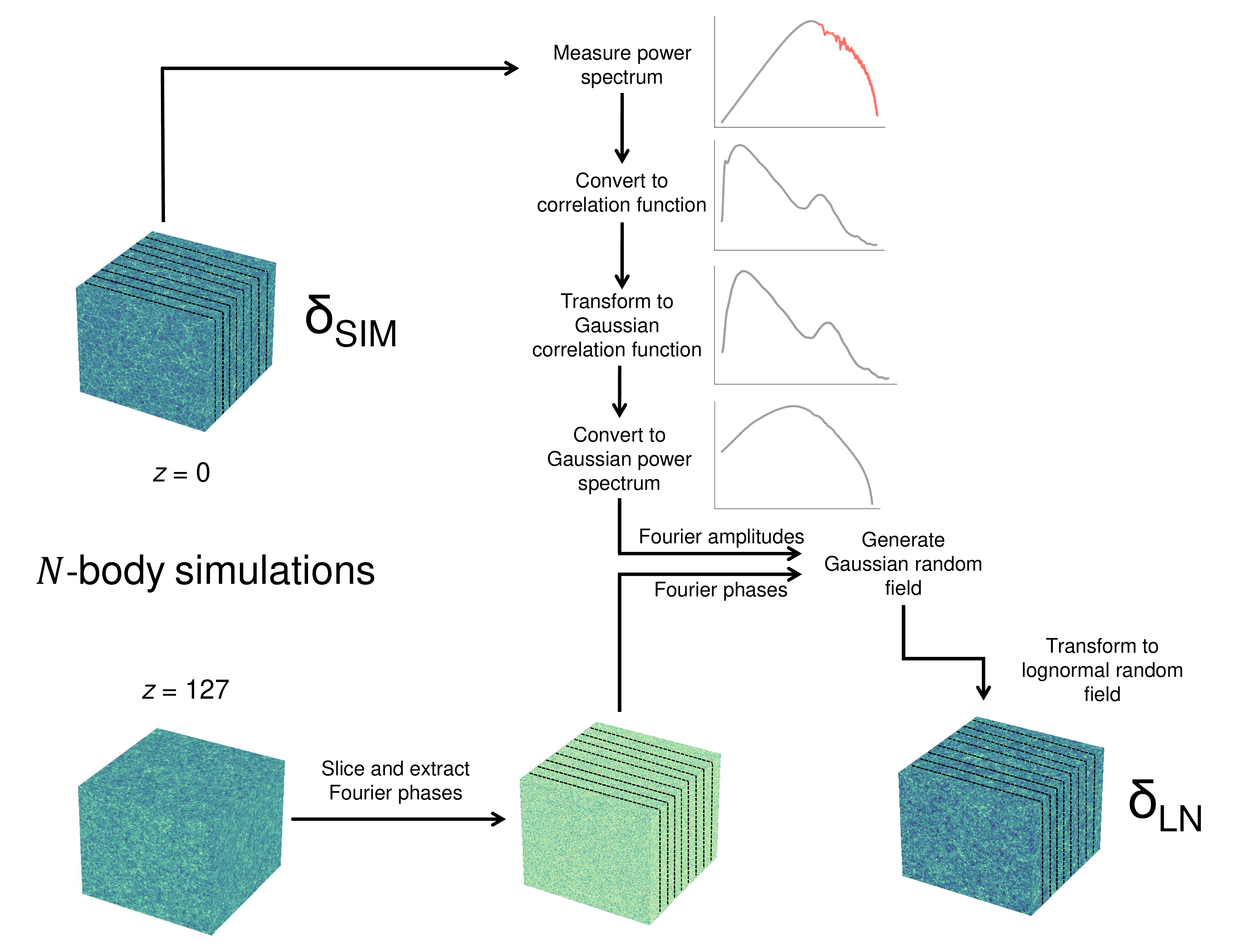}
    \caption{Flowchart of the steps to create the training data, as described in Sect.~\ref{sec:training_data}. We first measure the power spectrum of each slice of the $z=0$ boxes (in red in the top panel), which is concatenated with the theory power spectrum obtained using \textsc{CLASS} (\citealp{Blas11}; in grey in the top panel). We then generate a lognormal random field with this power spectrum, following \citet{Coles91, Percival04}; the term ``Gaussian correlation function'' indicates $\xi_{\rm{G}}(r)$ as in Eq.~(\ref{eq:g_corr_f}); the corresponding ``Gaussian power spectrum'' is obtained using Eq.~(\ref{eq:power}). Crucially, when generating the underlying Gaussian field, we use the Fourier phases of the initial conditions of the $N$-body simulation, which consist of a Gaussian random field at $z=127$. In this way, the lognormal field displays increased correlation with the $N$-body field. 
    The final training data consists of pairs of lognormal ($\delta_{\rm{LN}}$) and simulated ($\delta_{\rm{SIM}}$) density fields, with either low (side $N_{\rm{low}} = 128$) or high (side $N_{\rm{high}} = 512$) resolution, as explained in Sect.~\ref{sec:data}. The machine learning model employed to learn the mapping from $\delta_{\rm{LN}}$ to $\delta_{\rm{SIM}}$ is presented in Sect.~\ref{sec:iti_translation}.}
    \label{fig:flowchart}
\end{figure*}

In order to create the lognormal counterpart of the more realistic maps, we start by measuring the power spectrum of each simulation's slice at $z=0$, which we wish to impose on the lognormal fields. We recall here that the 2-D matter power spectrum $P(k)$ can be implicitly defined through the Fourier transform $\delta(\mathbf{k})$ of the matter density contrast $\delta(\mathbf{x})$\footnote{From now on, we use $\mathbf{k}$, $\mathbf{x}$ and $\mathbf{r}$ to indicate projected 2-D vectors.}, defined as in Eq.~(\ref{eq:overdensity}):
\begin{equation}
    \langle \delta(\mathbf{k}) \delta(\mathbf{k'}) \rangle = \left( 2\pi \right)^2 P(k) \delta_{\mathrm{D}}(\mathbf{k}+\mathbf{k'}) \ ,
    \label{eq:ps}
\end{equation}
where $\langle \cdot \rangle$ denotes an average over the whole Fourier space, $k=|\mathbf{k}|$, and $\delta_{\mathrm{D}}(\cdot)$ indicates the Dirac delta function \citep{Dodelson03}; this in turn yields the estimator
\begin{equation}
    \hat{P}(k) = \frac{1}{N_{\rm{modes}}(k)} \sum_{|\mathbf{k}|=k}  |\delta(\mathbf{k})|^2 \ ,
    \label{eq:ps_est}
\end{equation}
where $N_{\rm{modes}} (k)$ is the number of modes in each $k$ bin, and the sum is performed over all $\mathbf{k}$ vectors whose magnitude is $k$. The definition in Eq.~(\ref{eq:ps}) implies that $P(k)$ is the Fourier counterpart of the 2-D matter correlation function $\xi(r)$, with $r=|\mathbf{r}|$, i.e.\
\begin{equation}
    P(k) = \int  \xi(r) e^{-\mathrm{i} \mathbf{k} \cdot \mathbf{r}}  \dif^2 \mathbf{r} \ ,
    \label{eq:power}
\end{equation} 
where $\xi(r)$ is defined as 
\begin{equation}
    \xi(r) = \langle \delta(\mathbf{x}) \delta(\mathbf{x}+\mathbf{r}) \rangle \ ,
\end{equation}
with $\langle \cdot \rangle$ representing the average over all locations $\textbf{x}$ in the plane in this case. 

In order to generate a lognormal random field with a given power spectrum, we follow the procedure of \citet{Coles91, Percival04}. We start by converting the measured power spectrum to the matter correlation function $\xi_{\rm{LN}}(r)$, then we calculate the corresponding Gaussian correlation function,
\begin{equation}
    \xi_{\rm{G}}(r) = \ln \left[ 1+\xi_{\rm{LN}}(r) \right] \ ,
    \label{eq:g_corr_f}
\end{equation}
transform it back to Fourier space and create a Gaussian random field realisation $\delta_{\rm{G}}$ on a grid with this power spectrum and the required resolution ($N_{\rm{low}}$ or $N_{\rm{high}}$). It is well known that a zero-mean Gaussian field is entirely specified by the given power spectrum, which only depends on the absolute value of the Fourier coefficients: this means that the Fourier phases can be uniformly sampled from the $[ 0, 2\pi ]$ interval \citep{Coles00, Chiang00, Watts03}. Crucially, when generating the Gaussian random field, we employ the set of phases of the Gaussian initial conditions of the Quijote simulation realisation. In this way, the final lognormal density fields will have a high level of correlation with the density fields obtained from the simulations: while the amount of correlation is limited due to the evolution from $z=127$ to $z=0$, the Pearson correlation coefficient between pairs of maps can be as high as 0.5, if we smooth the fields with a Gaussian kernel on scales of about $50 \ h^{-1}$ Mpc, while it is consistent with 0 if using completely random phases. Therefore, we argue that our choice facilitates learning the mapping from random fields to $N$-body slices.

Finally, we obtain the lognormal field $\delta_{\rm{LN}}$ by calculating for each grid point
\begin{equation}
\delta_{\rm{LN}} = \exp{ \left( \delta_{\rm{G}} - \sigma^2_{\rm{G}}/2 \right) } -1 \,
\end{equation}
where $\sigma_{\rm{G}}$ is the standard deviation of the Gaussian field. For all these operations we employ the \textsc{Python} package \textsc{nbodykit} \citep{Hand18}. A flowchart representing the steps followed to produce the training data is reported in Fig.~\ref{fig:flowchart}.

\begin{figure*}
	\includegraphics[width=2\columnwidth]{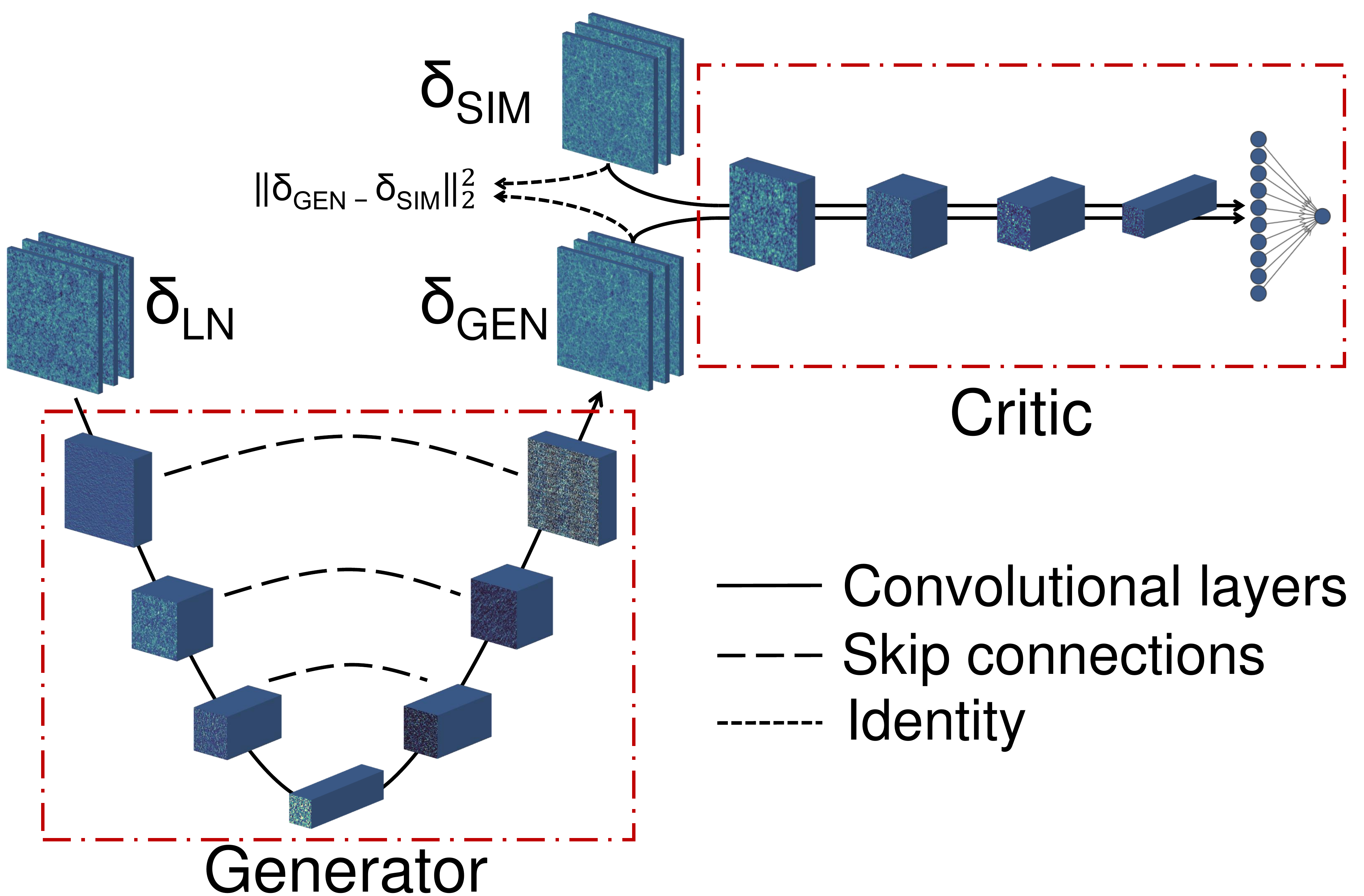}
    \caption{A representation of the generative model employed in this work, as described in Sect.~\ref{sec:iti_translation}. Following \citet{Isola17}, we have two convolutional neural networks, the generator (bottom left) and the critic (top right). We feed the lognormal maps through the generator, which is a U-net \citep{Ronneberger15}, that first downsamples and then upsamples each image using various convolutional layers, with all details reported in Appendix~\ref{app:model}. To improve the performance of the model, each upsampling step is concatenated with the output of a downsampling step, as indicated by the dashed lines (\textit{skip connections}). The output of the generator, dubbed $\delta_{\rm{GEN}}$, is then compared with the target data $\delta_{\rm{SIM}}$ by the critic network, which is again made of various convolutional layers, ending with a dense layer in order to have a single output. We chose this architecture based on those described in the literature \citep[e.g.][]{Isola17}; a full investigation over different architecture designs is beyond the scope of this paper. The critic and generator networks are trained together, minimising the loss function of Eq.~(\ref{eq:WGGAN-GP}). Note that in addition to the standard adversarial loss, we include a penalty term in the form of the mean squared error between the generated and target maps, which we found to significantly improve the performance of our model; this is indicated by the short-dashed lines (\textit{identity}).}
    \label{fig:architecture}
\end{figure*}

We observe two limitations due to the fact that we measure the power spectrum from a finite-resolution grid. First, by relying on the boxes only, we are capable of surveying only a limited range in $k$, namely no larger than $k \in [0.025 \ h \ \rm{Mpc ^{-1}}$, $1 \ h \ \rm{Mpc ^{-1}}]$ in the high-resolution case, and $k \in [0.025 \ h \ \rm{Mpc ^{-1}}$, $0.3 \ h \ \rm{Mpc ^{-1}}]$ in the low-resolution case. In order to access larger scales (i.e.\ lower $k$ values), we concatenate the measured power spectrum with the theoretical one obtained with \textsc{CLASS} \citep{Blas11} for $k \in [10^{-5} \ h \ \rm{Mpc ^{-1}}$, $0.025 \ h \ \rm{Mpc ^{-1}}]$: this makes the procedure outlined in the previous paragraphs more stable numerically.


Second, we observe a mismatch in power in the lognormal fields with respect to the imposed power spectrum. We attribute this discrepancy to the fact that when converting the Quijote initial conditions (obtained using second-order Lagrangian perturbation theory) to a density field, the mass assignment scheme and non-vanishing non-linearities arising from perturbation theory introduce extra spurious correlation in the phases. We correct for this effect, which actually introduces non-Gaussian features and is more pronounced at higher resolution, by iteratively rescaling the input power by the ratio of the output and target power at each $k$, until the mismatch across a sample of 100 random maps is smaller than $1 \%$ at all $k$ values. We checked that this iterative adaptation scheme effectively removes the power mismatch, leaving the final performance of the model unaffected at prediction time: the results presented in Sect.~\ref{sec:results} do not change significantly if, after training the model, we give it as an input a slice of a 3-D lognormal field generated with completely random Fourier phases. We further remark that this correction would not be necessary if we had access to a perfectly Gaussian density field of the initial conditions. 

We are left with pairs of square density field maps (dubbed $\delta_{\rm{LN}}$ and $\delta_{\rm{SIM}}$), which we use as the training (80\%), validation (10\%) and test (10\%) data, further discussed in the next section. This split is done at simulation level, so that the test, validation and training datasets are completely independent. Note that to reduce the correlations between slices coming from the same simulation cube we shift the pixels by a random amount along both the first and second axis, independently for each pair of maps, assuming periodic boundary conditions. It could also be possible to randomly rotate and flip the slices in order to augment the training data; while we found it is not needed in our setup, we defer further investigations to future work. Before feeding the pairs into the neural network architecture described in the next section, we additionally preprocess each map by calculating $\ln { \left( 1+ \delta\right)}$ to decrease the dynamic range of each density value $\delta$.

\subsection{Image-to-image translation}
\label{sec:iti_translation}
As discussed in Sect.~\ref{sec:intro}, machine learning generative techniques have extensively been applied to $N$-body simulations. In this work, we aim at mapping lognormal fields to more realistic fields, hence we employ the \textit{pix2pix} network structure, first proposed in \citet{Isola17}. The model is composed of two parts, as sketched in Fig.~\ref{fig:architecture}; all implementation details are reported in Appendix~\ref{app:model}. The first part is a U-net \citep{Ronneberger15}, which takes as an input a lognormal map $\delta_{\rm{LN}}$, obtained and preprocessed as described in Sect.~\ref{sec:training_data}. The map is passed through various convolutional layers to yield a compressed feature map, which is then upsampled back to the original resolution. Crucially, these upsampling steps are concatenated with the corresponding downsampled feature maps, which allow various scales to be accessible in the output map; removing these skip connections significantly impairs the performance of the model. We call the output map the \textit{generated} map $\delta_{\rm{GEN}}$.

We want the generated map to carry the same statistical information as the $\delta_{\rm{SIM}}$ density field. We tested that minimising a simple $\ell^1$ or $\ell^2$ norm between $\delta_{\rm{GEN}}$ and $\delta_{\rm{SIM}}$ is not sufficient to yield accurate results. For this reason, following \citet{Isola17}, we employ a second convolutional block as a discriminator, and express the loss in the framework of adversarial training. In the standard GAN framework \citep{Goodfellow14}, the generator network $G$ is trained together with the discriminator network $D$ until an equilibrium where neither $G$ or $D$ can improve their performance is reached: while $G$ attempts to generate realistic images, $D$ tries to distinguish between real and fake examples. Since we found this framework to be particularly unstable during training, we actually implemented the Wasserstein GAN with gradient penalty \citep[WGAN-GP,][]{Arjovsky17, Gulrajani17}, which we found superior both in performance and training stability. In this framework, a generator $G$ is trained alongside a critic $C$ to minimise the following cost function:
\begin{align}
    \nonumber & \mathbb{E}_{\delta_{\rm{GEN}}} \left [ C(\delta_{\rm{GEN}}) \right ] - \mathbb{E}_{\delta_{\rm{SIM}}} \left [C(\delta_{\rm{SIM}}) \right ] + \\ 
    & \ \ \ \ \ \ \ \ \ \ \ \ \ +  \lambda_1 \mathbb{E}_{\hat{\delta}} \left [ \left( ||\nabla_{\hat{\delta}} C(\hat{\delta}) ||_2 - 1 \right) ^2 \right ] + \lambda_2  || \delta_{\rm{GEN}} - \delta_{\rm{SIM}} ||^2_2 \ ,
\label{eq:WGGAN-GP}
\end{align}
where $\delta_{\rm{GEN}} = G(\delta_{\rm{LN}})$, $\mathbb{E}_{\delta_{\rm{GEN}}}$ and $\mathbb{E}_{\delta_{\rm{SIM}}}$ indicate the expectation value over samples of the generated and simulated maps (usually estimated through sample averages), respectively, $\hat{\delta}$ represents a linear combination of $\delta_{\rm{GEN}}$ and $\delta_{\rm{SIM}}$\footnote{In particular, $\hat{\delta} = \delta_{\rm{SIM}} + u(\delta_{\rm{GEN}} - \delta_{\rm{SIM}})$, with $u\sim U(0,1)$, where $U(0,1)$ indicates the uniform distribution between 0 and 1. This linear combination means that we are constraining the gradient norm to be 1 only along lines connecting real and fake data, which should be sufficient to guarantee good experimental results \citep{Gulrajani17}.}, $|| \cdot ||_2$ indicates the $\ell^2$ norm, and $\lambda_1$ and $\lambda_2$ are two positive hyperparameters that allow us to tune the amount of regularisation given by the gradient penalty and the $\ell^2$ norm, respectively. In short, Eq.~(\ref{eq:WGGAN-GP}) indicates that we wish to minimise the Wasserstein-1 (or earth mover) distance between the real data and generated data distributions, while constraining the gradient of the critic network to be close to unity; this is needed since the formulation of the Wasserstein distance as in the first two terms of Eq.~(\ref{eq:WGGAN-GP}) only holds when the critic is a 1-Lipschitz function, i.e.\ when its gradient is bound \citep[see][for more details]{Gulrajani17}. We observe that in the standard WGAN-GP formulation $\lambda_2=0$, while in our case we found it key to minimise the $\ell^2$ norm between simulated and generated maps as well in order to obtain improved results.

To train the networks, we use the Adam optimiser \citep{Kingma14} with learning rate $10^{-5}$; we set the additional Adam hyperparameters $\beta_1=0$ and $\beta_2=0.9$, following \citet{Gulrajani17}, and refer the reader to \citet{Kingma14} and \citet{Gulrajani17} for more details. We feed the data in batches of 32 pairs at each iteration, and train our model for 10 epochs (150 in the low-resolution case), where each epoch consists of feeding the entire training set through the network. For each batch, we update the critic parameters $n_{\rm{critic}}=10$ times, and the generator parameters only once. Multiple iterations of the critic are usually set to ensure its optimality while still allowing the generator to learn \citep{Arjovsky17, Gulrajani17}; in our work, we only explored $n_{\rm{critic}}=5$ and $n_{\rm{critic}}=10$, and used the latter since it showed slightly better results. Each epoch takes about 0.5 h (4 h) for the low (high) resolution case, on a Tesla P100 GPU; after training, mapping a lognormal map through the generator takes $\mathcal{O} (1 \ \rm{s})$ on the same hardware, and can be efficiently done in batches.

We save the model after each epoch. In order to select the best model amongst the saved ones, for each of them we run the statistical tests described in Sect.~\ref{sec:statistics}, and measure the mean percentage difference between the target and predicted maps for randomly-sampled maps of the validation set. The best model is chosen as the one which minimises the sum of the mean percentage differences over all tests; the results are then shown on maps from the test set. In the high-resolution case only, we actually found that the trained model can generate maps whose power spectrum is significantly different (more than 10\%) from the input and target ones, which we attribute to instabilities of the WGAN-GP framework; we show one such example in Appendix~\ref{sec:prob_spectra}. For this reason, we propose a ranking system that takes all the predictions from the test set, and orders them based on the mean difference between the input power spectrum and the predicted power spectrum, since they must match by construction. We select the best 100 maps according to this metric, and discuss possible ways to make the model more stable in Sect.~\ref{sec:conclusions}. We envision that in a realistic scenario where an arbitrary number of lognormal maps can be generated with the goal of producing $m$ augmented fields, one could iteratively generate augmented lognormal maps through our model and discard those whose precision is below a desired threshold, until all $m$ maps are produced.

We show the results of our best models in the next section. These models are found with $\lambda_1 = 100$ and $\lambda_2 = 10$; we defer a full grid search over these hyperparameters to future work.

\begin{figure*}
	\includegraphics[width=2\columnwidth]{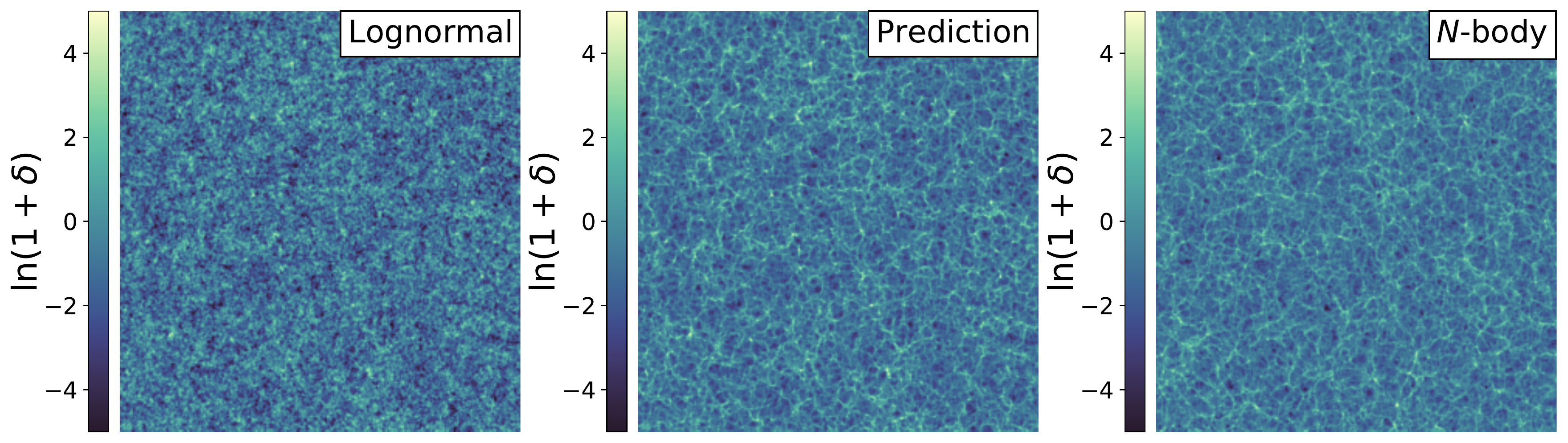}
    \caption{The lognormal (left) and $N$-body (right) density fields as in Fig.~\ref{fig:histo_with_density}, with the prediction of our model (middle) given the lognormal field. In these maps, we clipped the maximum and minimum values before applying the logarithm to reduce their dynamic range. The model is described in Sect.~\ref{sec:iti_translation}. We remark that we are not interested in an exact match of the middle and right panels, as we explain in Sect.~\ref{sec:vis_insp}, and thoroughly test that the predicted fields carry the same statistical information as the $N$-body maps from Sect.~\ref{sec:statistics}.}
    \label{fig:vis_insp}
\end{figure*}

\section{Results}
\label{sec:results}
In this section, we validate the performance of the trained model by comparing the statistics of the generated and simulated maps. In Appendix~\ref{sec:mode_collapse}, we also show that our model is not affected by mode collapse.

\subsection{Qualitative comparison}
\label{sec:vis_insp}
While the appearance of the maps is irrelevant for the purpose of our statistical analysis, a visual inspection is nonetheless useful to intuitively understand whether our model is on the right track to learn the $N$-body features. In Fig.~\ref{fig:vis_insp}, we show a lognormal map, its $N$-body counterpart and the prediction of our model given the lognormal map for the high-resolution case. 

Our goal is not to obtain an exact visual match between the model's prediction and the $N$-body map, given that we used the random phases of the $z=127$ simulations, which only have partial correlations with the $z=0$ slices. For the applications we focus on (discussed in Sect.~\ref{sec:conclusions}), the actual position of peaks and voids in the lognormal map is irrelevant, since it is dictated by the random sampling of the phases: we only aim to generate maps which carry the same statistical signal as the $N$-body maps on average, improving on the lognormal approximation. We observe that while the predicted field does not match the $N$-body pattern pixel by pixel, the model has learnt the correct morphology of the large-scale structure on top of the lognormal field.

\begin{figure*}
\centering
    \begin{subfigure}{0.22\linewidth}
        \includegraphics[width=\linewidth]{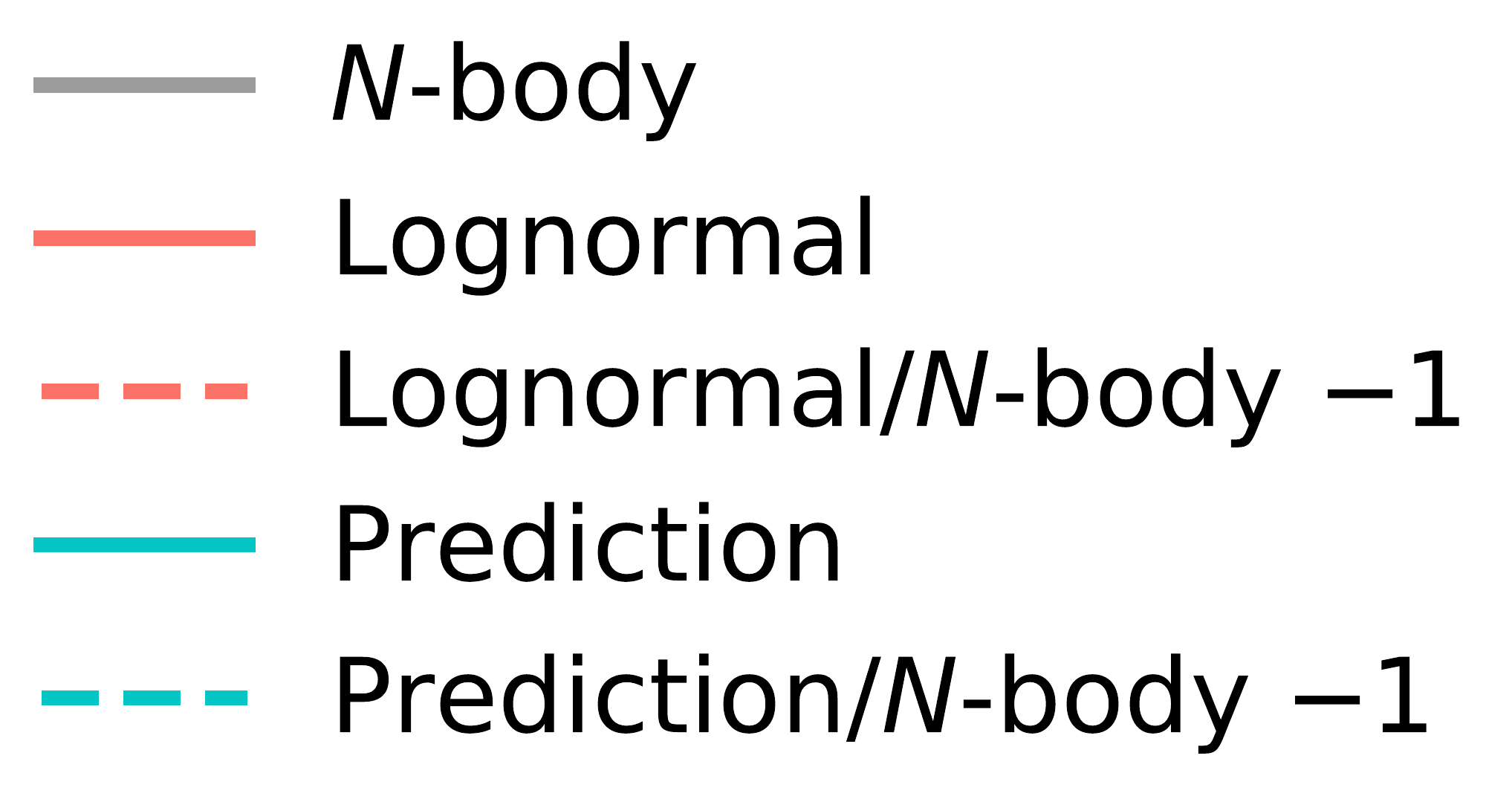}
        \vspace{2cm}
    \end{subfigure}
\hfil
    \begin{subfigure}{0.22\linewidth}
        \includegraphics[width=\linewidth]{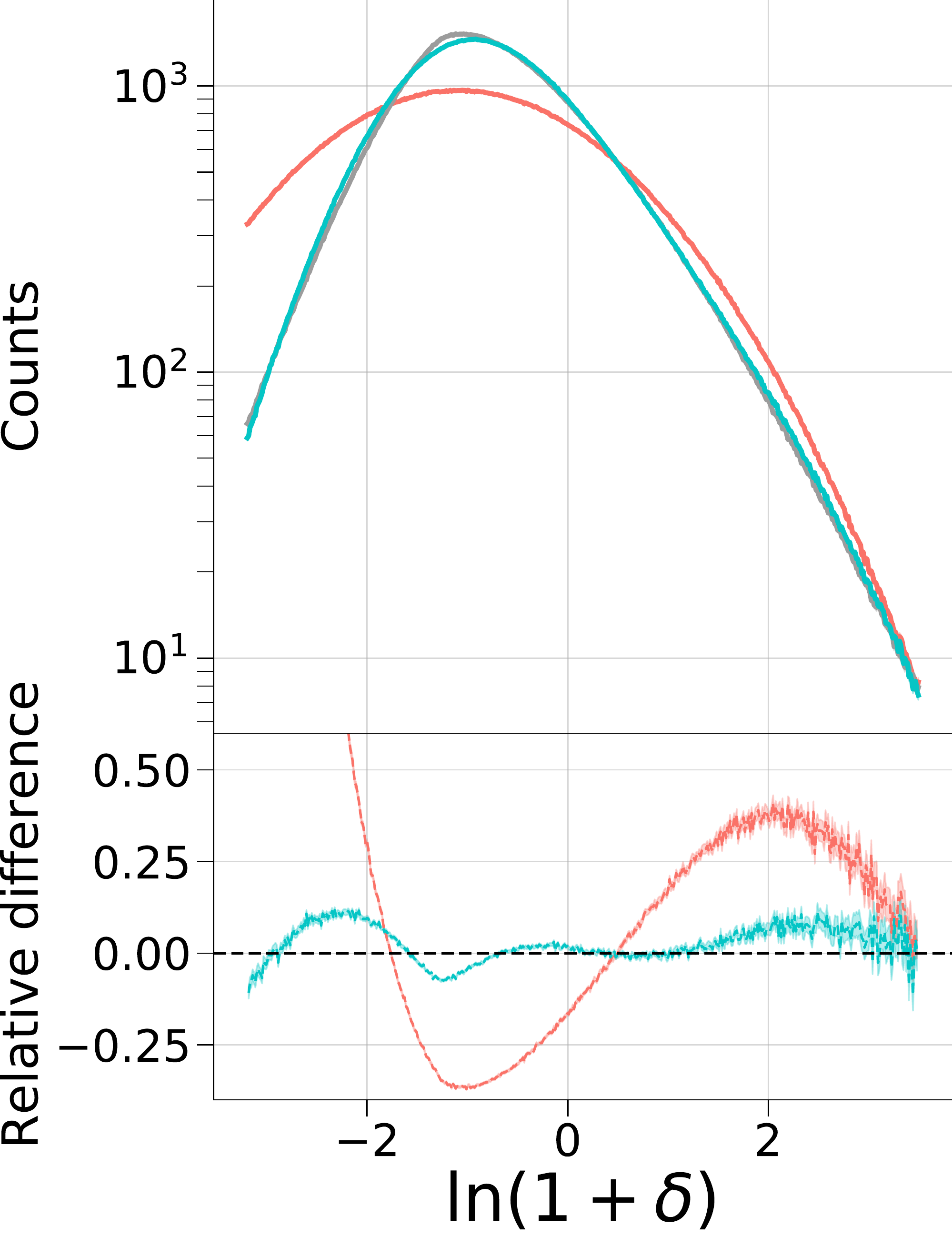}
    \caption{Pixel counts}
    \end{subfigure}
\hfil
    \begin{subfigure}{0.22\linewidth}
        \includegraphics[width=\linewidth]{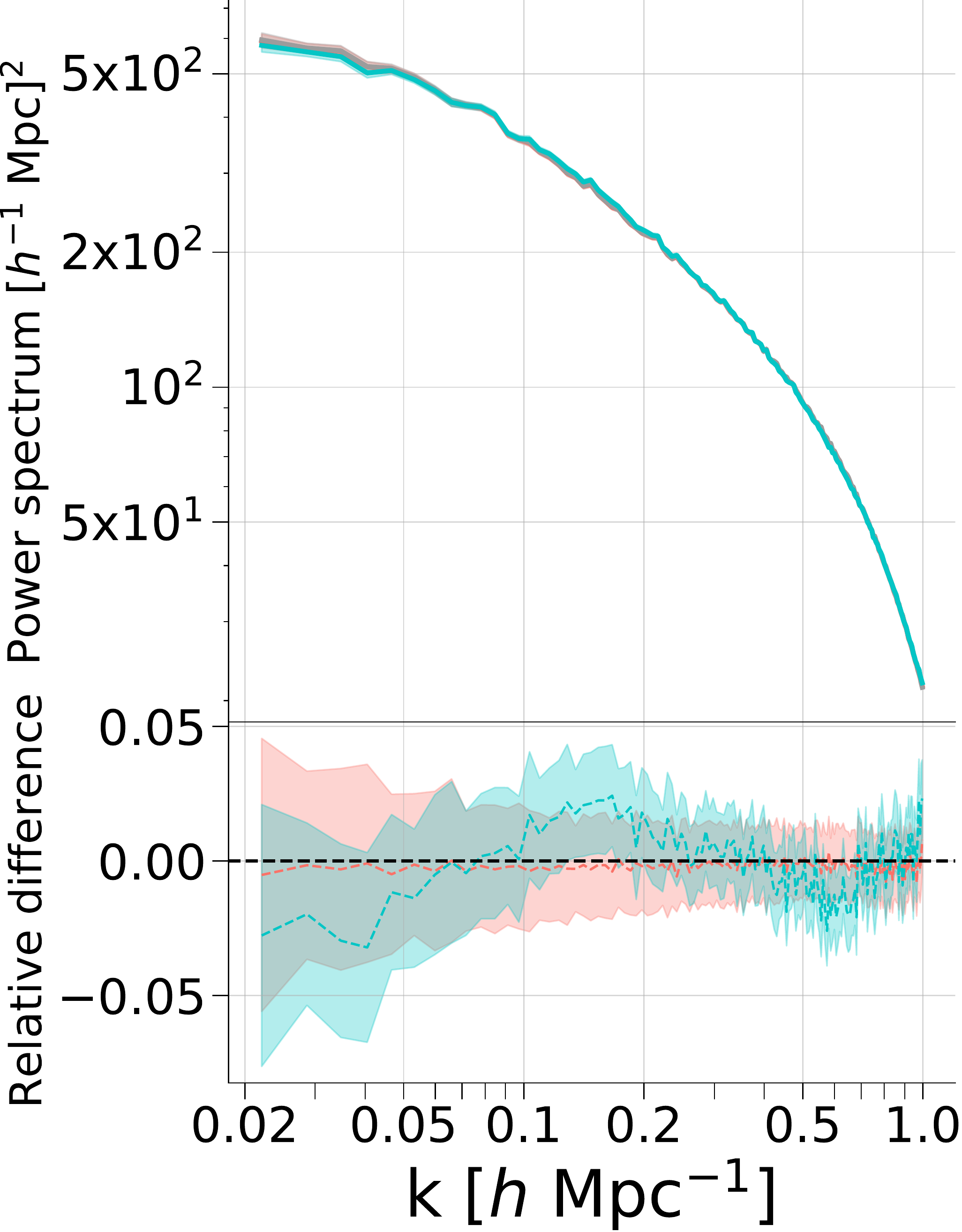}
    \caption{Power spectrum}
    \end{subfigure}
\hfil
    \begin{subfigure}{0.22\linewidth}
        \includegraphics[width=\linewidth]{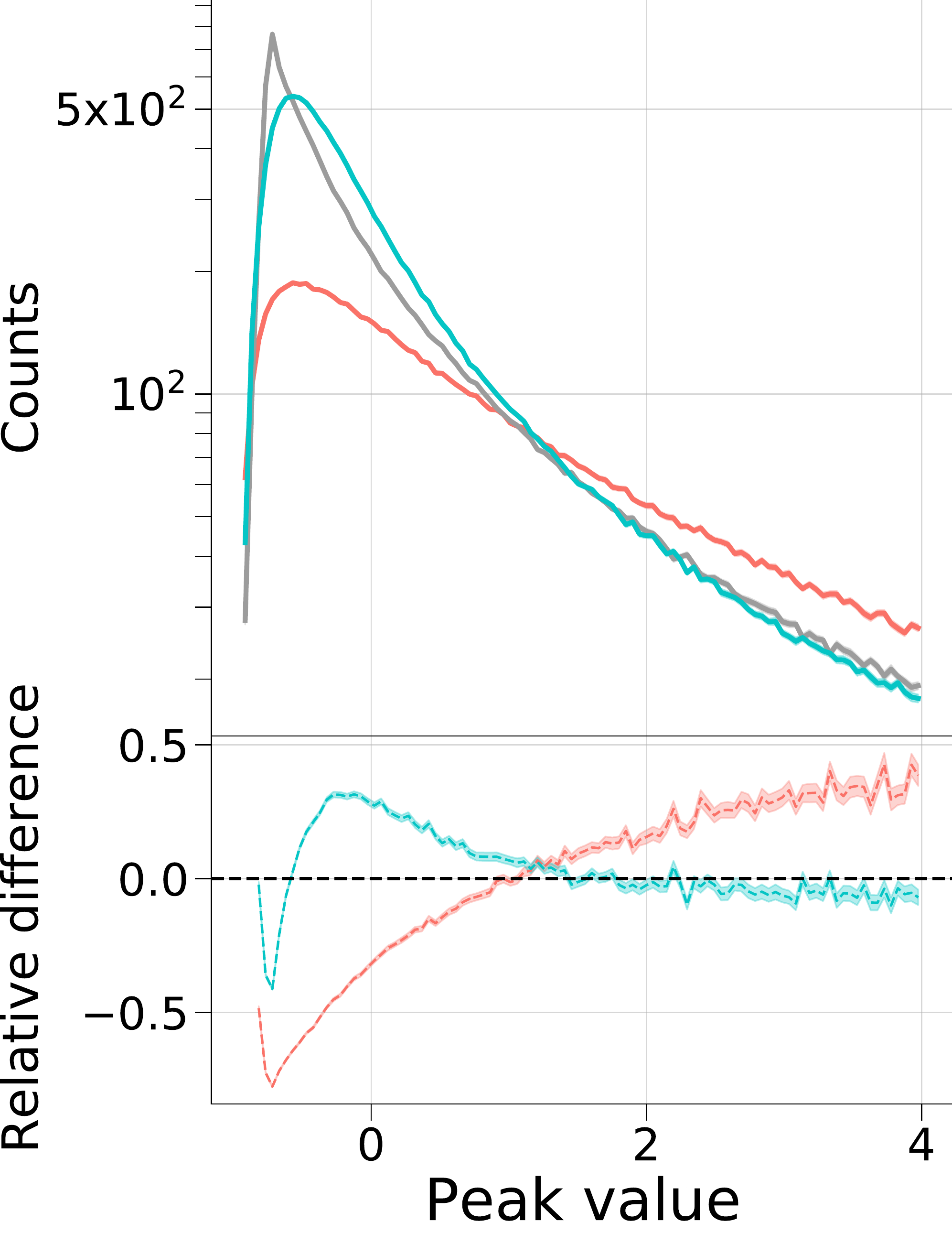}
    \caption{Peak counts}
    \end{subfigure}

    \begin{subfigure}{0.22\linewidth}
        \includegraphics[width=\linewidth]{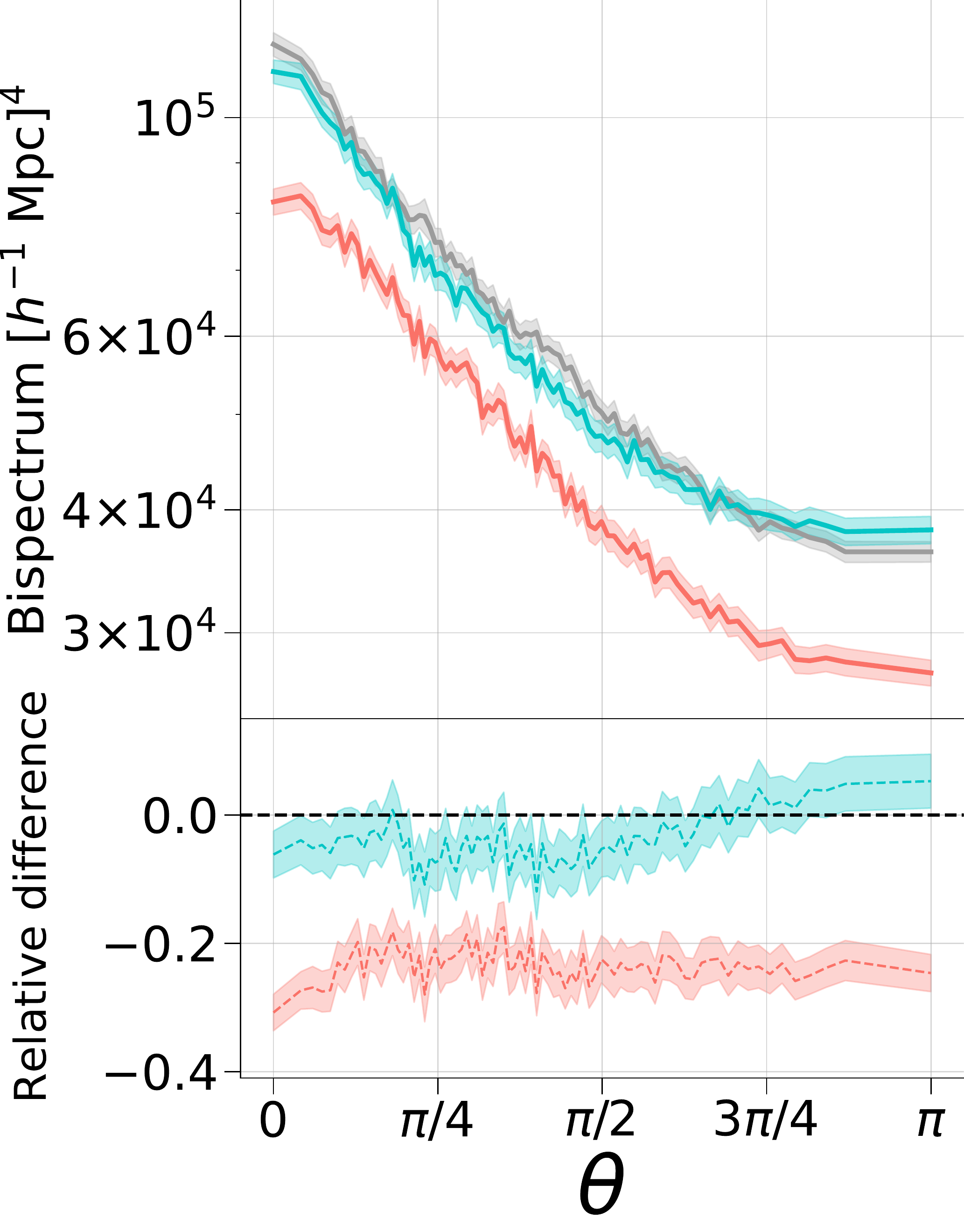}
    \caption{Bispectrum with $k_1=0.4 \ h \ \rm{Mpc}^{-1}$, $k_2=0.6 \ h \ \rm{Mpc}^{-1}$}
    \end{subfigure}
\hfil
    \begin{subfigure}{0.22\linewidth}
        \includegraphics[width=\linewidth]{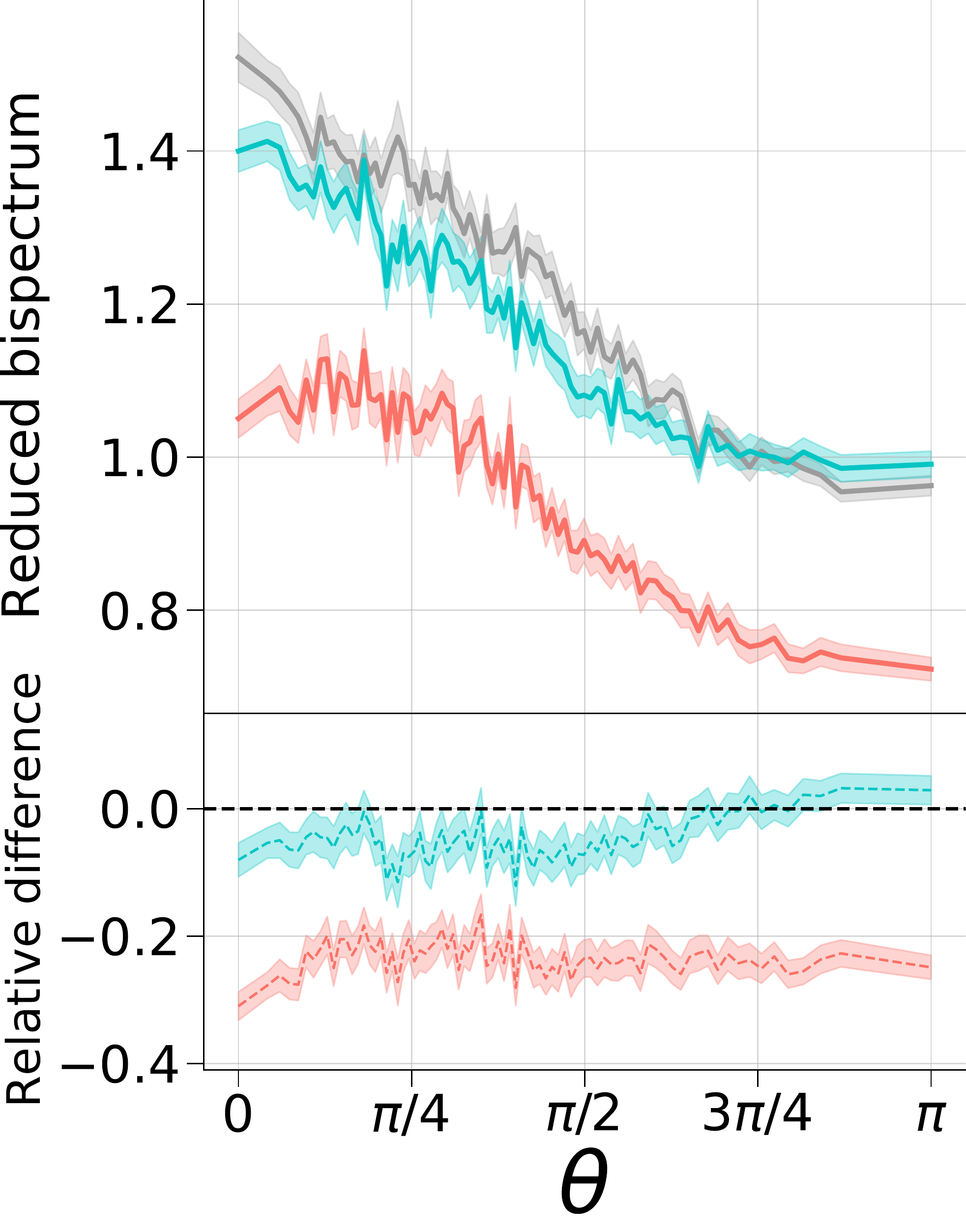}
    \caption{Reduced bispectrum with $k_1=0.4 \ h \ \rm{Mpc}^{-1}$, $k_2=0.6 \ h \ \rm{Mpc}^{-1}$}
    \end{subfigure}
\hfil
    \begin{subfigure}{0.22\linewidth}
        \includegraphics[width=\linewidth]{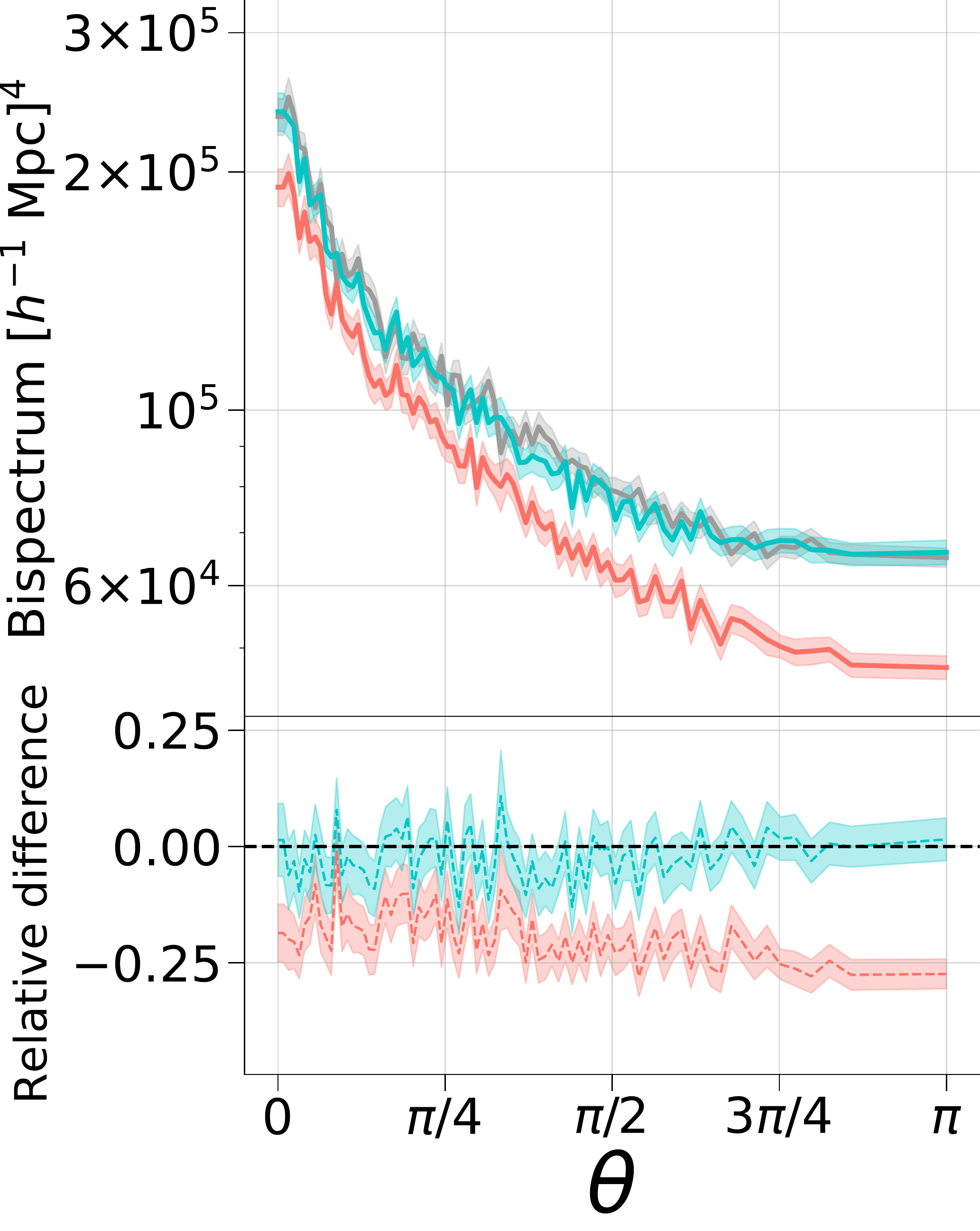}
    \caption{Bispectrum with $k_1=0.4 \ h \ \rm{Mpc}^{-1}$, $k_2=0.4 \ h \ \rm{Mpc}^{-1}$}
    \end{subfigure}
\hfil
    \begin{subfigure}{0.22\linewidth}
        \includegraphics[width=\linewidth]{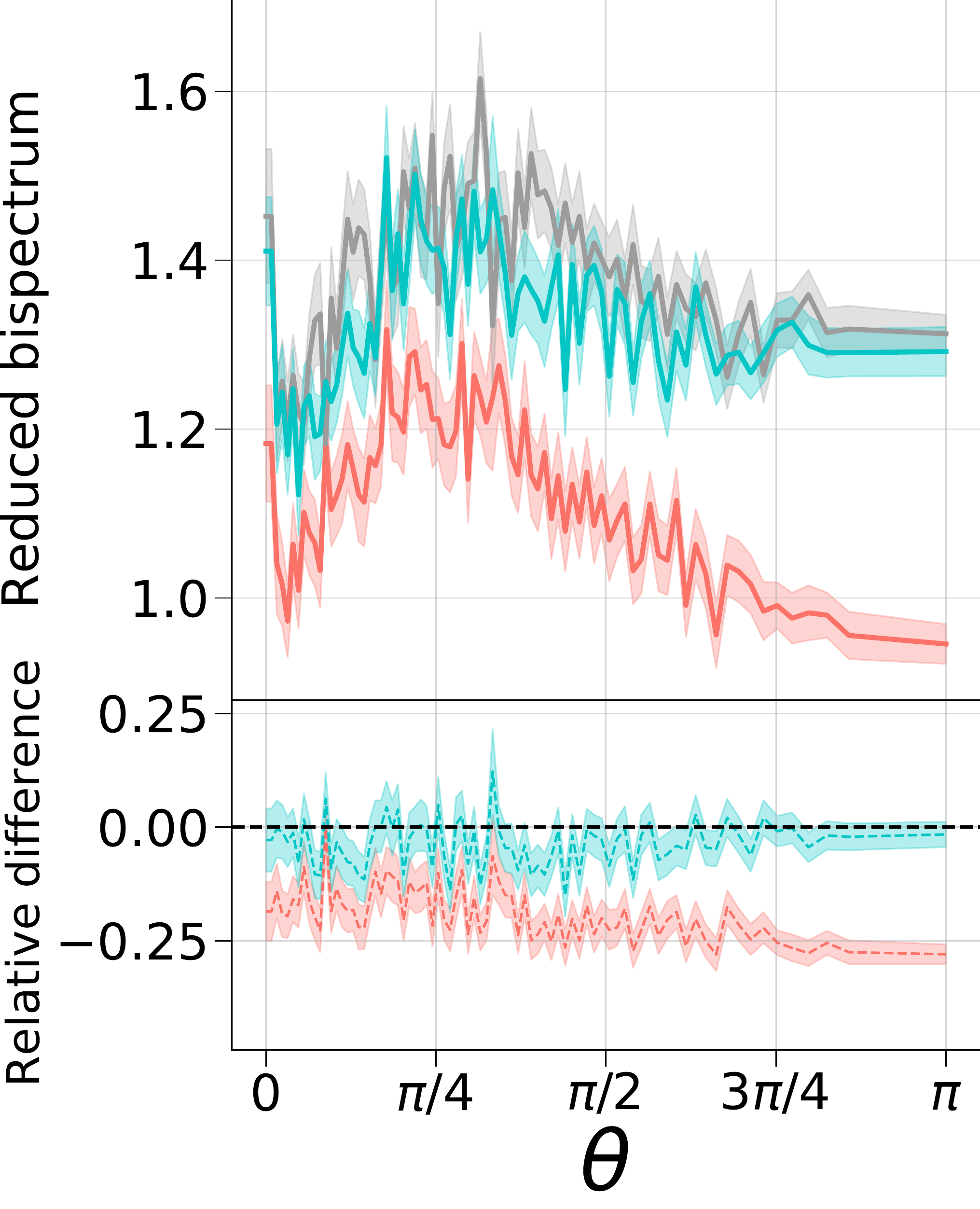}
    \caption{Reduced bispectrum with $k_1=0.4 \ h \ \rm{Mpc}^{-1}$, $k_2=0.4 \ h \ \rm{Mpc}^{-1}$}
    \end{subfigure}
\caption{Comparison of the statistical tests described in Sect.~\ref{sec:statistics} for the lognormal ($\delta_{\textrm{LN}}$, in red), $N$-body ($\delta_{\textrm{SIM}}$, in grey), and predicted ($\delta_{\textrm{GEN}}$, in cyan) maps, considering a resolution of $N_{\textrm{high}} = 512$. The performance is measured at the bottom of each panel by calculating the relative difference of $N$-body against predicted and lognormal maps (dashed lines). All solid lines indicate the mean values over 100 maps, and the error bars represent the error on the mean (or propagated error, in the case of the relative differences). We observe that, except for the range $\delta < 0$ in panel (a) and (c) and some individual $\theta$ values in panels (d)--(g), the prediction always matches the target statistics within the error bars, performing significantly better than the lognormal approximation.}
    \label{fig:512_z0}
\end{figure*}

\begin{figure*}
\centering
    \begin{subfigure}{0.22\linewidth}
        \includegraphics[width=\linewidth]{figures/legend.pdf}        \vspace{2cm}
    \end{subfigure}
\hfil
    \begin{subfigure}{0.22\linewidth}
        \includegraphics[width=\linewidth]{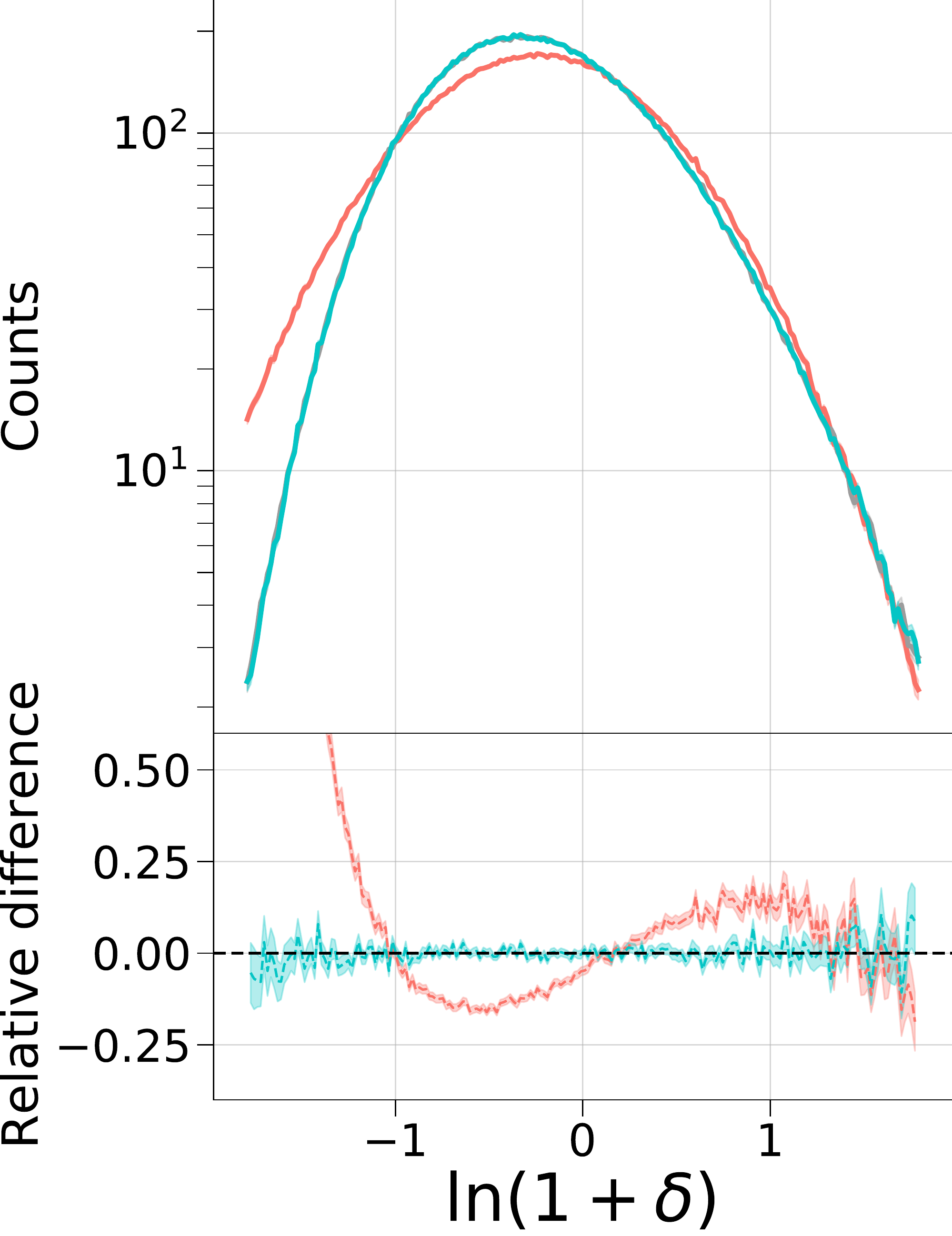}
    \caption{Pixel counts}
    \end{subfigure}
\hfil
    \begin{subfigure}{0.22\linewidth}
        \includegraphics[width=\linewidth]{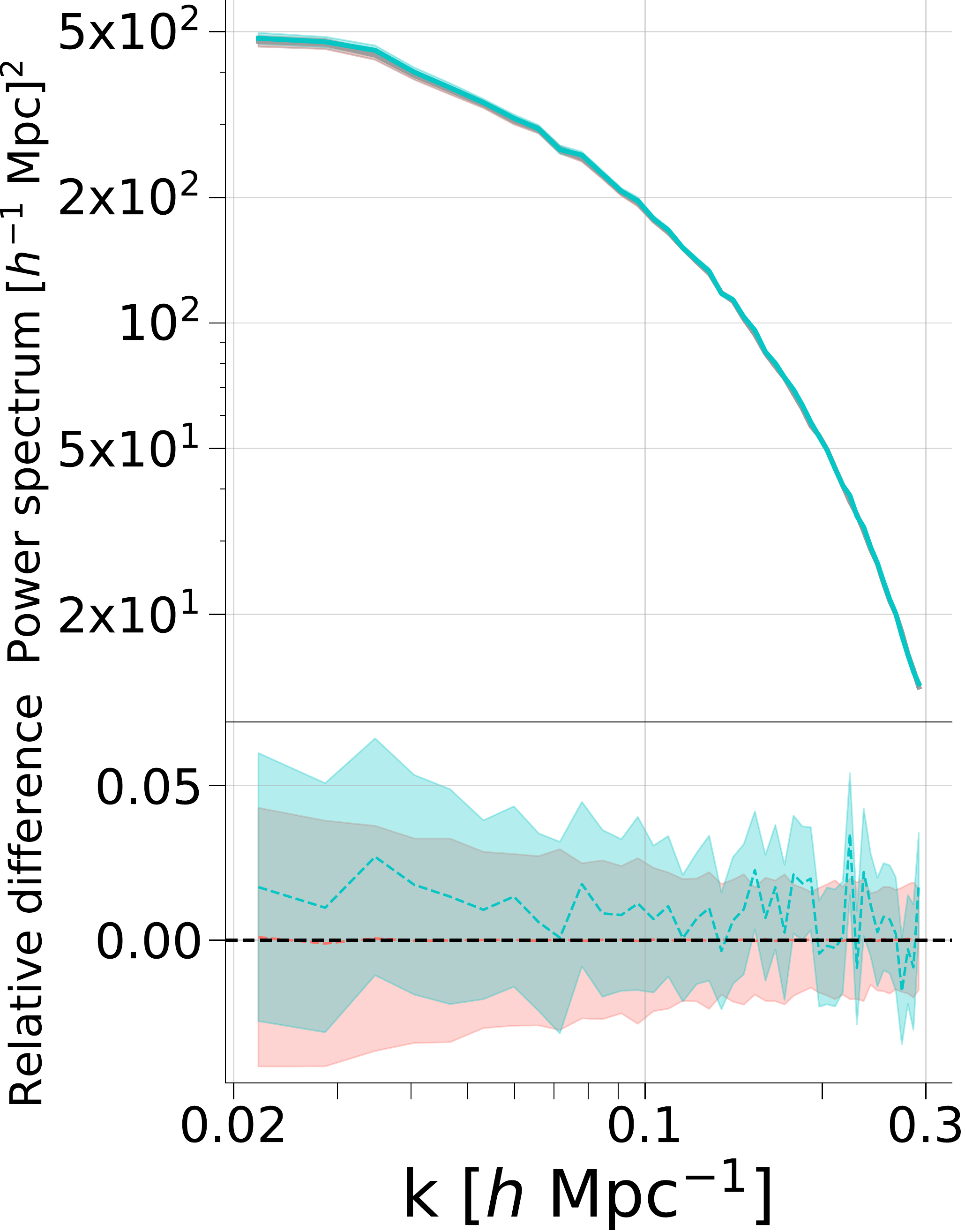}
    \caption{Power spectrum}
    \end{subfigure}
\hfil
    \begin{subfigure}{0.22\linewidth}
        \includegraphics[width=\linewidth]{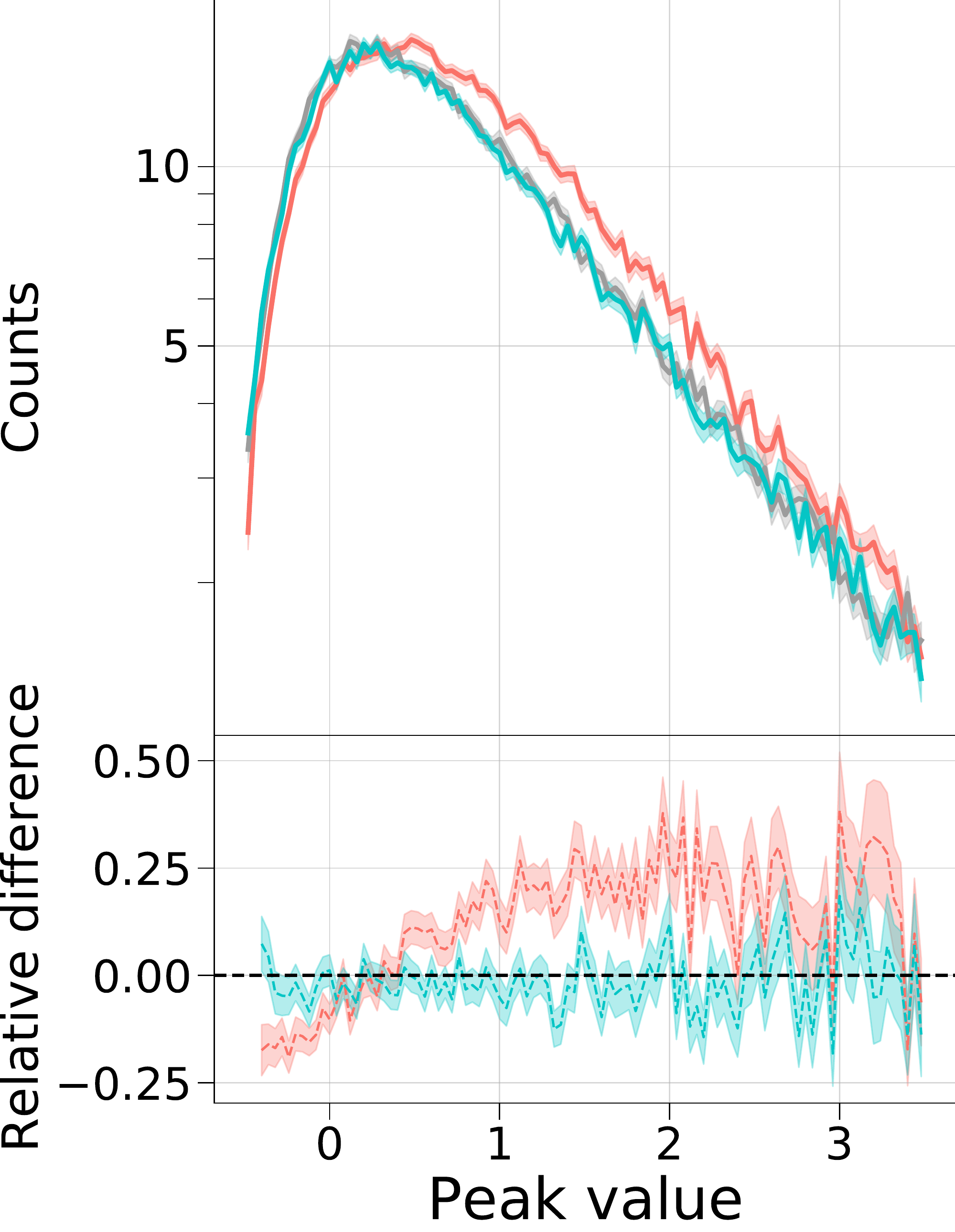}
    \caption{Peak counts}
    \end{subfigure}

    \begin{subfigure}{0.22\linewidth}
        \includegraphics[width=\linewidth, trim={0 0 0 0.2cm}, clip]{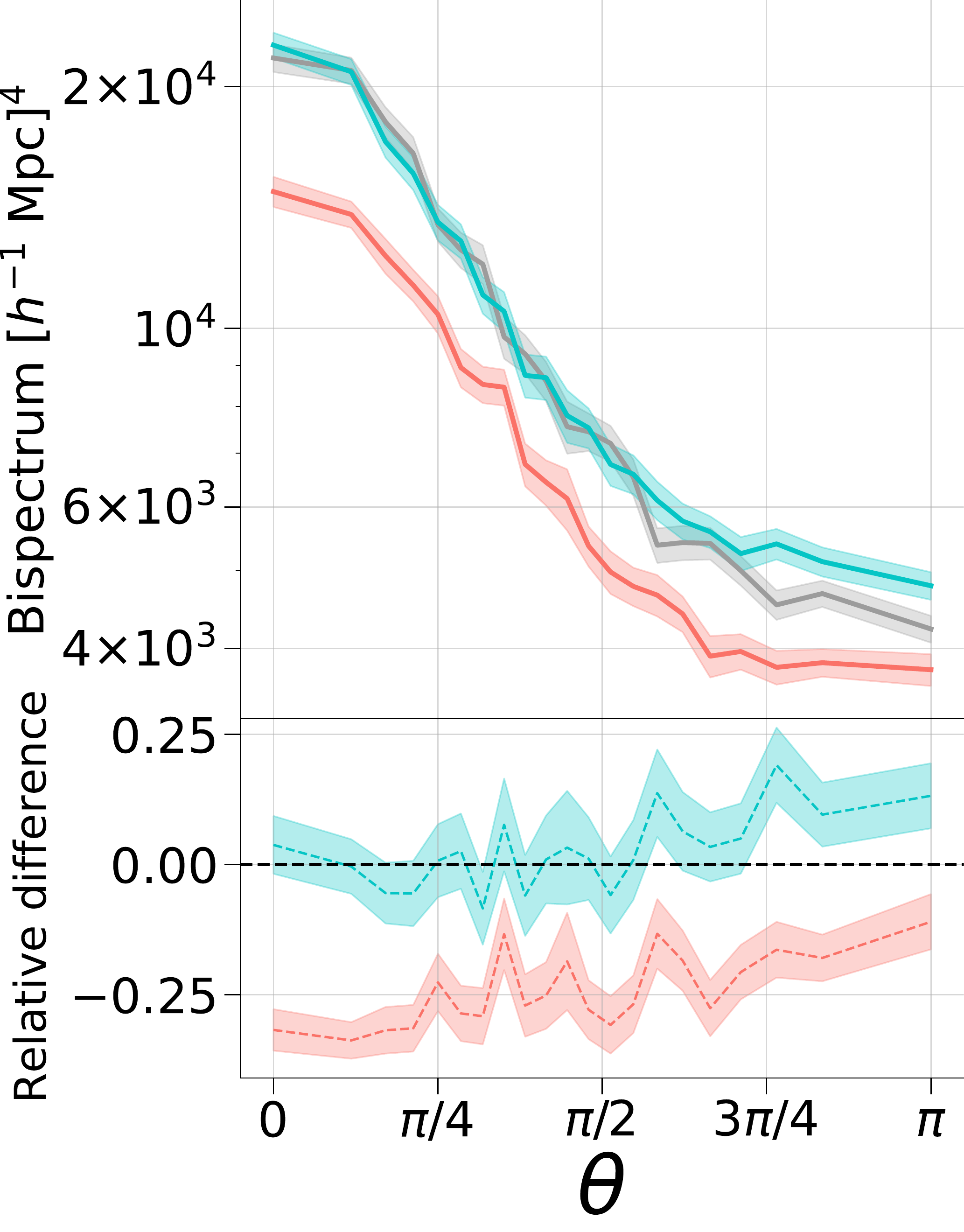}
    \caption{Bispectrum with $k_1=0.1 \ h \ \rm{Mpc}^{-1}$, $k_2=0.3 \ h \ \rm{Mpc}^{-1}$}
    \end{subfigure}
\hfil
    \begin{subfigure}{0.22\linewidth}
        \includegraphics[width=\linewidth]{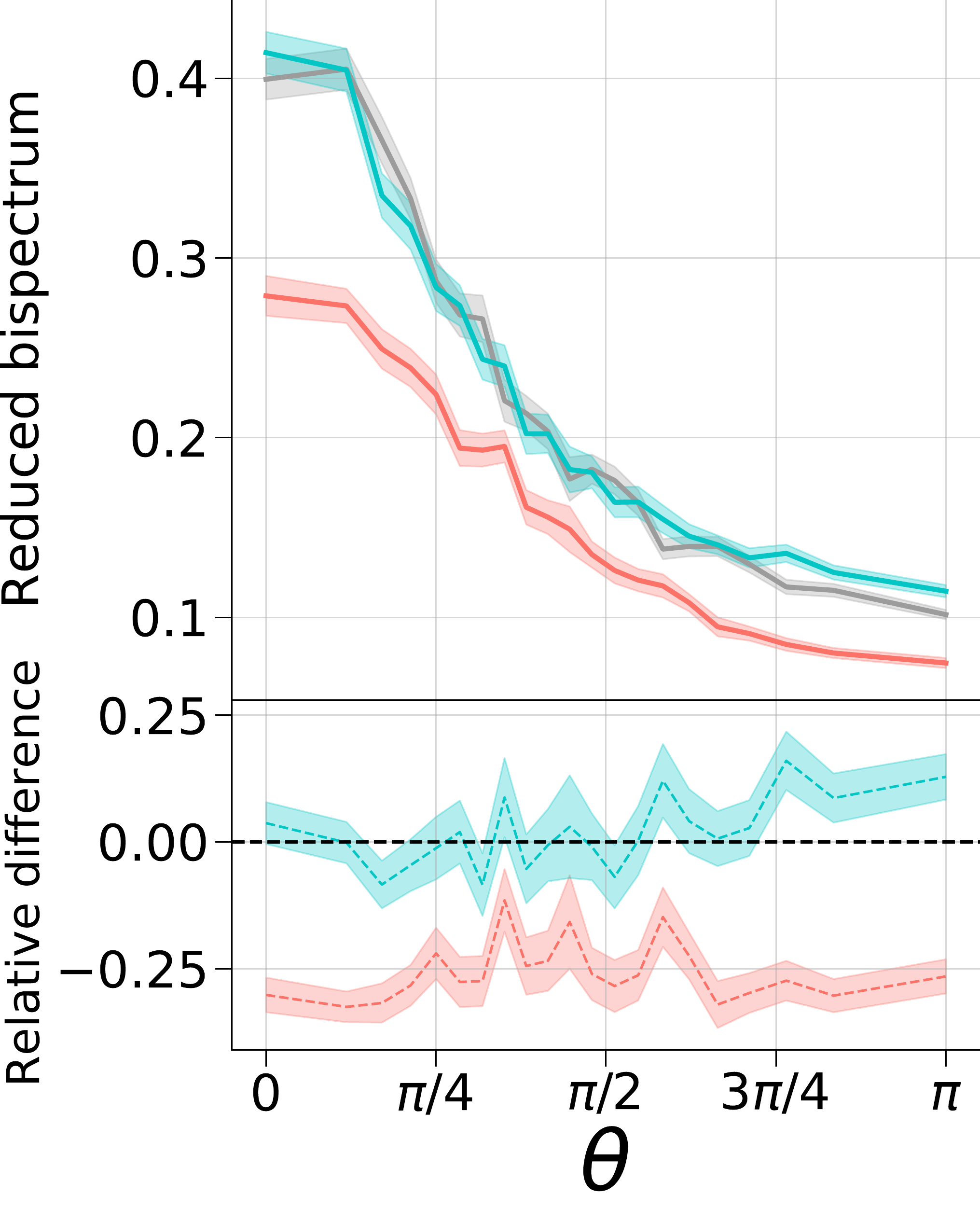}
    \caption{Reduced bispectrum with $k_1=0.1 \ h \ \rm{Mpc}^{-1}$, $k_2=0.3 \ h \ \rm{Mpc}^{-1}$}
    \end{subfigure}
\hfil
    \begin{subfigure}{0.22\linewidth}
        \includegraphics[width=\linewidth, trim={0 0 0 0.5cm},clip]{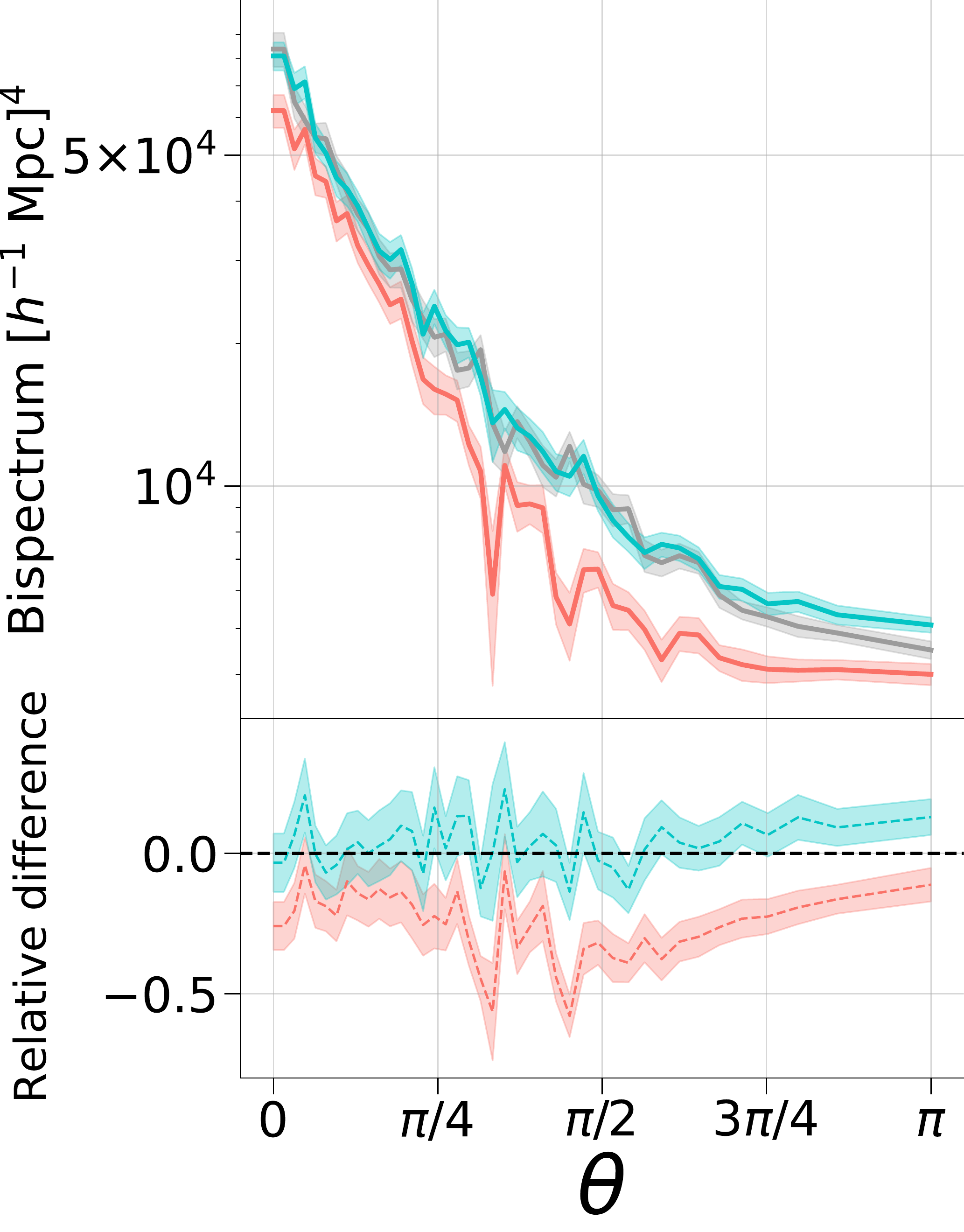}
    \caption{Bispectrum with $k_1=0.2 \ h \ \rm{Mpc}^{-1}$, $k_2=0.2 \ h \ \rm{Mpc}^{-1}$}
    \end{subfigure}
\hfil
    \begin{subfigure}{0.22\linewidth}
        \includegraphics[width=\linewidth, trim={0 0 0 0.2cm},clip]{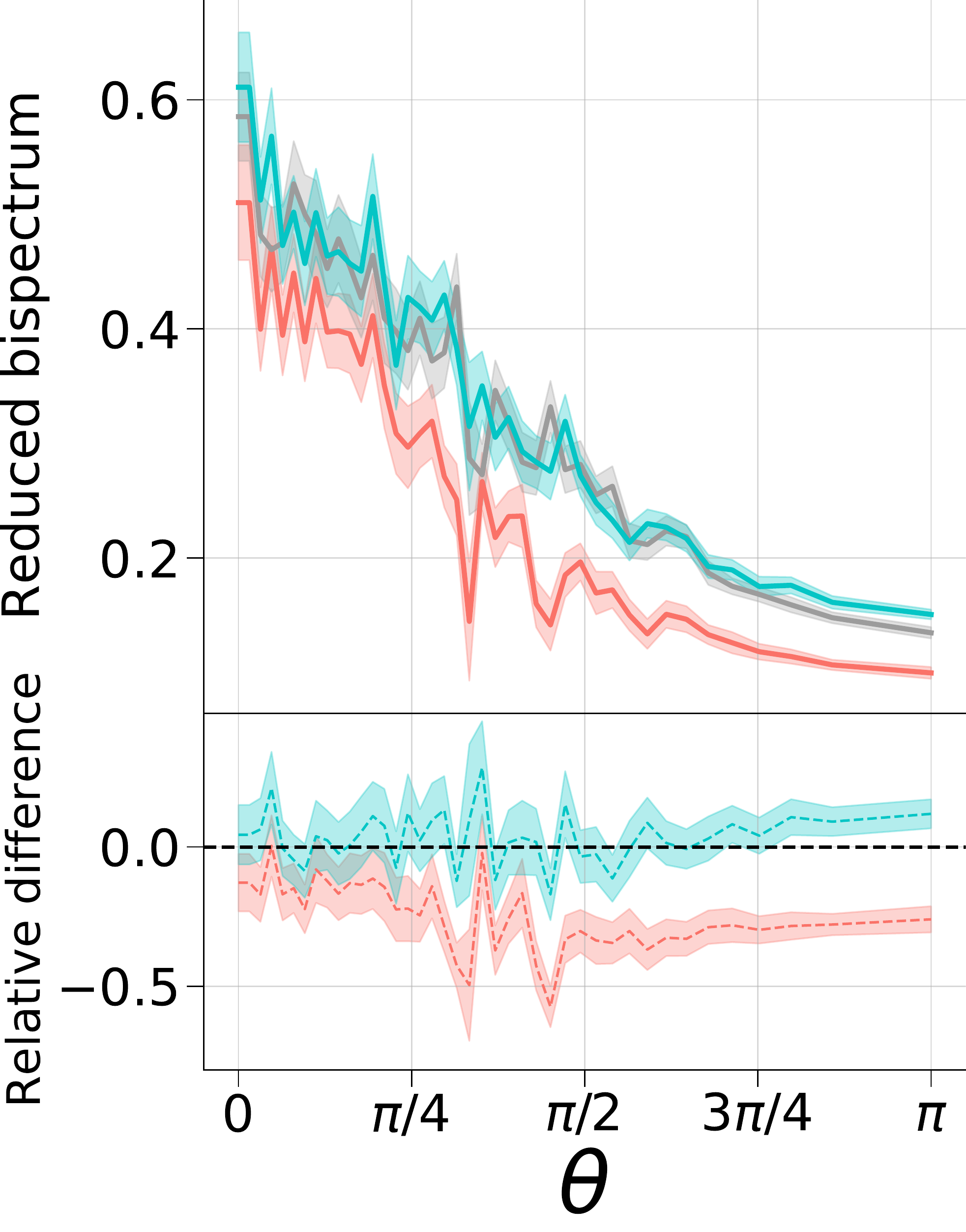}
    \caption{Reduced bispectrum with $k_1=0.2 \ h \ \rm{Mpc}^{-1}$, $k_2=0.2 \ h \ \rm{Mpc}^{-1}$}
    \end{subfigure}
\caption{Same as Fig.~\ref{fig:512_z0}, with a lower field resolution $N_{\rm{low}} = 128$. The solid lines indicate the mean values over 100 maps. We observe that the model's performance is almost always within the 5\% range, except for the bispectra, where significant differences are present at high $\theta$; we discuss these discrepancies in Sect.~\ref{sec:lowres}. Despite these differences, our model still outperforms the lognormal approximation.}
    \label{fig:128_z0}
\end{figure*}

\begin{figure*}
\centering
    \begin{subfigure}{0.22\linewidth}
        \includegraphics[width=\linewidth]{figures/legend.pdf}
        \vspace{2cm}
        \end{subfigure}
\hfil
    \begin{subfigure}{0.22\linewidth}
        \includegraphics[width=\linewidth]{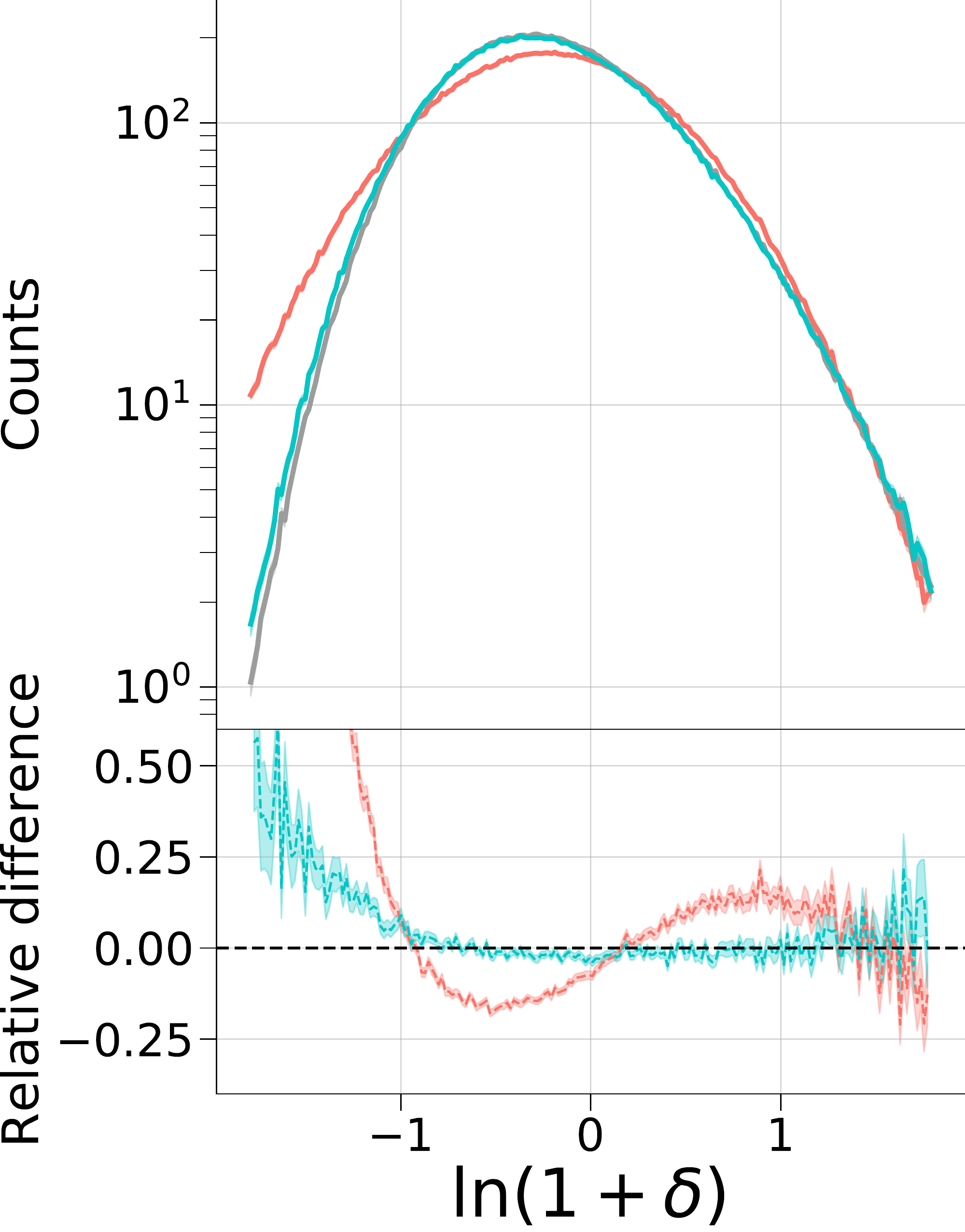}
    \caption{Pixel counts}
    \end{subfigure}
\hfil
    \begin{subfigure}{0.22\linewidth}
        \includegraphics[width=\linewidth]{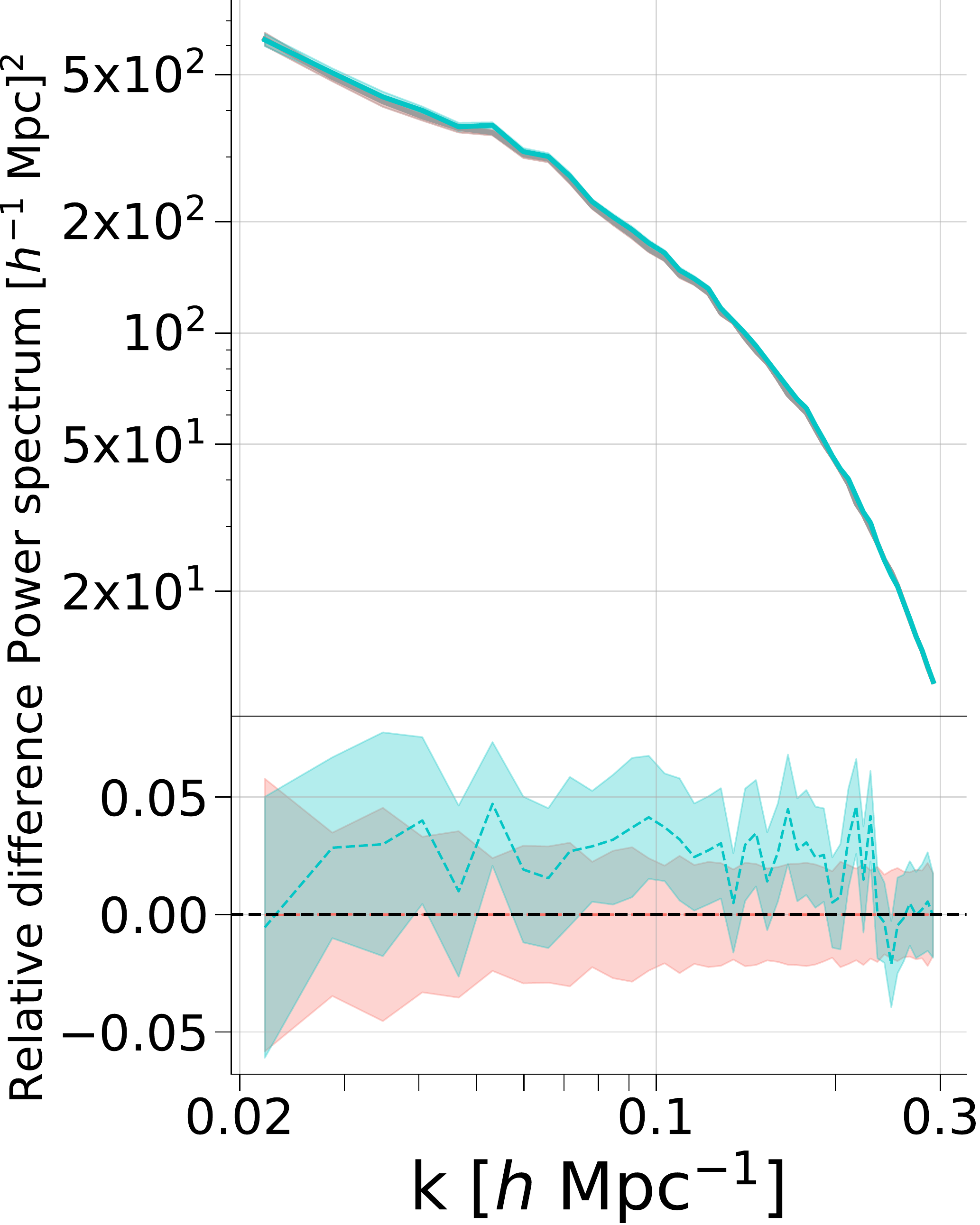}
    \caption{Power spectrum}
    \end{subfigure}
\hfil
    \begin{subfigure}{0.22\linewidth}
        \includegraphics[width=\linewidth]{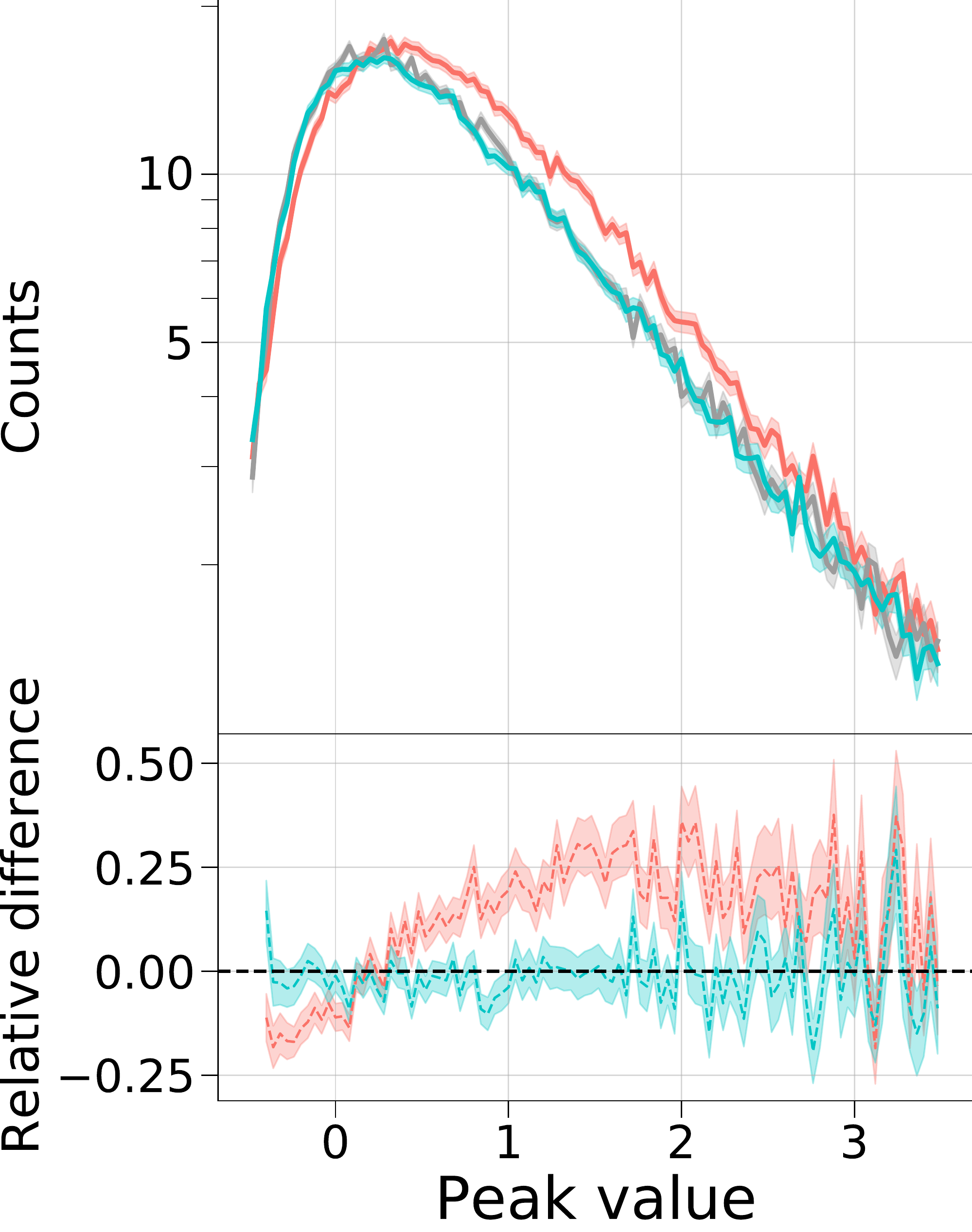}
    \caption{Peak counts}
    \end{subfigure}

    \begin{subfigure}{0.22\linewidth}
        \includegraphics[width=\linewidth]{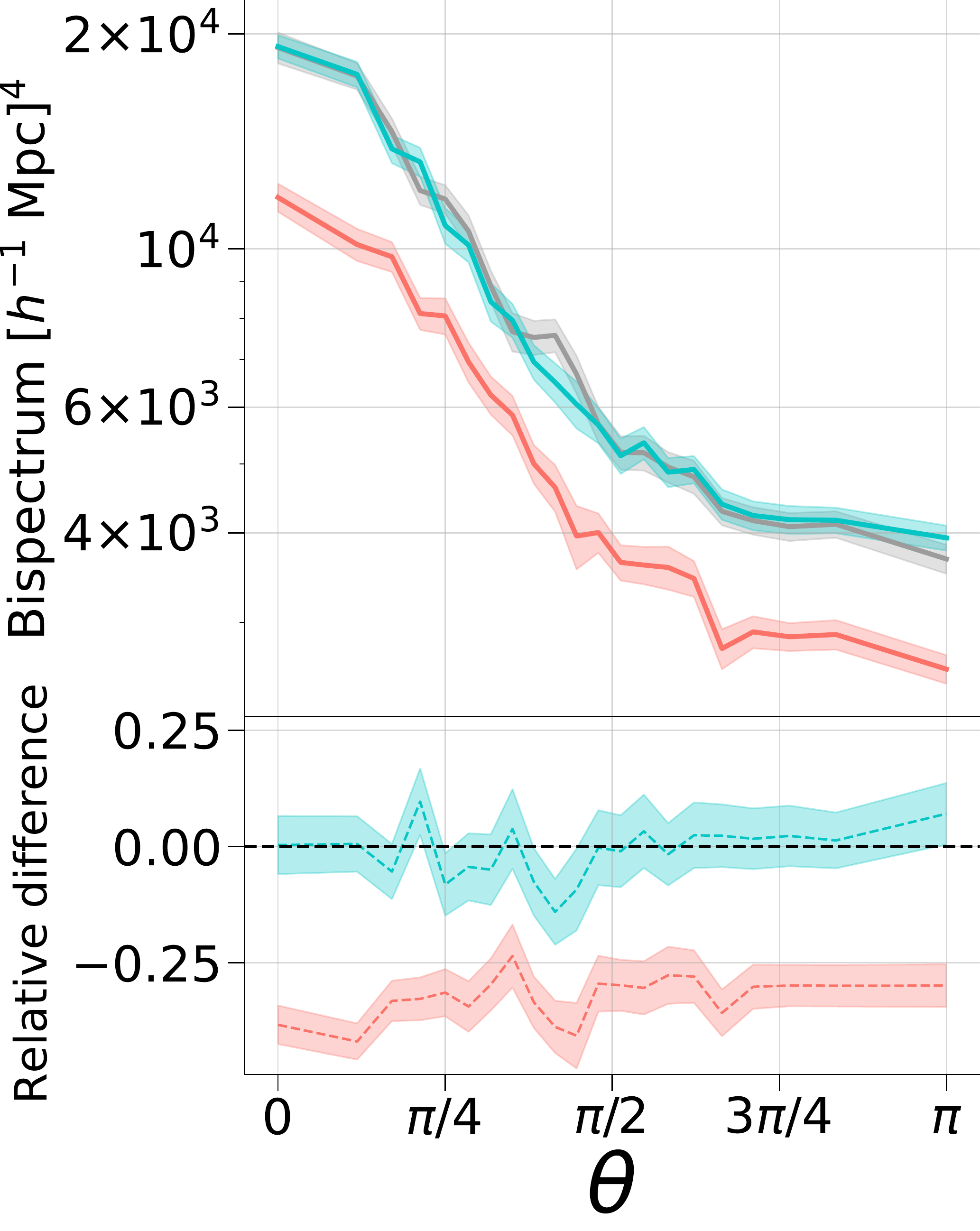}
    \caption{Bispectrum with $k_1=0.1 \ h \ \rm{Mpc}^{-1}$, $k_2=0.3 \ h \ \rm{Mpc}^{-1}$}
    \end{subfigure}
\hfil
    \begin{subfigure}{0.22\linewidth}
        \includegraphics[width=\linewidth]{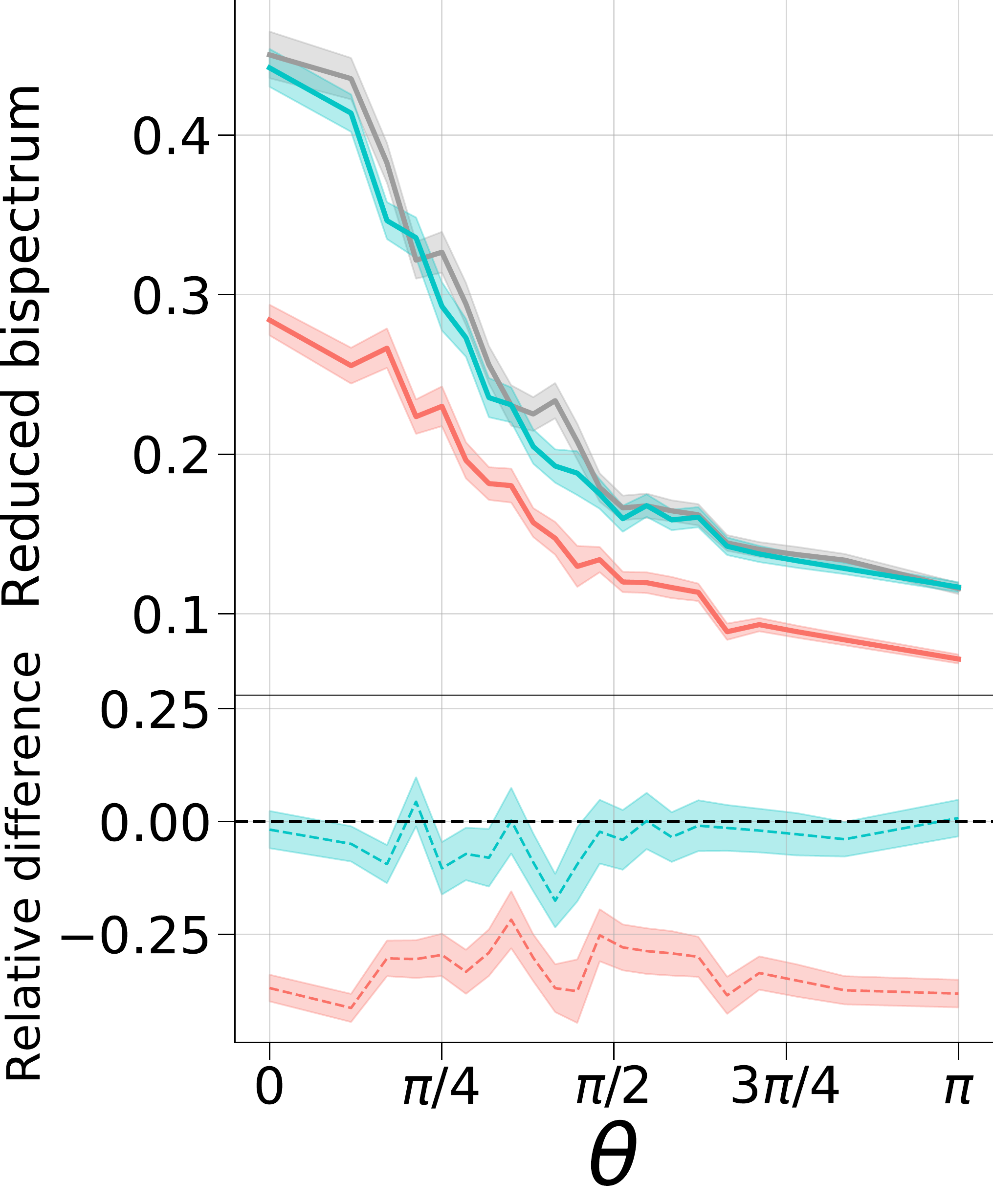}
    \caption{Reduced bispectrum with $k_1=0.1 \ h \ \rm{Mpc}^{-1}$, $k_2=0.3 \ h \ \rm{Mpc}^{-1}$}
    \end{subfigure}
\hfil
    \begin{subfigure}{0.22\linewidth}
        \includegraphics[width=\linewidth]{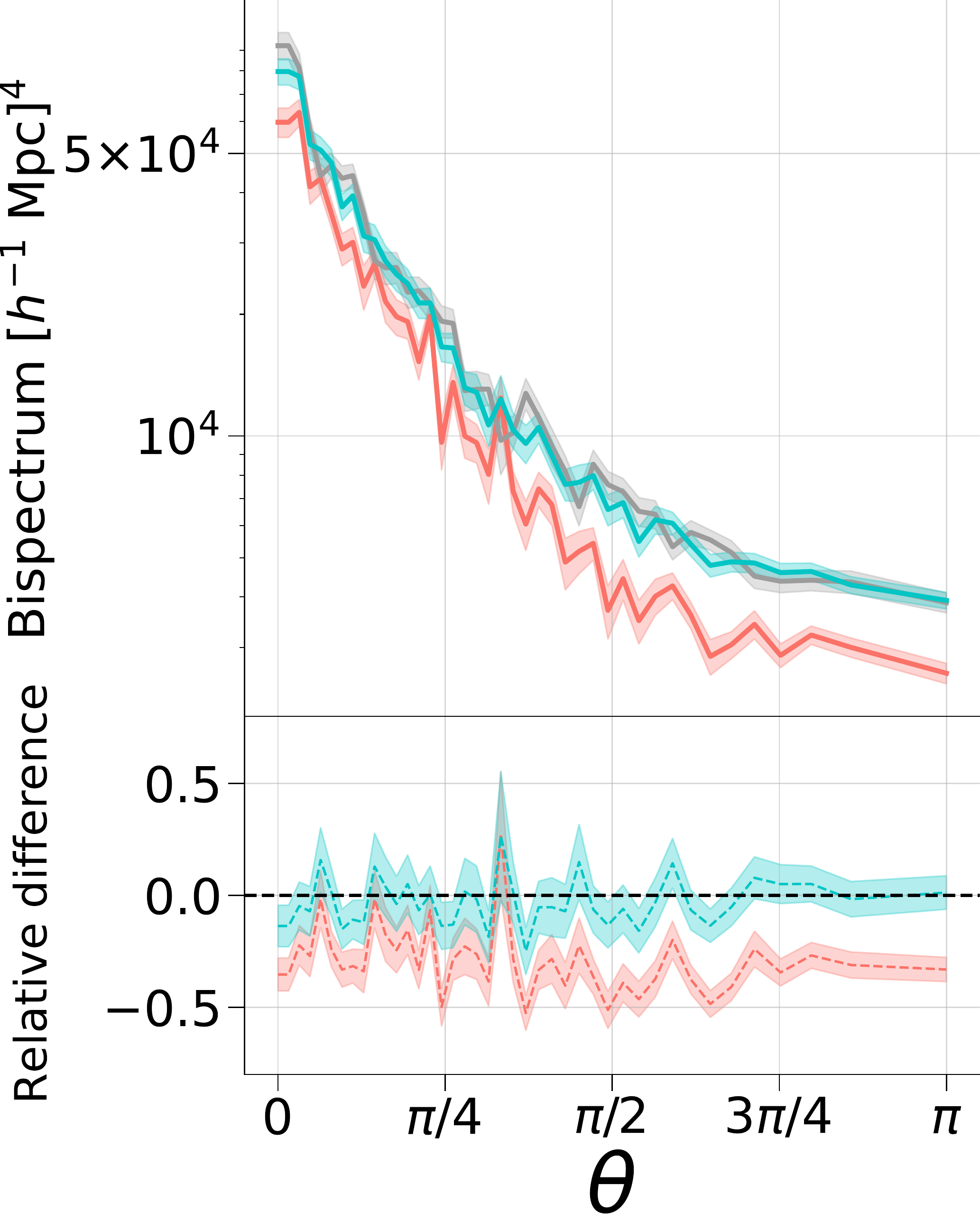}
    \caption{Bispectrum with $k_1=0.2 \ h \ \rm{Mpc}^{-1}$, $k_2=0.2 \ h \ \rm{Mpc}^{-1}$}
    \end{subfigure}
\hfil
    \begin{subfigure}{0.22\linewidth}
        \includegraphics[width=\linewidth]{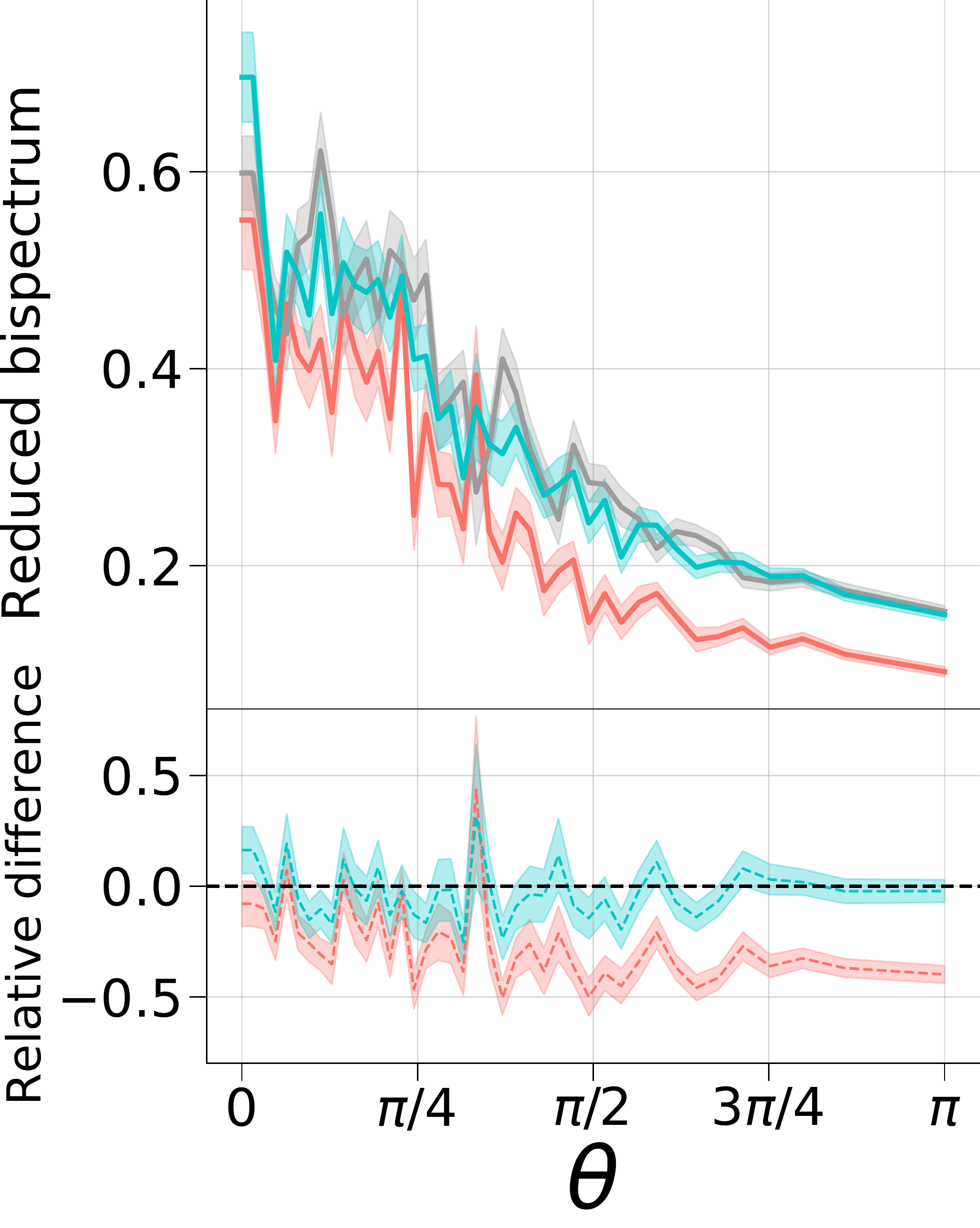}
    \caption{Reduced bispectrum with $k_1=0.2 \ h \ \rm{Mpc}^{-1}$, $k_2=0.2 \ h \ \rm{Mpc}^{-1}$}
    \end{subfigure}
\caption{Same as Fig.~\ref{fig:128_z0}, but for a model trained on latin-hypercube simulations and applied to a different cosmology, as described in Sect.~\ref{sec:zc_dep}. Except at low values of $\delta$ in panel (a), the results show good agreeement between the model predictions and the target $N$-body fields, demonstrating that our model has good generalisation performance across different cosmologies.}
    \label{fig:128_lh_lh}
\end{figure*}

\begin{figure*}
\centering
    \begin{subfigure}{0.22\linewidth}
        \includegraphics[width=\linewidth]{figures/legend.pdf}
        \vspace{2cm}
        \end{subfigure}
\hfil
    \begin{subfigure}{0.22\linewidth}
        \includegraphics[width=\linewidth]{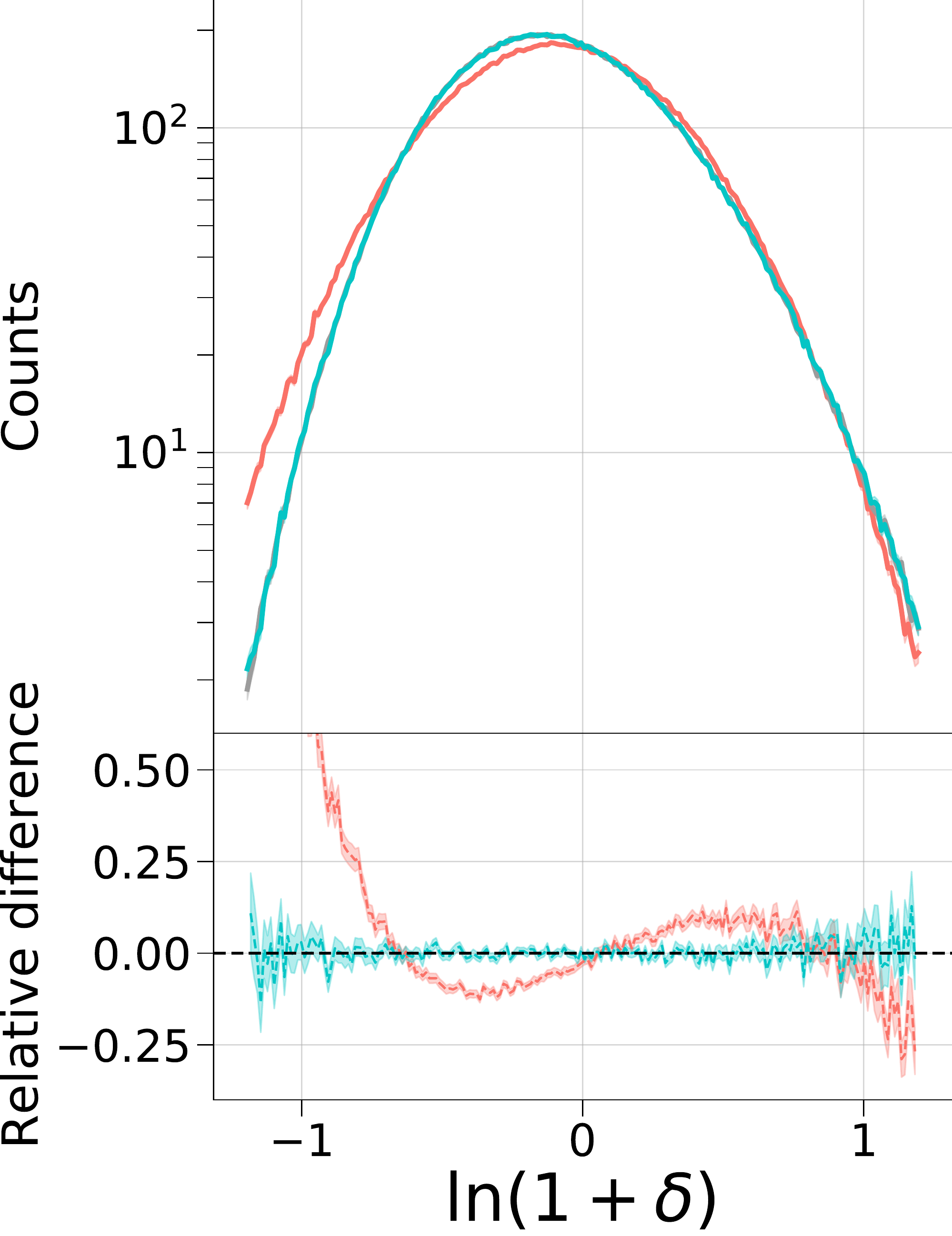}
    \caption{Pixel counts}
    \end{subfigure}
\hfil
    \begin{subfigure}{0.22\linewidth}
        \includegraphics[width=\linewidth]{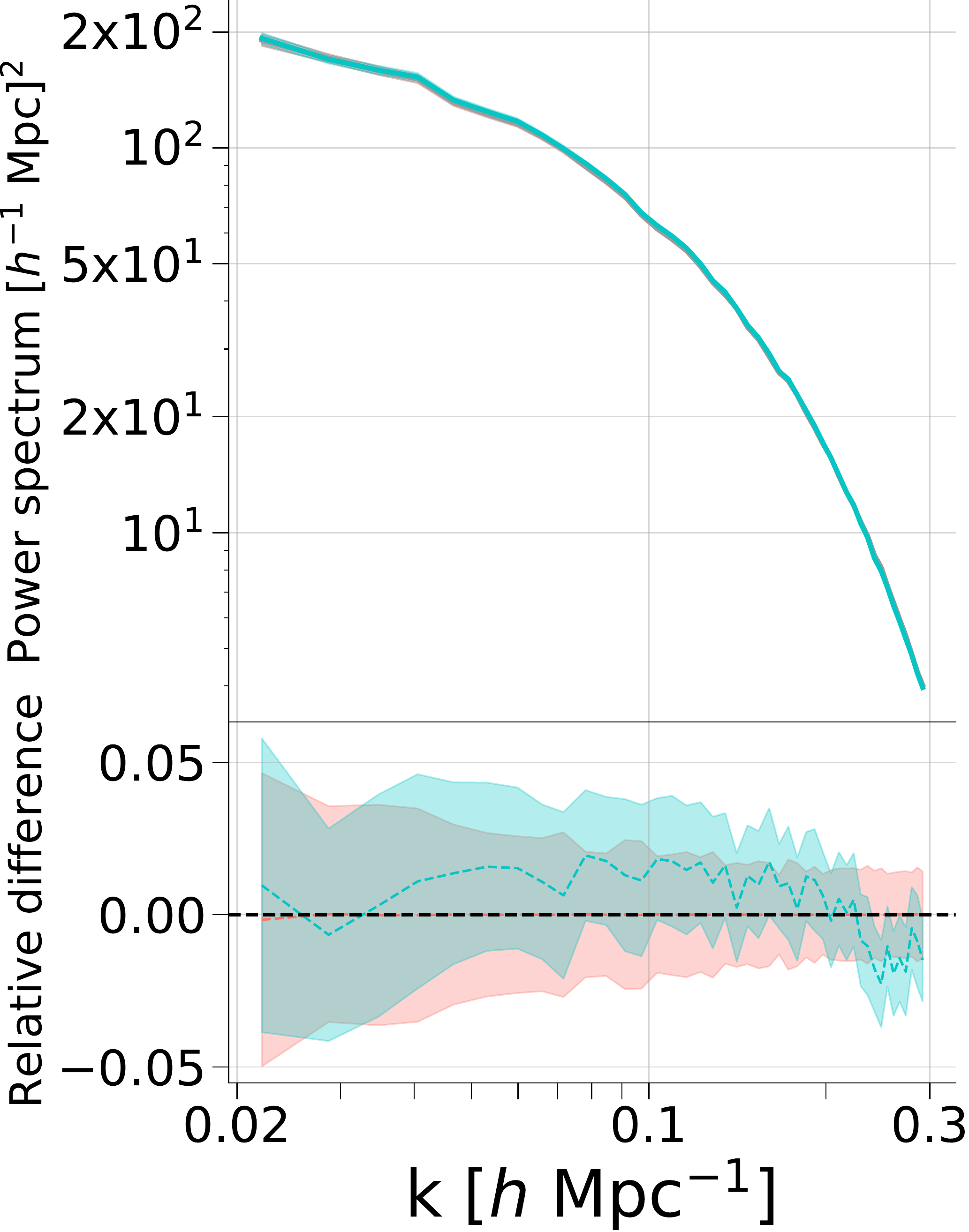}
    \caption{Power spectrum}
    \end{subfigure}
\hfil
    \begin{subfigure}{0.22\linewidth}
        \includegraphics[width=\linewidth]{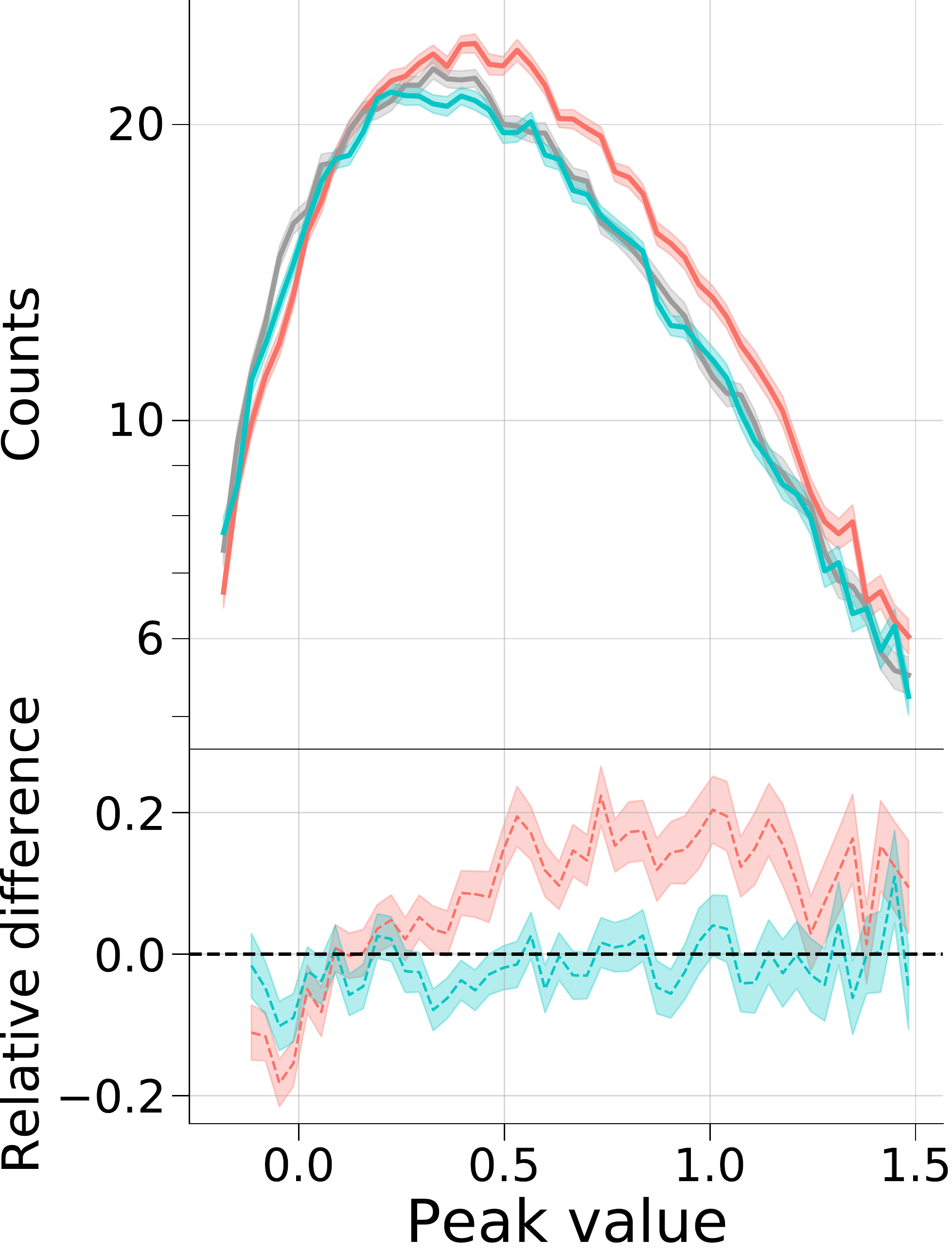}
    \caption{Peak counts}
    \end{subfigure}

    \begin{subfigure}{0.22\linewidth}
        \includegraphics[width=\linewidth]{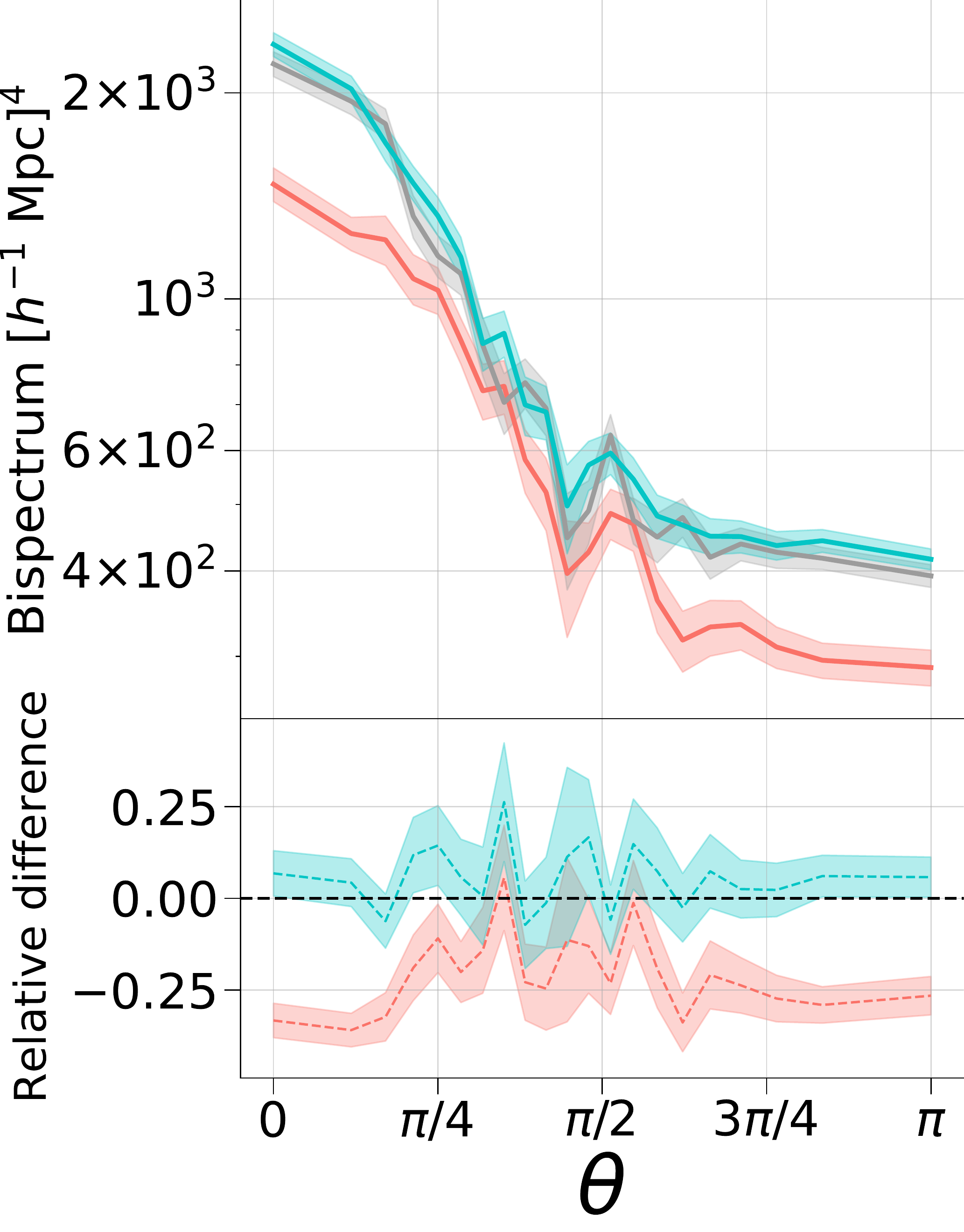}
    \caption{Bispectrum with $k_1=0.1 \ h \ \rm{Mpc}^{-1}$, $k_2=0.3 \ h \ \rm{Mpc}^{-1}$}
    \end{subfigure}
\hfil
    \begin{subfigure}{0.22\linewidth}
        \includegraphics[width=\linewidth]{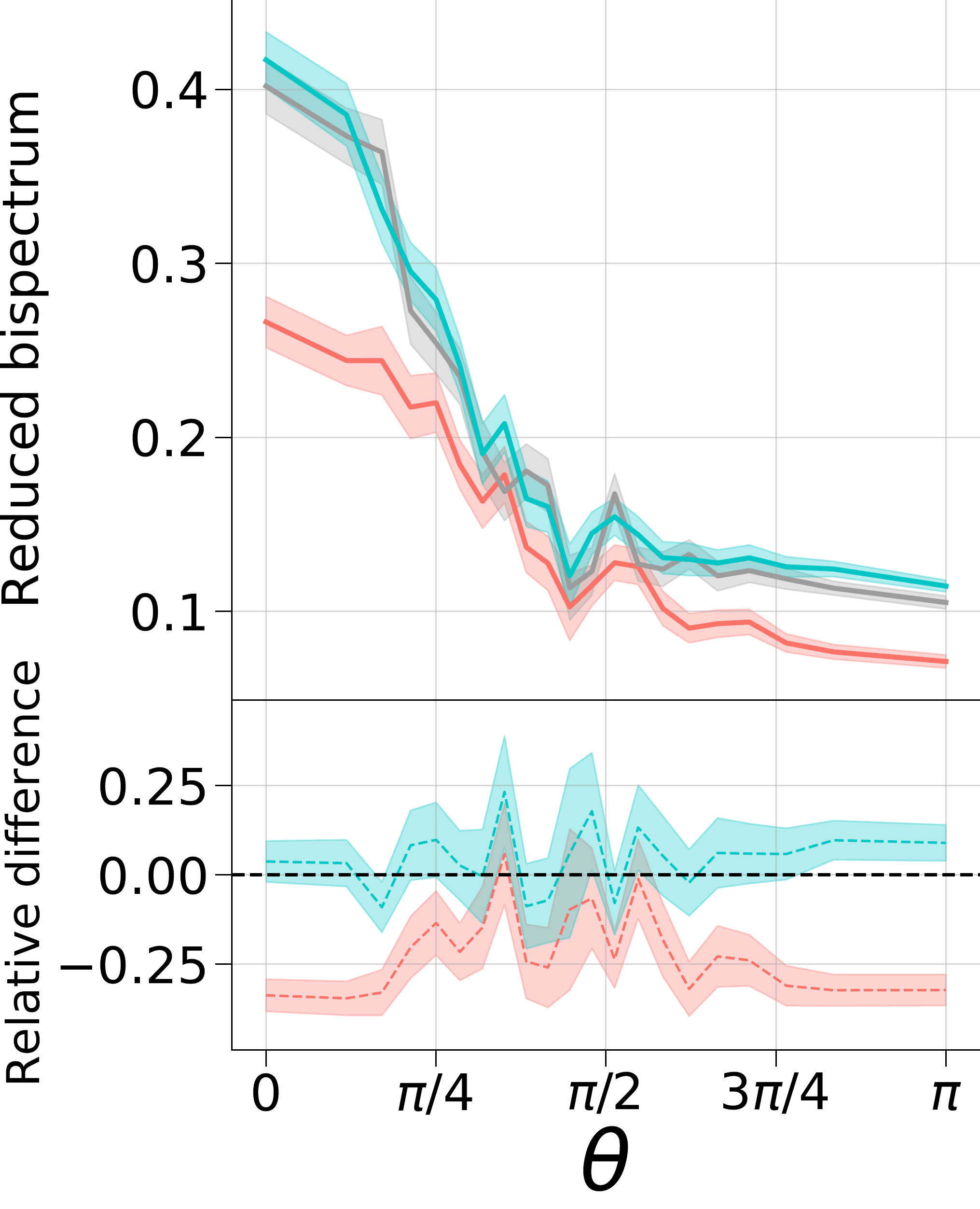}
    \caption{Reduced bispectrum with $k_1=0.1 \ h \ \rm{Mpc}^{-1}$, $k_2=0.3 \ h \ \rm{Mpc}^{-1}$}
    \end{subfigure}
\hfil
    \begin{subfigure}{0.22\linewidth}
        \includegraphics[width=\linewidth]{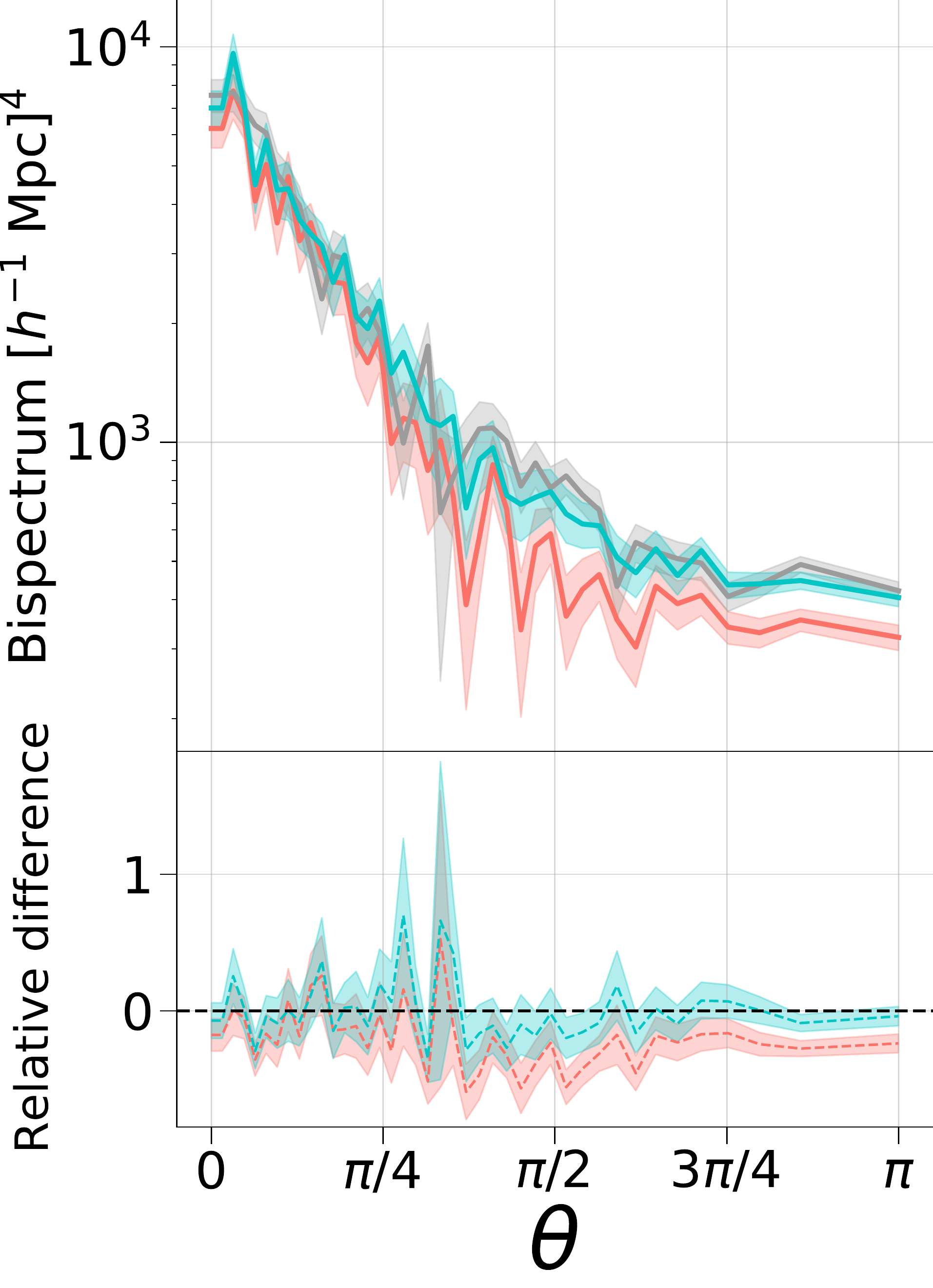}
    \caption{Bispectrum with $k_1=0.2 \ h \ \rm{Mpc}^{-1}$, $k_2=0.2 \ h \ \rm{Mpc}^{-1}$}
    \end{subfigure}
\hfil
    \begin{subfigure}{0.22\linewidth}
        \includegraphics[width=\linewidth]{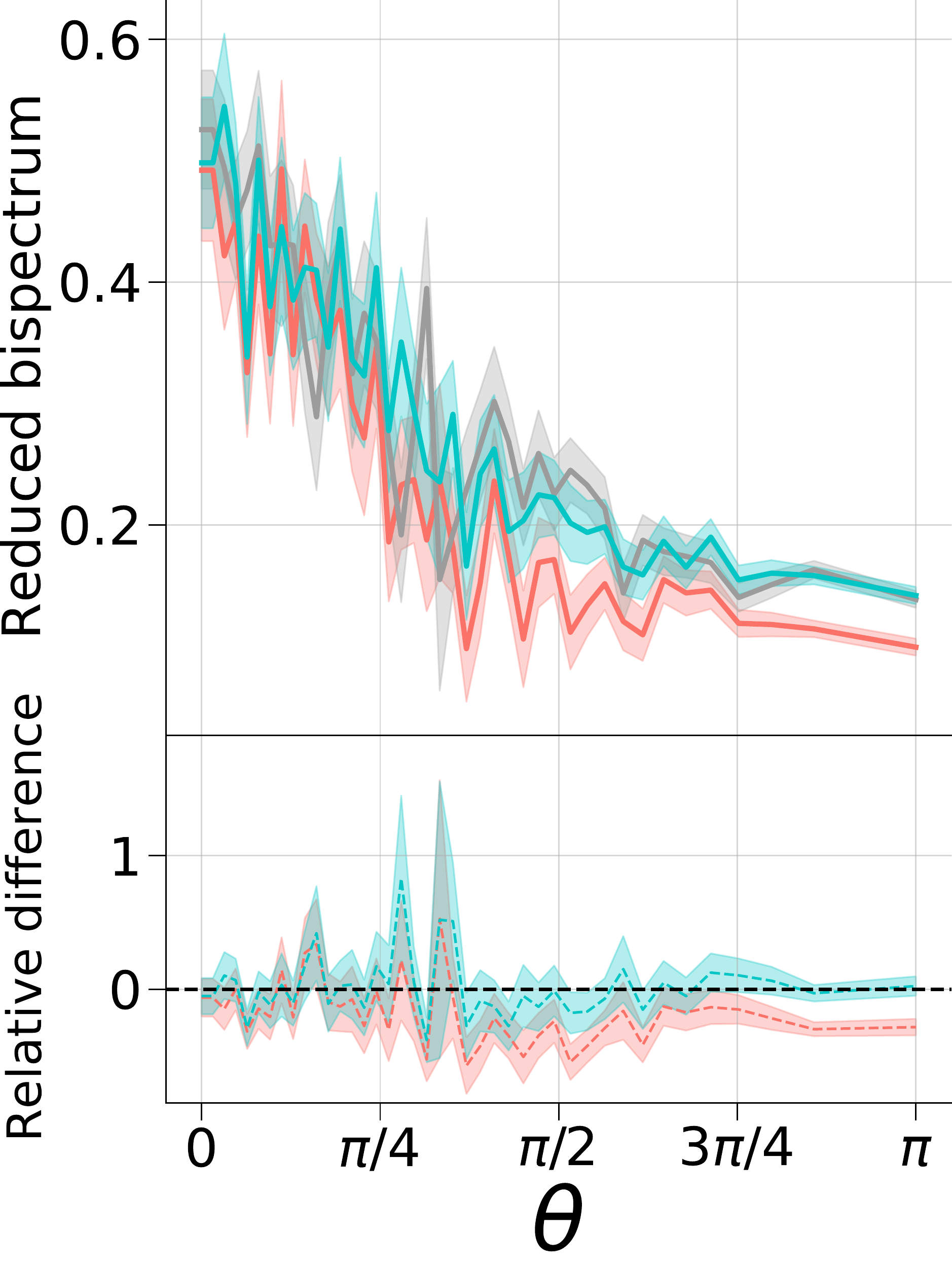}
    \caption{Reduced bispectrum with $k_1=0.2 \ h \ \rm{Mpc}^{-1}$, $k_2=0.2 \ h \ \rm{Mpc}^{-1}$}
    \end{subfigure}
\caption{Same as Fig.~\ref{fig:128_z0}, for a different model trained on data at redshift $z=1$. We observe a good overall performance of our model, which generally outperforms the lognormal approximation.
}
    \label{fig:128_z1}
\end{figure*}

\subsection{Statistics}
\label{sec:statistics}
While visual inspection of the generated maps against the target ones is a necessary zeroth-order test to provide intuition on whether the model was adequately trained, it is then fundamental to compare the summary statistics of interests and carefully quantify their agreement. We compare the generated and simulated fields through four different summary statistics, which we briefly describe here.

\subsubsection{Pixel counts}
The first test consists of binning the pixels of the generated and target density fields into a histogram. While the lognormal distribution is a good approximation of the simulated fields, there is a significant difference between the two (see e.g.\ Fig.~\ref{fig:histo_with_density}). We show in panel (a) of Fig.~\ref{fig:512_z0} and Fig.~\ref{fig:128_z0} the performance of our model with respect to the pixel counts for high and low resolution, respectively. 

\subsubsection{Power spectrum}
\label{subsec:ps}
We compare the power spectrum as defined in Eq.~(\ref{eq:ps}) for the simulated maps and the ones predicted by our model given the lognormal maps. While it could be argued that this is a trivial task (given that the input and output maps have the same power spectrum by construction), it is not obvious that our model does not modify the power spectrum information while learning the new density distribution, and as anticipated at the end of Sect.~\ref{sec:iti_translation} we actually found some failures of the trained model which yield discrepant power spectra in the high-resolution case. 
We therefore compute the power spectra and show the results in panel (b) of Fig.~\ref{fig:512_z0} and Fig.~\ref{fig:128_z0}, for high and low resolution, respectively. Since $N$-body simulations are mainly used to associate a covariance matrix to real measurements, we have also computed power spectrum covariance matrices for all datasets and models we considered. These are very similar for all cases, and while we do not show them for brevity, they further validate the performance of our model.

\subsubsection{Bispectrum}
To probe the non-Gaussian features of the density fields, we measure the matter bispectrum of the maps, i.e.\ the counterpart of the three-point matter correlation function in Fourier space. The matter bispectrum $B(k_1, k_2, k_3)$ for a 2-D field is defined implicitly as (see e.g.\ \citealp{Sefusatti06}):
\begin{equation}
    \langle \delta(\mathbf{k}_1) \delta(\mathbf{k}_2) \delta(\mathbf{k}_3) \rangle= (2\pi)^2 \delta_{\mathrm{D}}(\mathbf{k}_1+\mathbf{k}_2+\mathbf{k}_3) B(k_1, k_2, k_3) \ ,
\end{equation}
where $\delta(\mathbf{k})$ indicates the Fourier transform of the matter overdensity $\delta(\mathbf{x})$, $k_i = |\mathbf{k}_i|$, and all $\mathbf{k}_i$ vectors are in the plane of the simulation box slices. To further assess that our model correctly captured the information beyond the power spectrum, we also measure the reduced matter bispectrum $Q(k_1, k_2, k_3)$, see e.g.\ \citet{Liguori10}, defined as:
\begin{equation}
 Q(k_1, k_2, k_3) = \frac{B(k_1, k_2, k_3)}{P(k_1)P(k_2)+P(k_1)P(k_3)+P(k_2)P(k_3)} \ .
\end{equation}

We measure bispectra and reduced bispectra for different configurations depending on the resolution; different triangle configurations usually probe different inflationary models \citep{Liguori10}, and one must include as many configurations as possible to break degeneracies when inferring cosmological parameters \citep{Berge10}. Moreover, different bispectra configurations can shed light on the size of collapsing regions, as well as on the relative position of clusters and voids in the large-scale structure \citep{Munshi20}. 

We calculate bispectra and reduced bispectra based on an estimator of the binned bispectrum (see e.g.\ \citealp{Bucher16}); we consider the centroid of each bin as the value of $\mathbf{k}$ at which the bispectrum is evaluated. We report the results in panels (d)--(g) of Fig.~\ref{fig:512_z0} and Fig.~\ref{fig:128_z0} (for high and low resolution, respectively) as a function of the angle $\theta$ between the vectors $\mathbf{k}_1$ and $\mathbf{k}_2$.

\subsubsection{Peak counts}
To further assess whether the model has correctly learnt the most non-Gaussian features of the simulated density fields, we verify that the peak counts of the generated and target maps match within the error bars. A peak is defined as a density pixel which is higher than the 8 surrounding pixels. Peak count statistics have been shown to carry significant cosmological information, especially in weak lensing studies, as they trace the most dense regions \citep{Pires09, Pires12, Dietrich10, Marian11, Mainini14, Lin15a, Lin15b, Lin16c, Kacprzak16, Shan18, Martinet18, Harnois21}. We bin the peak values for both the simulated and target maps, and compare them in panel (c) of Fig.~\ref{fig:512_z0} and Fig.~\ref{fig:128_z0}, for high and low resolution, respectively.

\subsection{High resolution}
\label{sec:highres}
In Fig.~\ref{fig:512_z0} we compare the performance of the predictions of our model against the target maps, for the case with $512^2$ pixels. We run the statistical tests on 100 maps sampled from the test set as described at the end of Sect.~\ref{sec:iti_translation}; the solid lines show the mean values and the dashed areas represent the error on the mean. 

In panel (b), we show that the trained model is capable of preserving the correct power spectrum on all scales from 0.025 $h \ \rm{Mpc}^{-1}$ to 1 $h \ \rm{Mpc}^{-1}$, with percentage differences going no higher than 3\%, and always within the error bars. At the same time, the model improves on the lognormal approximation as far as the pixel counts and peak counts are concerned, with however significant differences in particular for $\delta < 0$ in the latter case. We believe that the performance in this case could be ameliorated by e.g.\ exploring different network architectures. In panels (d)-(g), we show the results for the (reduced) bispectrum, for $k_1=0.4 \ h \ \rm{Mpc}^{-1}$, $k_2=0.6 \ h \ \rm{Mpc}^{-1}$ --- panels (d) and (e) --- and for $k_1=0.4 \ h \ \rm{Mpc}^{-1}$, $k_2=0.4 \ h \ \rm{Mpc}^{-1}$ --- panels (f) and (g). The performance is very good overall, with the percentage difference between target and predicted maps being within the error bars except for a few individual values of $\theta$, significantly improving on the lognormal approximation.

\subsection{Low resolution}
\label{sec:lowres}
In Fig.~\ref{fig:128_z0} we compare the performance of the predictions of our model against the target maps, for the case with $128^2$ pixels. We use 100 randomly-sampled maps from the test set. We observe good agreement between predicted and target maps for the pixel counts, power spectrum and peak counts statistics, with the power spectrum in particular being almost always within 2\%. As far as the bispectra are concerned, we consider two configurations, one with $k_1=0.1 \ h \ \rm{Mpc}^{-1}$, $k_2=0.3 \ h \ \rm{Mpc}^{-1}$ --- panels (d) and (e) --- and for $k_1=0.2 \ h \ \rm{Mpc}^{-1}$, $k_2=0.2 \ h \ \rm{Mpc}^{-1}$ --- panels (f) and (g). Since the fields have a lower resolution, the scales we probe are larger than in Sect.~\ref{sec:highres}. We observe a good agreement overall, except at high $\theta$: we argue that to improve the performance of the model we could use the same ranking approach based on the power spectrum as in the high-resolution case.  

\subsection{Redshift and cosmology dependence}
\label{sec:zc_dep}
So far, we have shown the performance of our model on a given fiducial set of cosmological parameters and redshift, from which the training data were obtained. However, in order for the method to become practical, it is critical to assess whether the performance degrades when the model is tested on lognormal maps obtained with a different cosmology or at different $z$. Examples of good generalisation properties of machine learning models applied to cosmological problems include \citet{He19, Kaushal21, Shao22}.

We checked that the performance of our model does not degrade much when acting on fields with slightly different (within 2\%) values of $\Omega_{\rm{m}}$ and $\sigma_8$, even though a more complete analysis on bigger variations is required. We additionally verified that feeding our model, trained using maps at $z=0$, with lognormal maps at $z=0.5$ or $z=1.0$ does not yield satisfactory results, with percentage errors going well above 50\%. This failure is not unexpected: the different dynamic range of the lognormal maps at different redshifts highlights that our model is not capable of extrapolating to such different input values.

To overcome these limitations, we use the Quijote simulations run on a latin hypercube of the cosmological parameters, which are publicly available together with the simulations run at the fiducial cosmology \citep{VillaescusaNavarro20}. We consider 800 of such simulations at low resolution $N_{\rm{low}} = 128$ and $z=0$, and repeat the procedure described in Sect.~\ref{sec:training_data} to generate a dataset of highly-correlated pairs of lognormal and $N$-body fields. For each of the first 700 simulations, we keep 90 slices for training, 19 for validation and 19 for testing (not used in this instance), and train the same model described in Sect.~\ref{sec:iti_translation} for 150 epochs. Note that we do not provide explicitly the label corresponding to the different cosmological parameters during training. We select the best model according to the best performance on the summary statistics over the validation set, as in the fiducial case.

We test the best model by applying it to 100 lognormal maps generated at the fiducial cosmology (with results shown in Appendix~\ref{sec:cosmo_dependence_app}), as well as at the cosmological parameters of one random simulation from the test set ($\Omega_{\rm{m}} = 0.1987$, $\Omega_{\rm{b}} = 0.0446$, $h = 0.5601$, $n_{\rm{s}} = 1.1707$, $\sigma_8 = 0.7665$) in Fig.~\ref{fig:128_lh_lh}. These results, while slightly worse than those presented in Fig.~\ref{fig:128_z0} since e.g.\ the power spectrum shows up to 5\% discrepancies, indicate that the model trained on the latin-hypercube simulations has a good generalisation performance, which extends to cosmologies that were never shown at training or validation time. We plan to extend these encouraging results to higher resolutions and different redshifts in future work.

Other possible solutions to obtain a good generalisation performance include normalising the maps, either after or before training the model. For instance, one could rescale the field at $z=1$ through the linear growth factor \citep{Eisenstein99} to $z=0$, and then invert this transformation after feeding the map through the generator trained as above. However, this approach would ignore the non-linear scales, and directly dividing each pixel value by the linear growth factor could lead to unphysical fields with $\delta<-1$. Alternatively, instead of the dark matter overdensity field, we could consider the corresponding peak height field, calculated by measuring for each pixel $1.686/\sigma(M)$, where $\sigma(M)$ is the mass enclosed within a given scale; since the peak height is known to extrapolate better, having a weaker dependence on cosmological parameters \citep{Press74, Bond91, Percival05, Kravtsov12}, we expect a model trained on this field to have an improved generalisation performance.

Finally, we show that with our current setup we can successfully train a second model on data generated as described in Sect.~\ref{sec:training_data} with $z \neq 0$. We show the results for a model trained on fields at $z=1$, which have a lower contrast and less non-linearity than $z=0$, in Fig.~\ref{fig:128_z1} for the low-resolution case, with good performance overall. We also expect that it would be possible to train a conditional model by providing the redshift `label' together with the input lognormal map, thus obtaining a conditional WGAN-GP \citep[see e.g.][]{Mirza14, Yiu21}; such a model could be trained e.g.\ on maps with $z=0$ and $z=1$, and then used to predict maps at $z=0.5$, similarly to \citet{Chen20}. All these points indicate that it will be possible to make our model conditional on $z$ and different cosmological parameters; we defer these studies to future work.

\section{Conclusions}
\label{sec:conclusions}
In this paper, we employed the Quijote simulations as a starting point to train a machine learning model that is capable of transforming projected lognormal realisations of the dark matter density field to more realistic samples of the dark matter distribution. We employed state-of-the-art image-to-image translation techniques, combining convolutional neural networks and adversarial training, to learn such a model, and extensively validated its performance through a thorough set of statistical tests. We observed a significant reduction in the error of non-Gaussian features like peak counts and bispectra, from tens of percent for the pure lognormal model to no more than 10\% obtained by our model in most cases; the latter frequently shows an order of magnitude improvement over the former. Furthermore, the mapping is extremely fast, taking $\mathcal{O}(1 \ s)$ on a single GPU.

In order to avoid running large suites of $N$-body simulations, the proposed method has to generalise well to other redshifts and cosmologies. We demonstrated that it is possible to train a model on simulations run over a latin hypercube of cosmological parameters and have good performance on the fiducial cosmology as well as on unseen cosmologies. We outlined a few promising avenues to investigate in order to extend these results to different redshifts and to higher resolutions. Moreover, while in this work we trained different models for different resolutions of the density field, we also expect an improved model to be able to deal with a varying slice thickness.

We plan to extend this work to random fields on the sphere, and integrate it into the \textsc{FLASK} package developed in \citet{Xavier16}. We aim to extend our approach to spherical random fields by iteratively applying our model to square patches of the sky, thus providing the community with a tool to quickly generate realistic dark matter realisations that overcome the limitations of the lognormal approximation. We also plan to compare our approach to a direct generation of spherical fields by means of spherical convolutional layers, as proposed e.g.\ for mass maps in \citet{Yiu21}.

Additionally, we believe that the image-to-image technique outlined in this paper could be applied to augment analytical approximations to $N$-body simulations (like L-PICOLA, \citealp{Howlett15}, or FastPM, \citealp{Feng16}), as well as semi-analytic models of galaxies, which, in the same vein as lognormal random fields, provide a fast approximation to hydrodynamical simulations by modelling complicated baryonic processes \citep{White91, Kauffmann93, Cole94, Somerville99, Lacey01}. In such instances, one could e.g.\ train a model to learn the mapping between an $N$-body simulation augmented with semi-analytical models and the corresponding hydrodynamical simulation. We further plan to explore the possibility to employ the dataset described in this work to reduce the variance in the statistics of large-scale structure observables using a small number of expensive simulations \citep{Chartier20, Chartier21, Ding22}, as well as to replace our WGAN-GP model with either a possibly more stable GAN version \citep{Kwon21}, or with a more compact model, like the one proposed in the context of Lagrangian deep learning (LDL, \citealp{Dai21}), using graph neural networks (GNNs, see e.g.\ \citealp{Zhou18} for a review) or through normalising flows (e.g.\ FFJORD, \citealp{Grathwohl18}, or more recently TRENF, \citealp{Dai22}). This will be investigated in future work.


\section*{Acknowledgements}

We thank Ofer Lahav, Shirley Ho, John Shawe-Taylor and Ilya Feige for useful discussion and feedback on this work. We also thank the anonymous reviewer for insightful comments that led to important new tests. DP thanks William Coulton and Susan Pyne for their help with the bispectrum estimation, and Prabh Bhambra for his help with the peak count plot. DP was supported by the STFC UCL Centre for Doctoral Training in Data Intensive Science. DP was also supported by the UCL Provost’s Strategic Development Fund, and by a Swiss National Science Foundation (SNSF) Professorship grant (No. 202671). This work has been partially enabled by funding from the UCL Cosmoparticle Initiative. Part of the computation underlying this work has been enabled by the Alan Turing Institute Post-doctoral Enrichment Award. The authors are pleased to acknowledge that part of the work reported on in this paper was performed using the Princeton Research Computing resources at Princeton University which is consortium of groups led by the Princeton Institute for Computational Science and Engineering (PICSciE) and Office of Information Technology's Research Computing. We acknowledge the use of $\textsc{NumPy}$ \citep{Harris20}, $\textsc{Matplotlib}$ \citep{Hunter07}, $\textsc{TensorFlow}$ \citep{Abadi15}, and $\textsc{NN-SVG}$ \citep{LeNail19}.

\section*{Data Availability Statement}
The original data underlying this article (i.e.\ the Quijote simulations) are available through the \textit{globus} server, as detailed in the documentation 
\href{https://quijote-simulations.readthedocs.io/en/latest/access.html}{at this link}. The byproduct data obtained for this work will be shared on reasonable request to the corresponding author. The code to reproduce this work is publicly available in this GitHub repository (\verb+https://github.com/dpiras/leap_of_lognormal+, or \href{https://github.com/dpiras/leap_of_lognormal}{\faicon{github}}).



\bibliographystyle{mnras}
\bibliography{example} 




\appendix

\section{Model architecture}
\label{app:model}
In its basic formulation, a layer in a convolutional neural network (CNN; see e.g.\ \citealp{Fukushima08, Krizhevsky17}) is made of a certain number of square filters, each associated to learnable parameters, usually called weights. During training, each filter is convolved through each input data-point: this means that the dot product of the learnable weights and the input pixels is calculated, representing a single output for that particular filter. Repeating this operation while moving the filter across the input data creates an output map, which is then passed through an activation function to introduce non-linearities in the network. This operation is done for multiple filters, and each output map becomes the input to the following convolutional layer. Stacking convolutional layers allows one to extract progressively larger scales from the input data, and represents a more efficient implementation of a neural network with respect to standard dense layers when dealing with high-dimensional data like images \citep{LeCun89, Goodfellow16}. 

As anticipated, our model, depicted in Fig.~\ref{fig:architecture}, consists of two neural networks. The first neural network (the generator) contains four downsampling blocks, followed by four upsampling blocks. Each downsampling block first pads the input data assuming periodic boundary conditions, and then applies a convolution operation with 4x4 filters. There are 64 convolutional filters in the first place, and this number doubles for each block. Note that no pooling layers are present \citep{Yamaguchi90}, and we are able to reduce the dimensionality of the extracted feature maps by shifting each filter by two pixels in both directions; in other words, we set a stride of 2. The compressed map is then symmetrically upsampled using the transposed convolution operation \citep{Dumoulin18}. At each block, each feature map is concatenated with the corresponding downsampled map by simply stacking them along the last spatial axis; this is done in order to better learn the representations at each level \citep{Ronneberger15}. The activation function used after each downsampling layer is the rectified linear unit \citep[ReLU,][]{Glorot11}, while for the upsampling blocks we found the leaky ReLU \citep{Maas13} with $\alpha=0.3$ to perform better. A final convolutional layer with linear activation function outputs the generated map $\delta_{\rm{GEN}}$. Note that all downsampling and upsampling blocks include batch normalisation \citep{Ioffe15}, which during training subtracts the batch mean and divides by the batch standard deviation, in order to make the training procedure more stable. The second neural network is done similarly, with three downsampling blocks followed by two convolutional layers with Leaky ReLU as the activation function, and a final dense layer with a single output and a linear activation function. Input and output shapes for each layer are reported in Table~\ref{tab:architecture}. We implement our neural networks in \textsc{TensorFlow} \citep{Abadi15}, and will make the trained models available upon acceptance of this work.

\begin{table}
	\centering
	\caption{Size of each layer's output in the generator and the critic neural networks, detailed in Sect.~\ref{sec:iti_translation} and Appendix~\ref{app:model}, for the high-resolution case. The low-resolution architecture is built analogously.}
	\label{tab:architecture}
	\begin{tabular}{c||c||c} 
		\hline
		 & \textbf{Size} & \textbf{Comments} \\
		\hline \hline
		 & 512$\times$512$\times$1 & Input size; $\delta_{\rm{LN}}$ \\
		 & 256$\times$256$\times$64 & \\
		 & 128$\times$128$\times$128 &\\
		 & 64$\times$64$\times$256 &\\
		 \smash{\raisebox{-1.48ex}{ \textbf{Generator}}} & 32$\times$32$\times$512 &\\
         & 64$\times$64$\times$512 & First upsampling step \\
		 & 128$\times$128$\times$256 & \\
		 & 256$\times$256$\times$128 &\\
		 & 512$\times$512$\times$64 &\\
		 & 512$\times$512$\times$1 & Linear activation; $\delta_{\rm{GEN}}$\\
		\hline \hline
		 & 512$\times$512$\times$1 & Input size; either $\delta_{\rm{SIM}}$ or $\delta_{\rm{GEN}}$ \\
		 & 256$\times$256$\times$32 & \\
         & 128$\times$128$\times$64 &\\
		\smash{\raisebox{-1.2ex}{ \textbf{Critic}}} & 64$\times$64$\times$128 &\\
		 & 63$\times$63$\times$512 &\\
		 & 62$\times$62$\times$1 &\\
		 & 3844 & Flattening; input to dense layer\\
		 & 1 &\\		 
		\hline
	\end{tabular}
\end{table}

\section{High-resolution model failures}
\label{sec:prob_spectra}
\begin{figure*}[b!]
	\includegraphics[width=2\columnwidth]{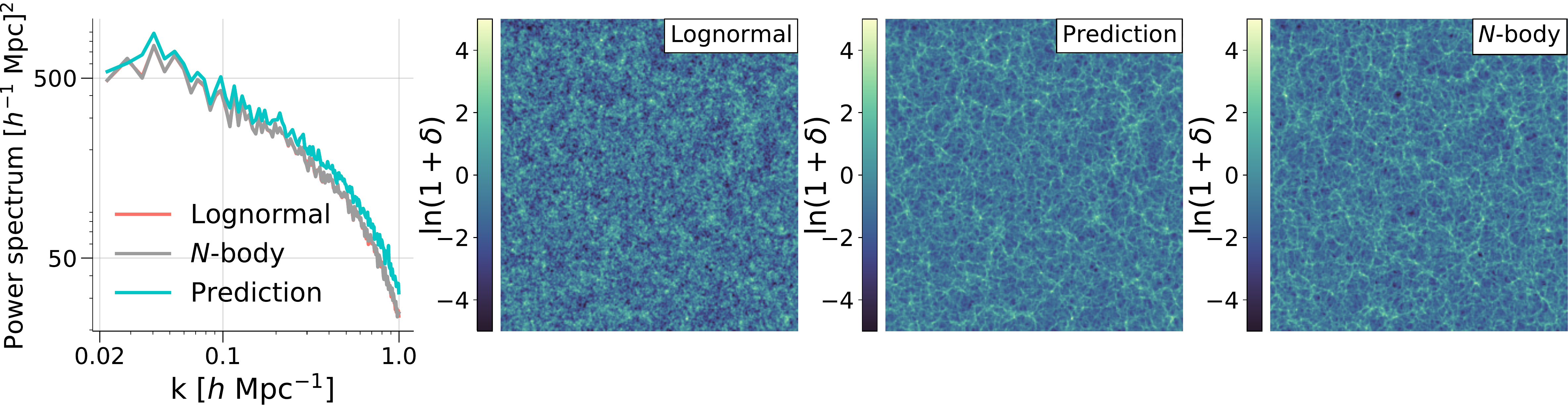}
    \caption{High-resolution example where the power spectrum, shown in the left panel for the three fields in the other panels, is significantly in disagreement (more than 20\%) between the prediction (cyan) and the $N$-body (grey). The lognormal power spectrum (red) agrees with the $N$-body one by construction. Given these failure modes, we employed the ranking system described in Sect.~\ref{sec:iti_translation} to select the predictions whose power spectrum agrees with their lognormal counterpart.}
    \label{fig:fail}
\end{figure*}

In Fig.~\ref{fig:fail}, we show an example field in the high-resolution case for which the prediction has a power spectrum with an average 20\% disagreement with respect to the expected one. We attribute such problems to instabilities in the WGAN-GP model we considered, and describe a possible ranking system that addresses this problem in Sect.~\ref{sec:iti_translation}.

\section{Mode collapse}
\label{sec:mode_collapse}
As we explained in Sect.~\ref{sec:intro} and Sect.~\ref{sec:iti_translation}, our choice of providing the model a lognormal field as input, as well as the choice of the WGAN-GP loss function, should prevent the model from memorising the training set, or only focus on single modes of the data. Following \citet{Mustafa19}, we provide evidence that the generator is capable of producing diverse maps and is not affected by mode collapse.

\begin{figure*}
\centering
    \begin{subfigure}{\linewidth}
        \includegraphics[width=\linewidth]{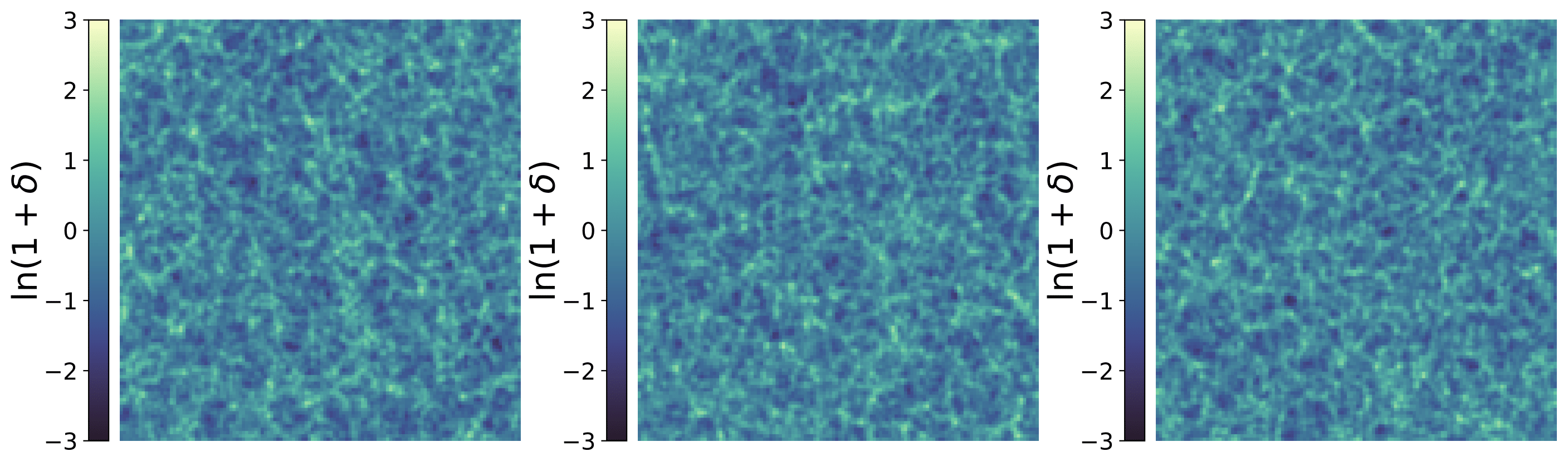}
        \caption{Random predictions from the test set. $N_{\rm{low}} = 128$, $z=0$, fiducial cosmology.}
        \end{subfigure}
\hfil
    \begin{subfigure}{\linewidth}
        \includegraphics[width=\linewidth]{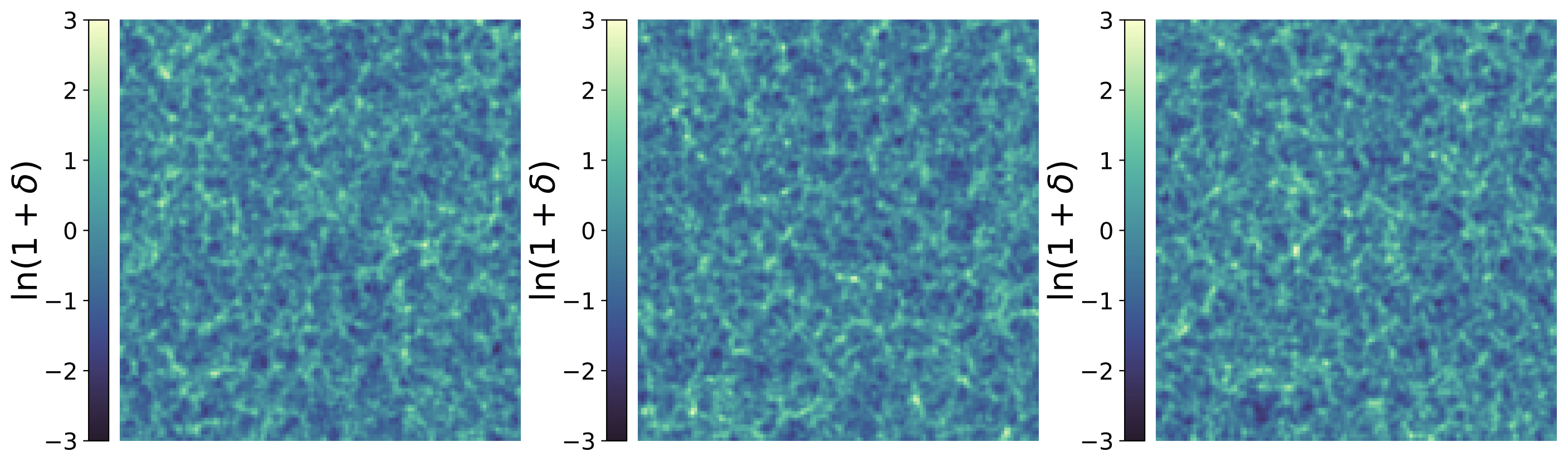}
    \caption{Closest maps in the training set. $N_{\rm{low}} = 128$, $z=0$, fiducial cosmology.}
    \end{subfigure}
\caption{\textit{First row}: three random predictions from the test set, in the low-resolution case at $z=0$. \textit{Second row}: each panel represents the field in the training set which is closest to the corresponding panel in the first row according to pixel-wise distance. As can be seen, while the maps have the same texture (and summary statistics in agreement, not shown here for brevity), they are visually different, thus confirming that our model is not affected by mode collapse.}
    \label{fig:mode_collapse}
\end{figure*}

In Fig.~\ref{fig:mode_collapse}, we show three random predictions from the test set, and the closest map in the training set according to pixel-wise distance; we focus on low-resolution data only to limit the computational cost. Despite showing similar texture, and having summary statistics in agreement (not shown for brevity), the maps are clearly different, thus confirming that our model is immune to mode collapse.

\section{Latin-hypercube model applied to fiducial cosmology}
\label{sec:cosmo_dependence_app}
\begin{figure*}[h]
\centering
    \begin{subfigure}{0.22\linewidth}
        \includegraphics[width=\linewidth]{figures/legend.pdf}
        \vspace{2cm}
        \end{subfigure}
\hfil
    \begin{subfigure}{0.22\linewidth}
        \includegraphics[width=\linewidth]{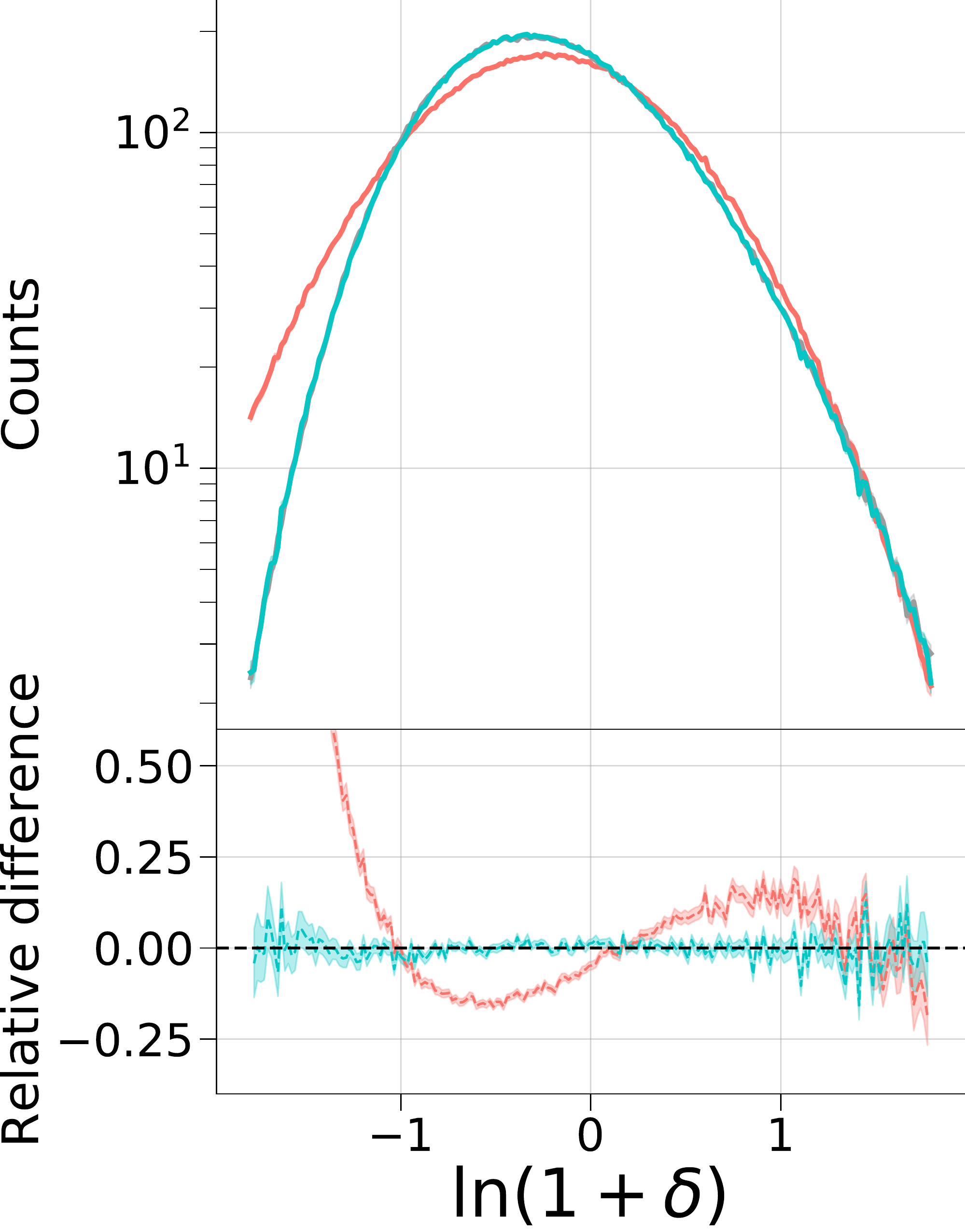}
    \caption{Pixel counts}
    \end{subfigure}
\hfil
    \begin{subfigure}{0.22\linewidth}
        \includegraphics[width=\linewidth]{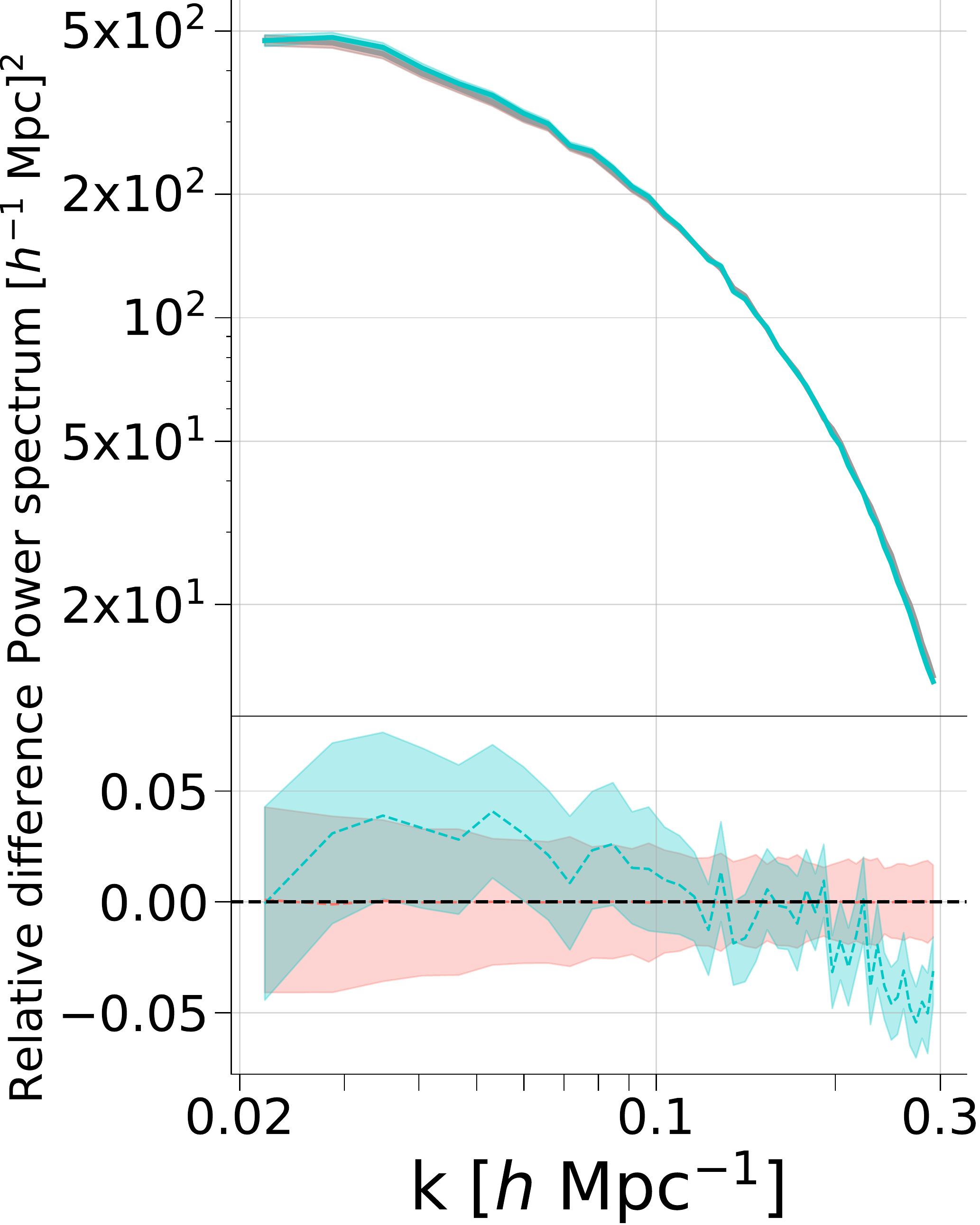}
    \caption{Power spectrum}
    \end{subfigure}
\hfil
    \begin{subfigure}{0.22\linewidth}
        \includegraphics[width=\linewidth]{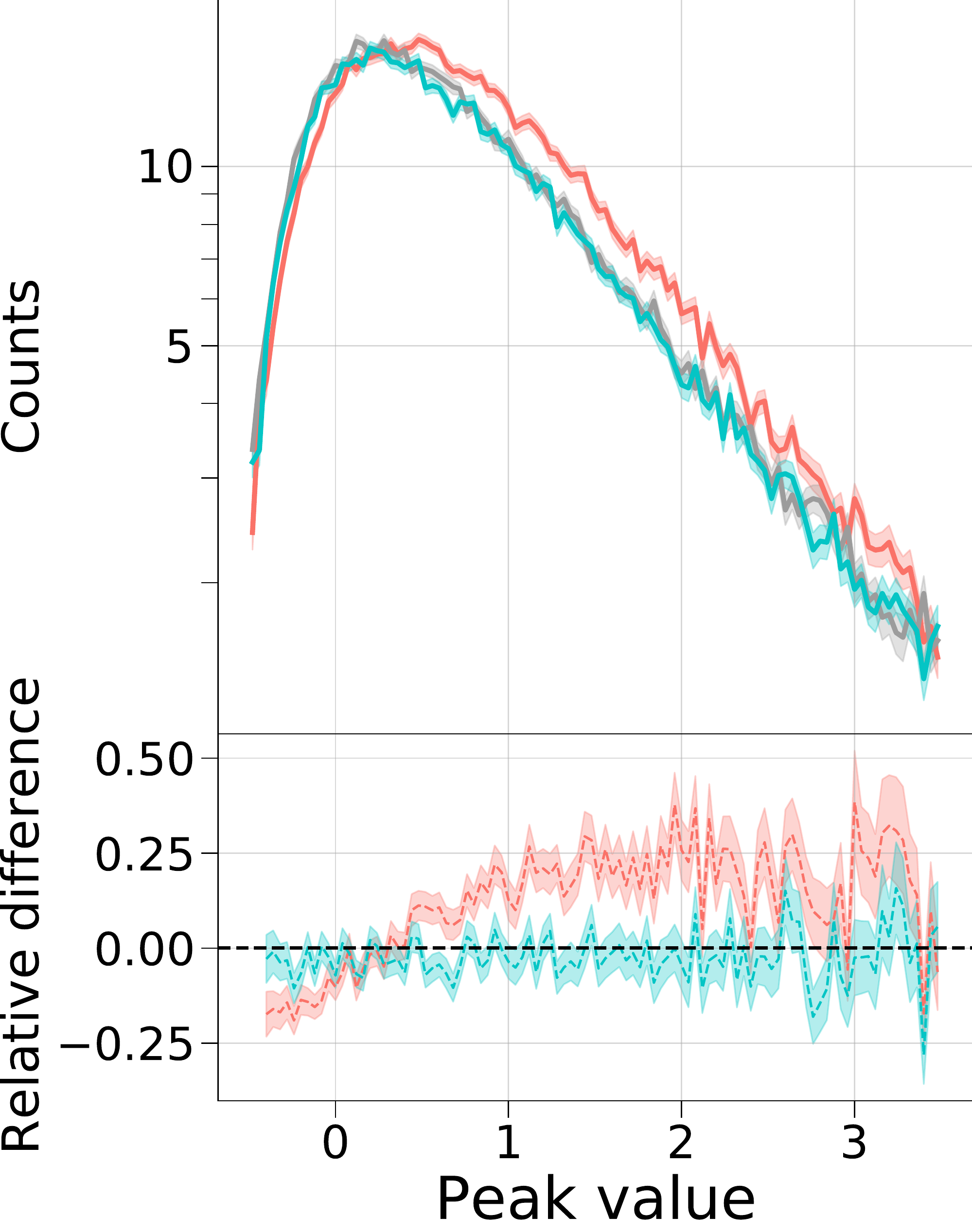}
    \caption{Peak counts}
    \end{subfigure}

    \begin{subfigure}{0.22\linewidth}
        \includegraphics[width=\linewidth]{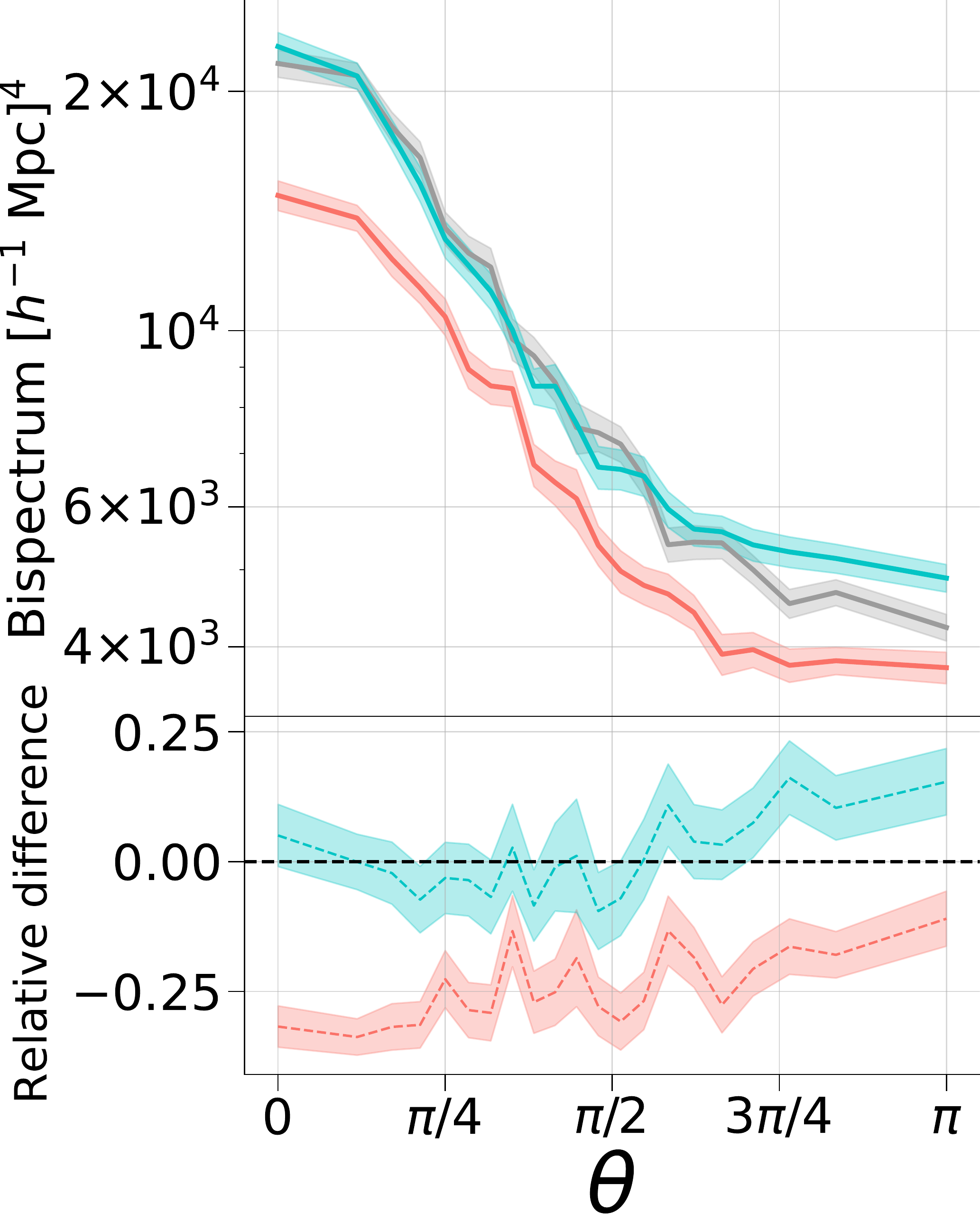}
    \caption{Bispectrum with $k_1=0.1 \ h \ \rm{Mpc}^{-1}$, $k_2=0.3 \ h \ \rm{Mpc}^{-1}$}
    \end{subfigure}
\hfil
    \begin{subfigure}{0.22\linewidth}
        \includegraphics[width=\linewidth]{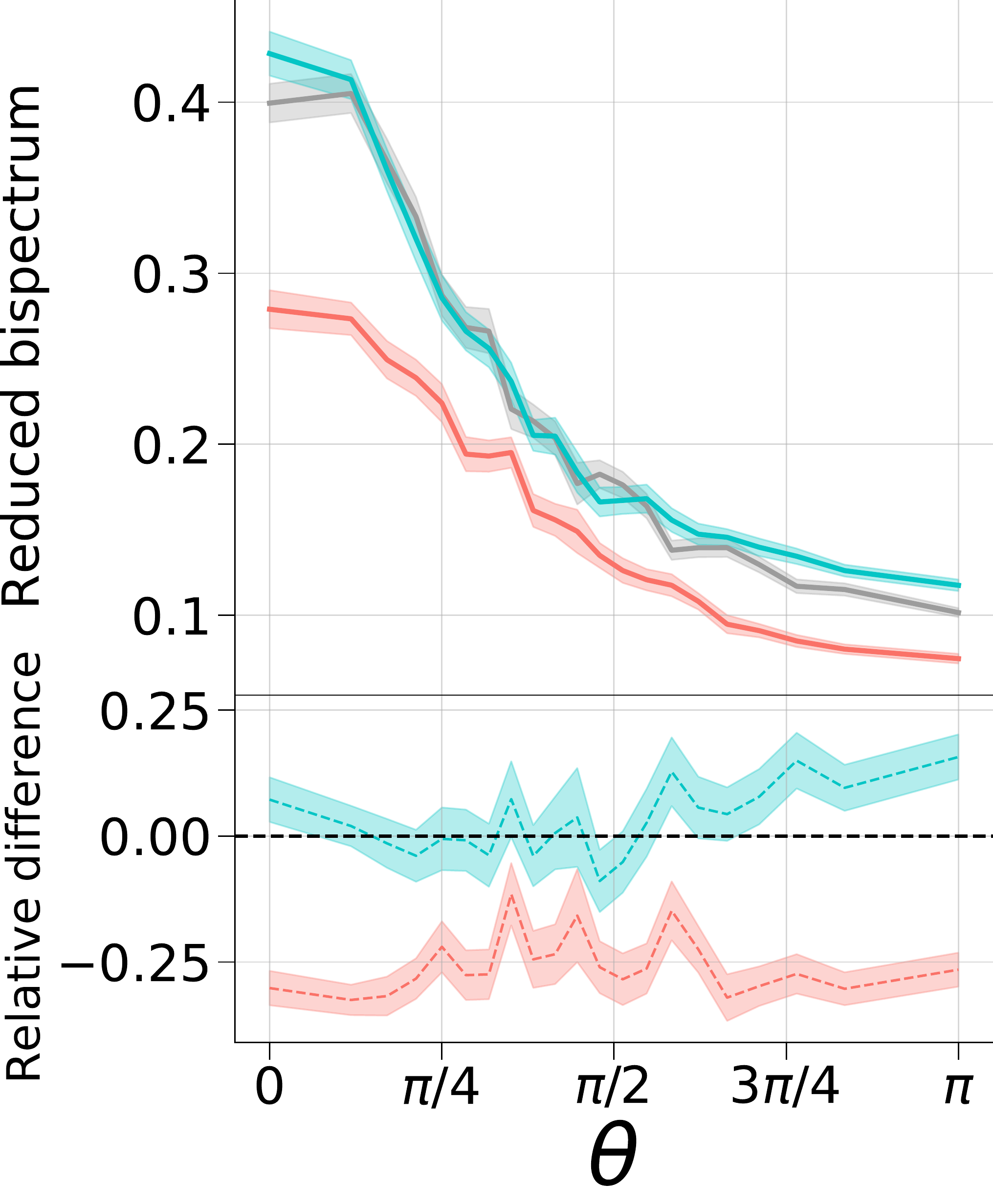}
    \caption{Reduced bispectrum with $k_1=0.1 \ h \ \rm{Mpc}^{-1}$, $k_2=0.3 \ h \ \rm{Mpc}^{-1}$}
    \end{subfigure}
\hfil
    \begin{subfigure}{0.22\linewidth}
        \includegraphics[width=\linewidth]{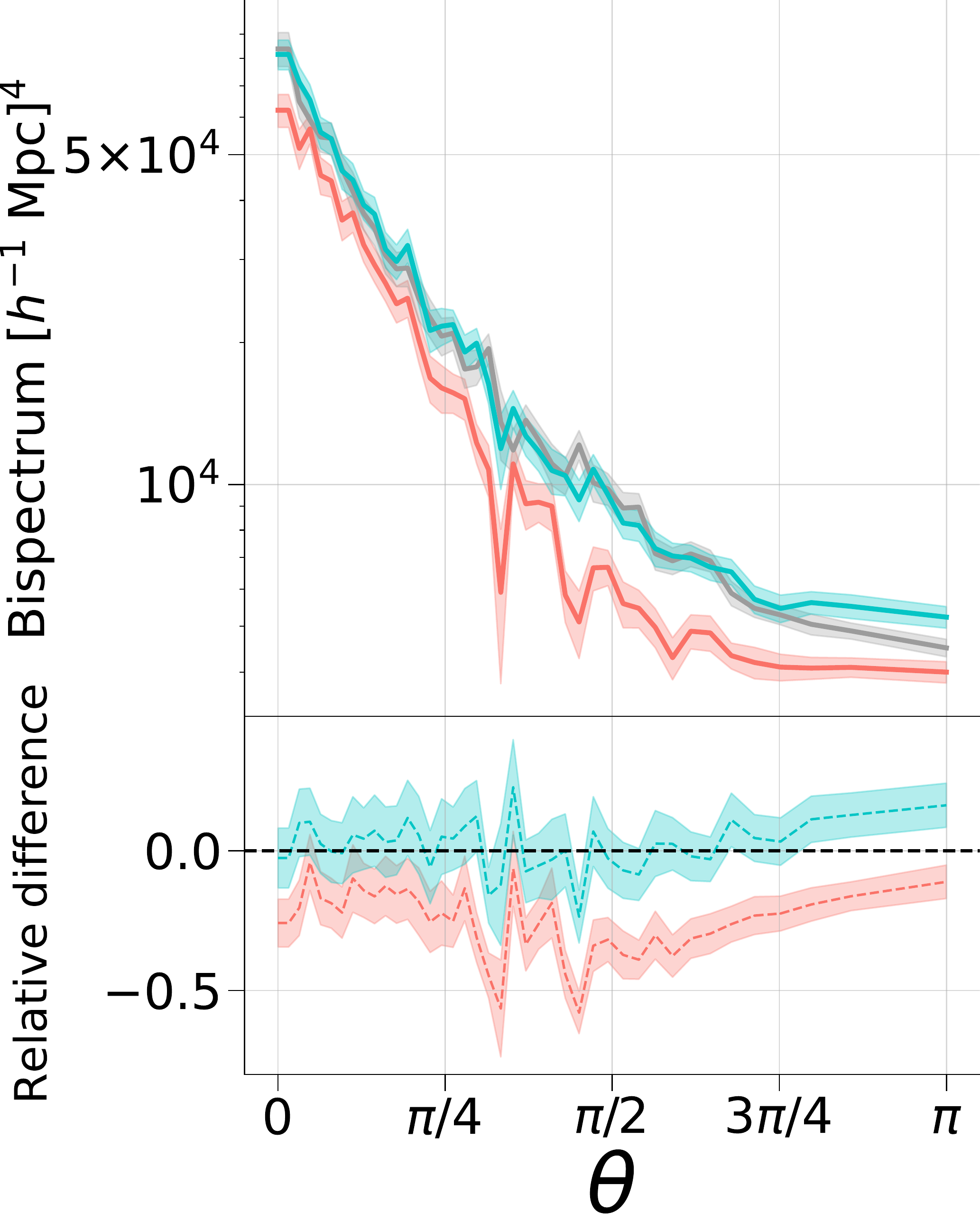}
    \caption{Bispectrum with $k_1=0.2 \ h \ \rm{Mpc}^{-1}$, $k_2=0.2 \ h \ \rm{Mpc}^{-1}$}
    \end{subfigure}
\hfil
    \begin{subfigure}{0.22\linewidth}
        \includegraphics[width=\linewidth]{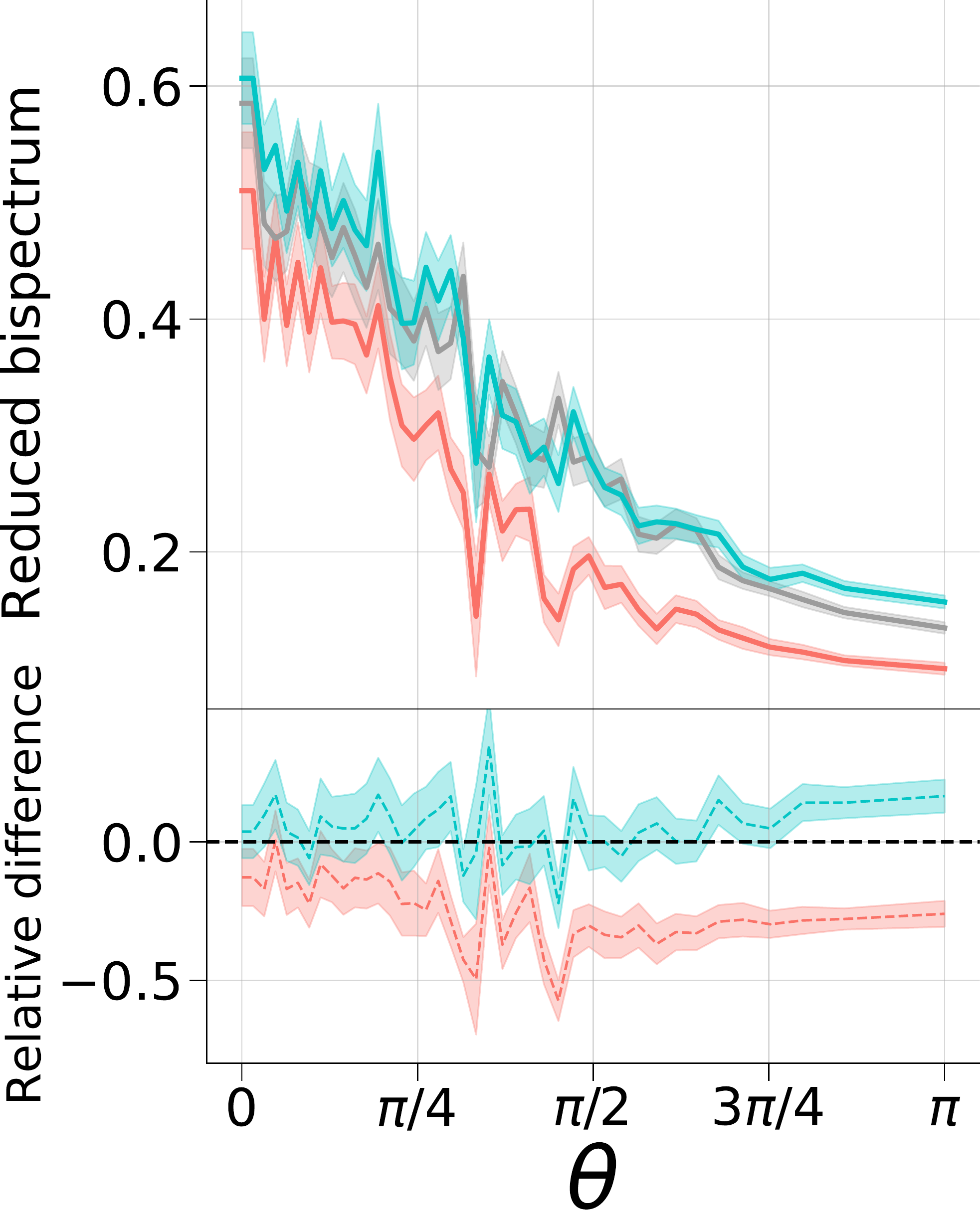}
    \caption{Reduced bispectrum with $k_1=0.2 \ h \ \rm{Mpc}^{-1}$, $k_2=0.2 \ h \ \rm{Mpc}^{-1}$}
    \end{subfigure}
\caption{Same as Fig.~\ref{fig:128_z0}, but in this instance the model was trained on latin-hypercube simulations, as described in Sect.~\ref{sec:zc_dep}. Despite a small degradation in performance, visible e.g. at high $k$ for the power spectrum, there is good agreement overall, further validating our approach.}
    \label{fig:128_lh_fid}
\end{figure*}

In Fig~\ref{fig:128_lh_fid} we show the summary statistics results for the model trained on latin-hypercube simulations and applied to fields at the fiducial cosmology, as described in Sect.~\ref{sec:zc_dep}. The redshift is fixed at $z=0$.

\bsp	
\label{lastpage}
\end{document}